\documentclass[usenatbib]{mnras}

\usepackage{graphicx}	
\usepackage{amsmath}	
\usepackage{amssymb}	
\usepackage{multicol}        
\usepackage{bm}		
\usepackage{pdflscape}	
\usepackage{longtable}

\newcommand{\kms}{\,km\,s$^{-1}$} 
\def\bl{\hbox{$B_l$}}
\newcommand{\pstar}{\hbox{$P_{\rm rot}$}}
\def\vsini{\hbox{$v \sin i$}}
\def\mstar{\hbox{$M_{\star}$}} 
\def\rstar{\hbox{$R_{\star}$}} 
\def\msun{\hbox{${\rm M}_{\odot}$}} 
\def\rsun{\hbox{${\rm R}_{\odot}$}} 

\usepackage[T1]{fontenc}
\usepackage{ae,aecompl}

\usepackage{txfonts}

\title[SPIRou Input Catalog: Activity]{\textit{SPIRou Input Catalog: Activity, Rotation and Magnetic Field of Cool Dwarfs}}

\author[C. Moutou et al]{\parbox[t]{\textwidth}{C. Moutou$^{1,2}$, E.M. H\'ebrard$^3$, J. Morin$^{4}$, L. Malo$^{5}$, P. Fouqu\'e $^{1,6}$, A. Torres-Rivas$^{1,7}$, E. Martioli$^{8}$, X. Delfosse$^{9}$, E. Artigau$^{5}$, R. Doyon$^{5}$}
\vspace{0.5cm} \\
$^{1}$ CFHT Corporation, 65-1238 Mamalahoa Hwy, Kamuela, Hawaii 96743, USA\\
$^2$ Aix Marseille Universit\'e, CNRS, LAM, Laboratoire d'Astrophysique de Marseille, Marseille, France\\
$^{3}$ Department of Physics and Astronomy, York University, 4700 Keele St., Toronto, Ontario, M3J 1P3 Canada\\
$^{4}$ LUPM, Universit\'e de Montpellier, CNRS, Place Eug\`ene Bataillon, F-34095 Montpellier, France\\
$^{5}$ Institute for Research on Exoplanets, D\'epartement de physique, Universit\'e de Montr\'eal, CP 6128, Succursale Centre-Ville, Montr\'eal, Quebec H3C 3J7, Canada \\
$^{6}$ Universit\'e de Toulouse, UPS-OMP, IRAP, 14 avenue Belin, F-31400 Toulouse, France\\
$^{7}$ D\'epartement de physique, Universit\'e de Sherbrooke, 2500 boulevard de l'Universit\'e, Sherbrooke, Quebec J1K 2R1, Canada \\
$^{8}$ Laboratorio Nacional de Astrofisica (LNA/MCTI), Rua Estados Unidos, 154, Itajuba, MG, Brazil\\
$^{9}$ Universit\'e Grenoble Alpes, CNRS, IPAG, F-38000 Grenoble, France }

\date{}

\pubyear{2017}

\begin{document}
\label{firstpage}
\pagerange{\pageref{firstpage}--\pageref{lastpage}}
\maketitle

\begin{abstract}
Based on optical high-resolution spectra obtained with CFHT/ESPaDOnS, we present new measurements of activity and magnetic field proxies of 442 low-mass K5-M7 dwarfs. The objects were analysed as potential targets to search for planetary-mass companions with the new spectropolarimeter and high-precision velocimeter, SPIRou. We have analysed their high-resolution spectra in an homogeneous way: circular polarisation, chromospheric features, and Zeeman broadening of the FeH infrared line. The complex relationship between these activity indicators is analysed: while no strong connection is found between the large-scale and small-scale magnetic fields, the latter relates with the non-thermal flux originating in the chromosphere.

We then examine the relationship between various activity diagnostics
and the optical radial-velocity jitter available in the literature, especially for planet host stars. We use this to derive for all stars an activity merit function (higher for quieter stars) with the goal of identifying the most favorable stars where the radial-velocity jitter is low enough for planet searches.  We find that the main contributors to the RV jitter are the large-scale magnetic field and the chromospheric non-thermal emission.

In addition, three stars (GJ~1289, GJ~793, and GJ~251) have been followed along their rotation using the spectropolarimetric mode, and we derive their magnetic topology. These very slow rotators are good representatives of future SPIRou targets. They are compared to other stars where the magnetic topology is also known. The poloidal component of the magnetic field is predominent in all three stars. 
\end{abstract}




\section{Introduction} 
Due to their low mass, M dwarfs are favorable to exoplanet searches with the radial-velocity (RV) method. The main reason is that the RV signal of a planet of a given mass and period increases with decreasing stellar mass. In addition, for a given surface equilibrium planet temperature, the orbital period is much shorter when the parent star is a small, low-luminosity star. Thus, telluric planets in the habitable zone of their parent stars have a more prominent RV signal when this host is an M dwarf compared to any other spectral type. 

In addition, such planets are seemingly frequent in the solar vicinity: radial-velocity survey of a hundred M dwarfs \citep{Bonfils13} showed that 36\% (resp., 52\%) M dwarfs have a planet in the mass range of 1 to 10 Earth mass and for orbital periods of 1-10 days (resp., 10-100 days) . Using a different method and a separate target sample, the $Kepler$ survey has measured a planet occurrence rate of 2.5$\pm$0.2 per M star, in the radius range of 1-4 Earth radii and period less than 200 days \citep{Dressing15} and a fraction of $\sim$ 50\% of M stars having a 1-2 Earth radii planet. The comparison of these occurrence rates depends on the mass-radius relationship of these planets, but they agree qualitatively and point to an abundant population of exoplanets, mostly of small size or mass.

It is, however, expected that exoplanet searches around M dwarfs are highly impacted by the surface activity of the parent stars. The stellar modulation may mimic a planetary signal, as shown by, \emph{e.g.}, \citet{Bonfils07} or \citet{Robertson14,Robertson15} or it may affect the mere detection of the planetary signal, as recently demonstrated by \citet{Dumusque17}. Furthermore, the jitter on M dwarfs, when not properly filtered out, results in major deviations in the measurement of orbital periods, planet minimum mass and/or eccentricity of the detected planets, as modeled by \citet{Andersen15}. 
Thus, exoplanet RV surveys aiming at M host stars require a thorough understanding of the processes that induce intrinsic stellar RV modulation. 

Covering a range of mass from 0.08 \msun\ to about 0.50 \msun, \emph{i.e.}, a factor of 6 in mass, and having the longest evolution history of all stars, the M dwarfs may encompass very different types of stellar surfaces and be dominated by a wide variety of phenomena. This can be particularly true in the transition from partly to fully convective interiors at the mass of 0.35\msun\ \citep{Chabrier00}. Many previous studies have explored the activity features of M dwarfs, including X-ray observations \citep{Stelzer13}, photometric variations \citep{Newton16}, long-term RV variations \citep{DaSilva11}, H$\alpha$, CaII, and rotation measurements \citep{Reiners2012a,Suarez15,Astudillo17b, Maldonado17, Scandariato17}, UV emission \citep{Shkolnik14}, surface magnetic field modulus \citep{Reiners07, Shulyak14} and large-scale magnetic field geometry \citep[][and references therein]{Donati06a,Morin10}. This richness of observed activity features illustrates the complexity of phenomena of magnetic origin in M dwarfs and offers complementary constraints to dynamo and convection modeling.

Concerning magnetic field measurements, we may have two different and complementary diagnostics: \begin{itemize}
\item  The Zeeman broadening in unpolarized spectra is sensitive to the magnetic field modulus but almost insensitive to the field spatial distribution or orientation. Modelling based on Zeeman broadening generally assesses a quantity called "magnetic flux" which corresponds to the product of the local magnetic field modulus $B$ with the filling factor $f$ in a simple model where a fraction $f$ of the surface is covered by magnetic regions of uniform modulus $B$.
\item  Zeeman-induced polarisation in spectral lines is sensitive to the vector properties of the magnetic fields. But due to the cancellation of signatures originating from neighbouring regions of opposite polarities, it can only probe the large-scale component of stellar magnetic fields. Using a time-series of polarised spectra sampling at least one rotation period, it is possible to recover information on the large-scale magnetic topology of the star, see section~3.3.
\end{itemize}

Whereas M dwarfs exhibiting the highest amplitude of activity have been more studied in magnetic-field explorations, the exoplanet RV searches will tend to focus on the intermediate to low activity stars, where the RV jitter should have the lowest impact on planet detection. For instance, large-scale magnetic field observations of fully convective M dwarfs have mainly focused on rapid rotators (\pstar~<~6~d) so far. Among those, they have characterized the coexistence of two types of magnetic topologies: strong axial dipole and weak multipolar field \citep{Morin08a, Morin08b, Morin10}. Large-scale magnetic fields of slowly-rotating fully-convective stars remain more poorly constrained, even though their characterization would extend the understanding of dynamo processes \citep{Morin10,Wright16}, provide further constraints on the evolution of stellar rotation \citep{Newton16}, and would permit a better definition of the habitable zone around mid to late-type M dwarfs. Recently, \citet{Hebrard16} have explored the large-scale magnetic properties of quiet M stars and derived a description of the RV jitter as a function of other activity proxies. These studies have shown some connection between the brightness features and the magnetically active regions, but no one-to-one relation, as well as a pseudo-rotational modulation of the RV jitter rather than a purely rotational behaviour. Such detailed investigations on a small number of M stars have thus shown complex spatio-temporal properties and require a broader exploration. 

\begin{figure}
\centering
\includegraphics[width=\columnwidth]{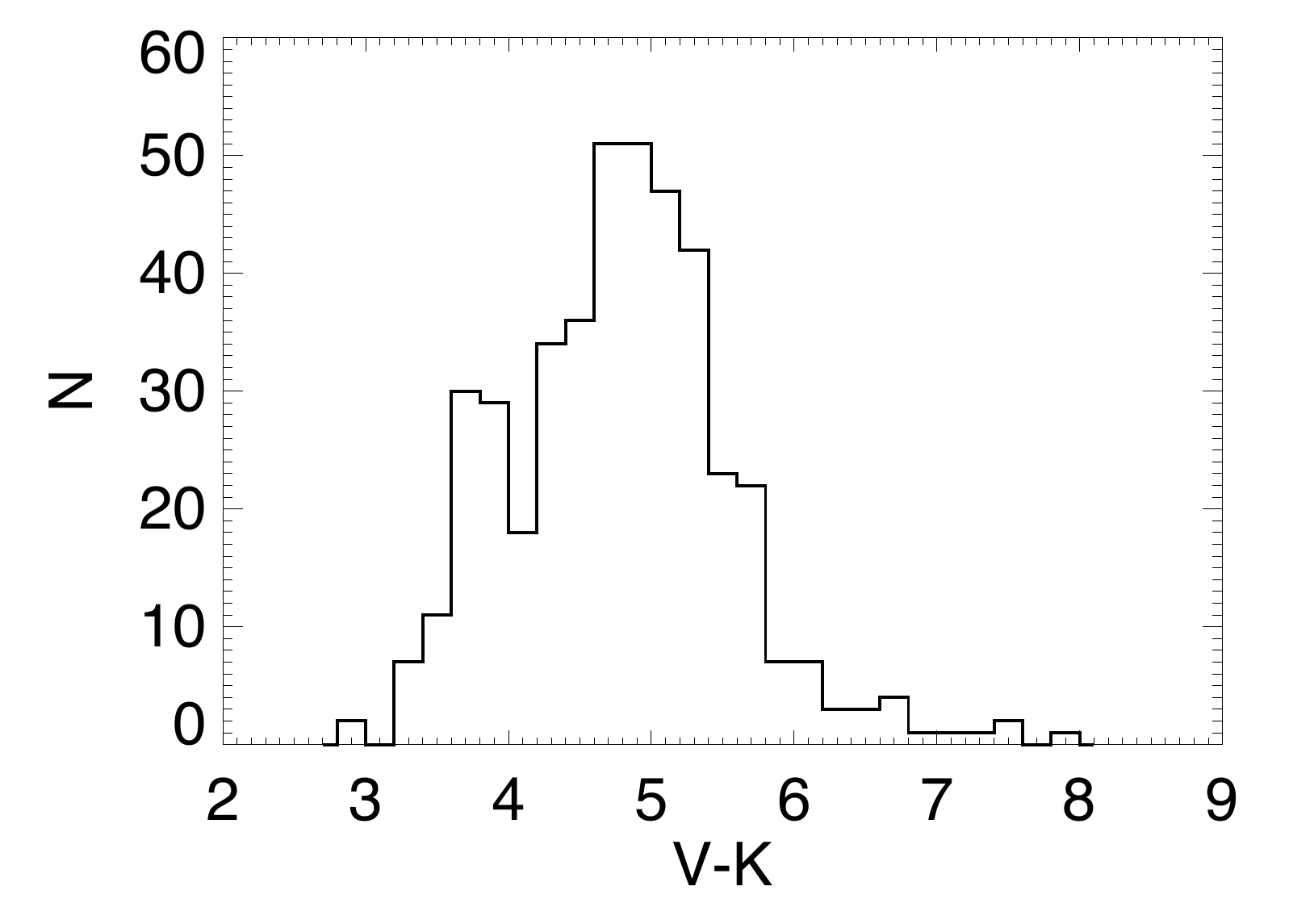}
\caption{The histogram of $V-K$ values in the sample. }
\label{VK}
\end{figure}

In the coming years, the new spectropolarimeter SPIRou\footnote{http://spirou.irap.omp.eu/} will be installed at the Canada-France-Hawaii Telescope atop Maunakea. SPIRou \citep{Artigau11} will be the ideal instrument to study the stellar properties of M dwarfs and search for their planetary companions by combining polarimetric measurements of the stellar magnetic field, the velocimetric precision required for planet searches, and the wide near-infrared simultaneous coverage of the YJHK bands. In preparation to the planet-search survey that will be conducted with SPIRou, we have collected and analysed all ESPaDOnS data available on M dwarfs. Data collection, catalog mining and fundamental parameters are described in companion papers (Malo et al and Fouqu\'e et al, in prep.). In this paper, we investigate the activity features and magnetic properties of the data sample, with the goals of improving our understanding of physical processes at play at the surface and in the atmosphere of M dwarfs that could generate RV jitter and hamper planet detection. By combining and relating several types of observed features of M-dwarf magnetic fields, we attempt to establish a merit function of activity that will allow us to sort and select the best possible targets for planet detection using the RV method and SPIRou. We also enlarge the picture of M-star topology exploration by adding three new slow-rotating M stars having their magnetic topology characterized.

The paper is organized as follows: in section 2, we describe the stellar and data samples and collected observations. In section 3, we show our data analysis methods. Results are discussed in section 4. Further discussion and conclusions are given in section 5.

\section{Sample and observations}\label{sec:obs}
All spectra of our sample of 442 stars correspond to cool stars in the solar vicinity. 
The origin of the data is diverse: 1) the exploratory part of the Coolsnap program\footnote{program IDs 14BF13/B07/C27, 15AF04/B02, 15BB07/C21/F13, 16AF25, 16BC27/F27 and 17AC30, P.I. E. Martioli, L. Malo and P. Fouqu\'e}, 2) the CFHT/ESPaDOnS archives\footnote{http://www.cadc-ccda.hia-iha.nrc-cnrc.gc.ca/en/cfht/} in the spectropolarimetric mode, 3) the CFHT/ESPaDOnS archives in the spectroscopic mode, and 4) the follow-up part of Coolsnap. 
The Coolsnap program is a dedicated observing project using spectropolarimetric mode with CFHT/ESPaDOnS, targeting about a hundred M main-sequence stars  with two visits per star. Table \ref{tab:sample} summarizes the stellar and data samples.

Figure \ref{VK} shows the distribution of the $V-K$ index in our sample. The $K$ magnitude is actually obtained in the 2MASS $K_s$ filter \citep{Skrutskie06}, while $V$ magnitudes are in the Johnson system (the compilation of magnitudes is described in the forthcoming paper by Fouqu\'e et al, in prep.). The histogram peaks at a value of 4.8 which corresponds to spectral types of M3-M4 \citep{Pecaut13} or masses of about 0.35\msun, or effective temperatures of $\sim$3300K \citep[see mass-luminosity and mass-radius relations as in ][]{Delfosse98,Baraffe98,Benedict16}. The star with the largest colour index is GJ~3622 ($V-K$ = 7.858) which was originally observed and studied by \citet{Morin10}. 

While the data origin is diverse, the homogeneity of the data lies in the use of the same instrument CFHT/ESPaDOnS, providing the wide optical range of 367 - 1050 nm in a single shot, at 65,000 -- 68,000 resolving power; data processing and analysis are also homogeneous. The sample of spectra, however, is heterogeneous in signal-to-noise ratio, number of spectra per star and temporal sampling. The choice of ESPaDoNS is mainly motivated by the spectro-polarimetric mode, a unique way to obtain the circular polarisation of stellar lines \citep{Donati03c}, and by the extended wavelength range towards the red, well adapted to M-star observations.

Data acquired in the "Star+Sky" spectroscopic mode is a single-exposure spectrum where the sky contribution is subtracted from the star spectrum. In the polarimetric mode, four sub-exposures are taken in a different polarimeter configuration to measure the circularly polarized spectra and remove all spurious polarization signatures \citep{Donati97b}. The unpolarized spectrum is the average of all four intensity spectra.

Although large-scale magnetic-field detections are obviously not available from spectra in "Star+Sky" mode, including those data allowed to significantly widen our sample while still allowing a large number of measurements: chromospheric emission indices, projected rotational velocity, radial velocity, fundamental parameters and Zeeman broadening proxy, when the SNR is sufficient.  Table \ref{tab:sample} gives a summary of main properties of our data sample. 
 For additional description of the stellar sample, we refer the reader to papers in the series by Malo et al (in prep.) for SPIRou planet-search program target selection, and by Fouqu\'e et al (in prep.), focusing on the determination of stellar fundamental parameters.

Consequently, as a follow-up to the Coolsnap exploratory program, we focused on two low-mass stars with slow rotation rates for which the magnetic field topologies were determined: GJ~1289 and GJ~793. They are the only stars with multiple visits observed in the polarimetric mode of ESPaDOnS for which the magnetic topology has not been published yet. Spectropolarimetric observations were collected from August to October 2016, using ESPaDOnS\footnote{GJ~1289 and GJ~793: program 16BF15, P.I. J. Morin}. Finally, we added the analysis of GJ~251. This star is one of the standards of the calibration plan of CFHT/ESPaDOnS\footnote{GJ~251: programs Q78 of each semester}, and as such, is regularly observed since 2014. The spectropolarimetric observations of GJ~251 are considered here as "Coolsnap follow-up" (Table \ref{tab:sample}).

\begin{table}
    \centering
    \caption{Summary of data collection and how the 1878 spectra distribute in the various programs and modes. S/N range corresponds to values per 2.6~km/s bin at 809 nm. }
    \label{tab:sample}
    \begin{tabular}{llccc}
    \hline
Program         & Mode         & Number   & Number of   &  S/N  \\
 Type           &              & of stars &spectra/star &  range\\
 \hline
Coolsnap-explore& polarimetric &  113      & 1-4         &   30-540  \\
Coolsnap-follow & polarimetric &    3      & 18,20,27    &   200-300 \\
Archives        & polarimetric &   75      & 1-110        &   10-1000 \\
Archives        & Star+Sky     &  298      & 1-15        &   10-450  \\
\hline
Total           &  mixed       &  442     & 1-110       &   10-1000 \\
\hline
    \end{tabular}
\end{table}

GJ 1289 is an M4.5 dwarf of 3100~K. Two preliminary observations within the Coolsnap-explore program in September 2014 and July 2015 showed two clear magnetic detections in the Stokes $V$ profile. A total of 18 were then collected over 2.5 months in 2016. Exposure times of 4$\times$380s or 4$\times$600s were used, depending on the external conditions.

GJ 793 is an M3 dwarf of 3400~K. Early observations in August and September 2014 similarly showed detections of the magnetic field. A total of 20 were collected over 2.5 months in 2016. Exposures times of 4$\times$150 to 4$\times$210s were used. 

GJ 251 is of slightly later type than GJ~793, with an estimated temperature of 3300~K. As it is observed in the context of the calibration plan of ESPaDOnS, the observation sampling is regular but infrequent and data are spread from Sept 2014 to March 2016. Since GJ~251 is a slow rotator (see section 4.4.3), such sampling is well adapted. The spectropolarimetric observations of GJ~251 have been obtained with exposures of 4$\times$60s.

The three stars are good representatives of the future SPIRou planet-search targets. The detailed journal of spectro-polarimetric observations for these three stars is shown in Table~\ref{tab:journal_obs}.

\section{Data analysis}
\subsection{Data reduction}
The data extraction of ESPaDOnS spectra is carried out with \textsc{\small{Libre-Esprit}}, a fully automated dedicated pipeline that performs bias, flat-field and wavelength calibrations prior to optimal extraction of the spectra. The initial procedure is described in \cite{Donati97b}. The radial velocity reference frame of the extracted spectra is first calibrated on the ThAr lamp and then, more precisely, on the telluric lines, providing an instrumental RV precision of 20m/s $rms$ \citep{Moutou07}.

Least-Squares Deconvolution (LSD, \citealt{Donati97b}) is then applied to all the observations, to take advantage of the large number of lines in the spectrum and increase the signal-to-noise ratio (SNR) by a multiplex gain of the order of 10. We used a mask of atomic lines computed with an \textsc{\small{Atlas}} local thermodynamic equilibrium (LTE) model of the stellar atmosphere \citep{Kurucz93}. The final mask contains about 4000 moderate to strong atomic lines with a known Land\'e factor. This set of lines spans a wavelength range from 350~nm to 1082~nm. The use of atomic lines only for the LSD masks relies on former studies of early and mid M dwarfs \citep{Donati06a}.

It is important to note that the use of a single mask over such a wide range of spectral characteristics is not optimal; in particular, the multiplex gain is not maximized for spectra corresponding to the latest type M dwarfs. Building up the collection of line lists with Land\'e factors and reliable line amplitudes, including molecular species and in various stellar atmospheres would be beneficial in a future work to this large-sample analysis. On the positive side, the mask we used in this study is the same one used in previous analyses of M stars observed with ESPaDOnS or NARVAL \citep[\emph{e.g.},][]{Morin08a}, which insures some homogeneity.

\begin{table*}
\begin{center}
\caption{Position and widths of passbands used to measure the activity indices. All numbers are in nm. For CaII HK, $tr$ is meant for triangular bandpass of base 0.109nm. Other indices use rectangular bandpasses.}
\label{tab:index}
\begin{tabular}{lllll}
\hline
Index & Line 1 [nm]  &   Line 2 [nm] & Line 3 [nm]& Continuum \\
\hline
 CaII HK &396.8469 $tr$ 0.109&393.3663 $tr$ 0.109&- &  400.107$\pm$1\\
 HeI     & 587.562$\pm$0.04& -& -&586.9-0.25, 588.1-0.25  \\
 NaD     & 589.592$\pm$0.05 & 588.995$\pm$0.05 & -&580.5$\pm$0.5, 609.0$\pm$1 \\
 H$_\alpha$& 656.2808$\pm$0.16&- &- & 655.087$\pm$0.054, 658.031$\pm$0.437\\
 KI      & 766.490$\pm$0.05 & 769.896$\pm$0.05 &  - & 761.95$\pm$0.1, 773.50$\pm$0.1    \\
 NaI IR  & 818.326$\pm$0.025 & 819.482$\pm$0.025 &  - & 814.1$\pm$0.1, 820.7$\pm$0.1    \\
CaII IRT &849.802$\pm$0.1 & 854.209$\pm$0.1 & 866.214$\pm$0.1 & 847.58$\pm$0.25, 870.49$\pm$0.25  \\
 \hline
\end{tabular}
\end{center}
\end{table*}

\subsection{Spectroscopic index measurements}
After correcting from the star radial velocity, we measured the spectroscopic tracers of chromospheric or photospheric activity in spectra with reference positions and widths as summarized in Table \ref{tab:index}. The width of the emission features was deliberately chosen wider than in the literature \citep[\emph{e.g.}][]{DaSilva11} because of the very strong emitters included in our sample, whose emission lines were twice wider than the bandpasses generally in use for quiet stars. Also, we chose for the CaII H and K line continuum to use the window around 400.107~nm only, and not the continuum window around 390.107~nm, in order to reduce the noise increasing in the bluest part of the continuum. The $S$ index was then calibrated with measurements from the literature, although this index is expected to vary in time, resulting in a significant dispersion. The calibration is shown in the Appendix (Fig. \ref{fig:calib}); it is based on literature values from the HARPS M-dwarf survey \citep{Astudillo17b}. 

Other indices were measured in similar ways with their respective continuum domains  optimized against telluric absorption and major stellar blends (Table \ref{tab:index}): H$_\alpha$, the 590-nm NaI doublet (NaD), the 587-nm HeI line, the 767-nm KI doublet (KI), the 819-nm NaI IR doublet (NaI IR), and the 850-nm CaII infrared triplet (CaII IRT). No attempt was made to calibrate these indices to literature values. 

From the $S$ index, we derived the log($R'_{HK}$) by correcting for the photospheric contribution and an estimate of the rotation period as proposed by \citet{Astudillo17a} (their equation 12). This method has the caveat that the rotation periods shorter than about 10 days cannot be derived due to degeneracy because CaII HK chromospheric emission reaches its saturation level. We applied a threshold in log($R'_{HK}$) of -4.5, meaning that a rotation period is deduced only for spectra with log($R'_{HK}$)$<$-4.5.
For other indices, however, we did not correct for the photospheric contribution as was done in other studies \citep[\emph{e.g.}][]{Martinez11}.
We thus do not expect to find similar relationships between chromospheric lines as those obtained when the basal component is removed.

\begin{figure*}
	\includegraphics[width=\columnwidth]{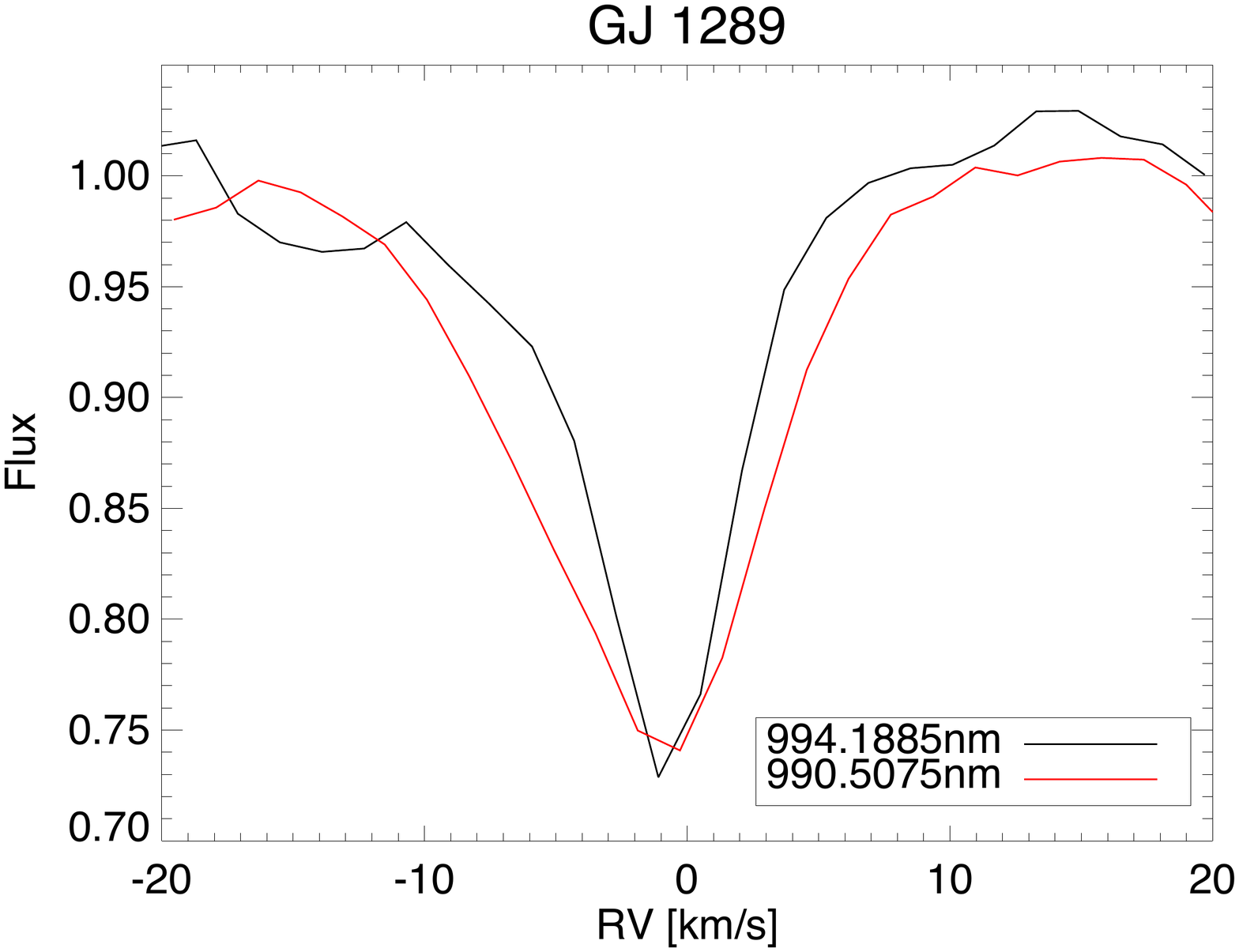}
	\includegraphics[width=\columnwidth]{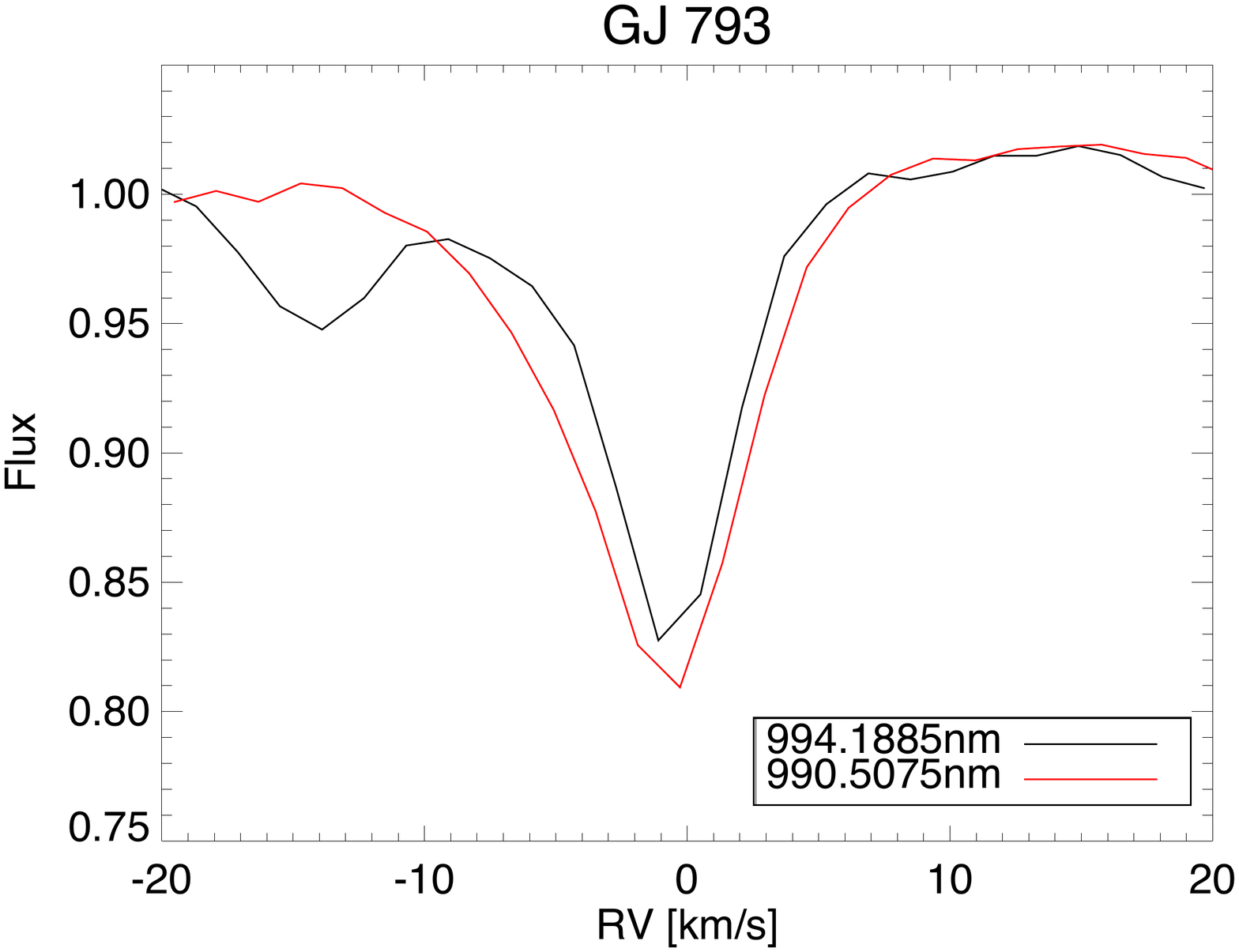}
	\includegraphics[width=\columnwidth]{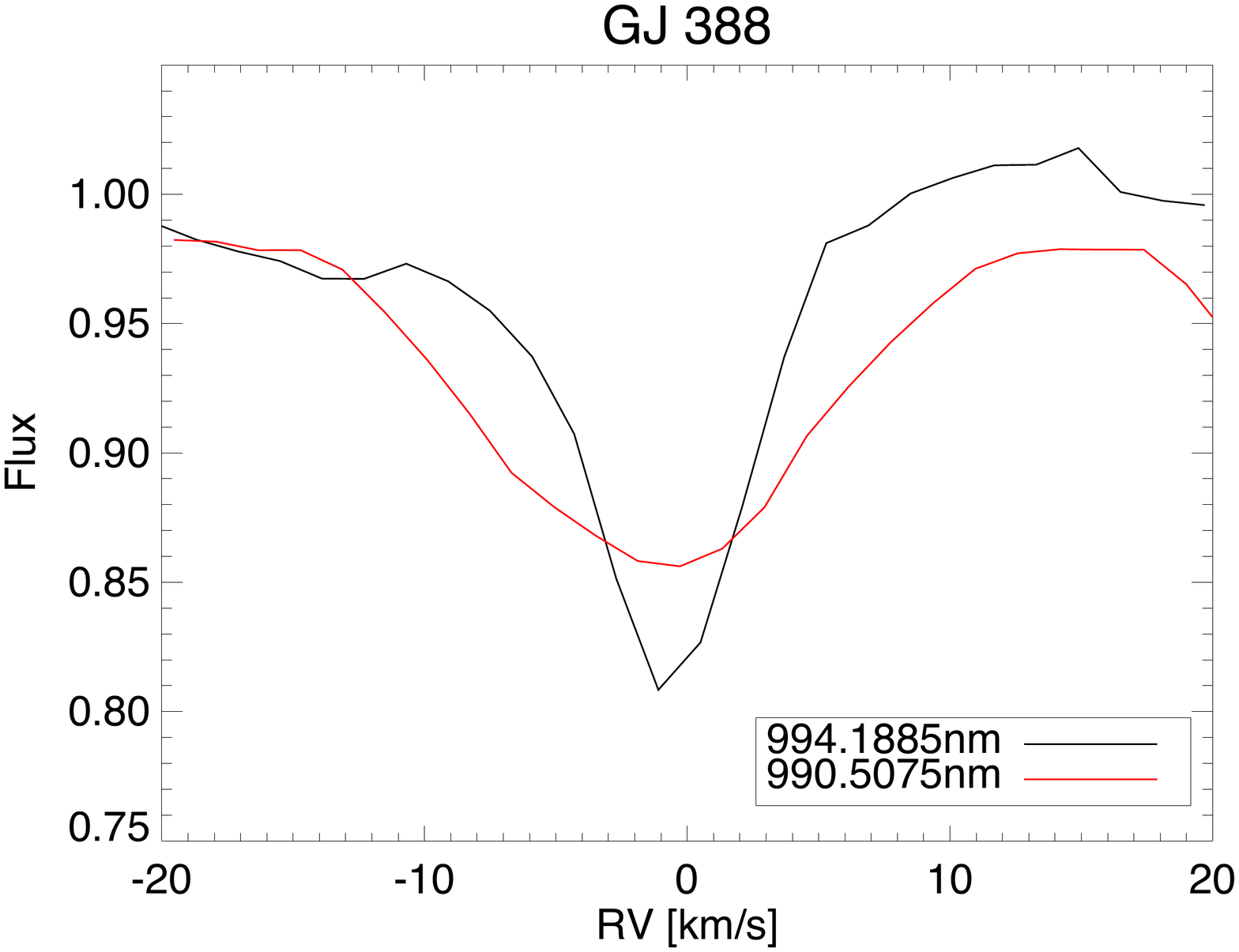}
	\includegraphics[width=\columnwidth]{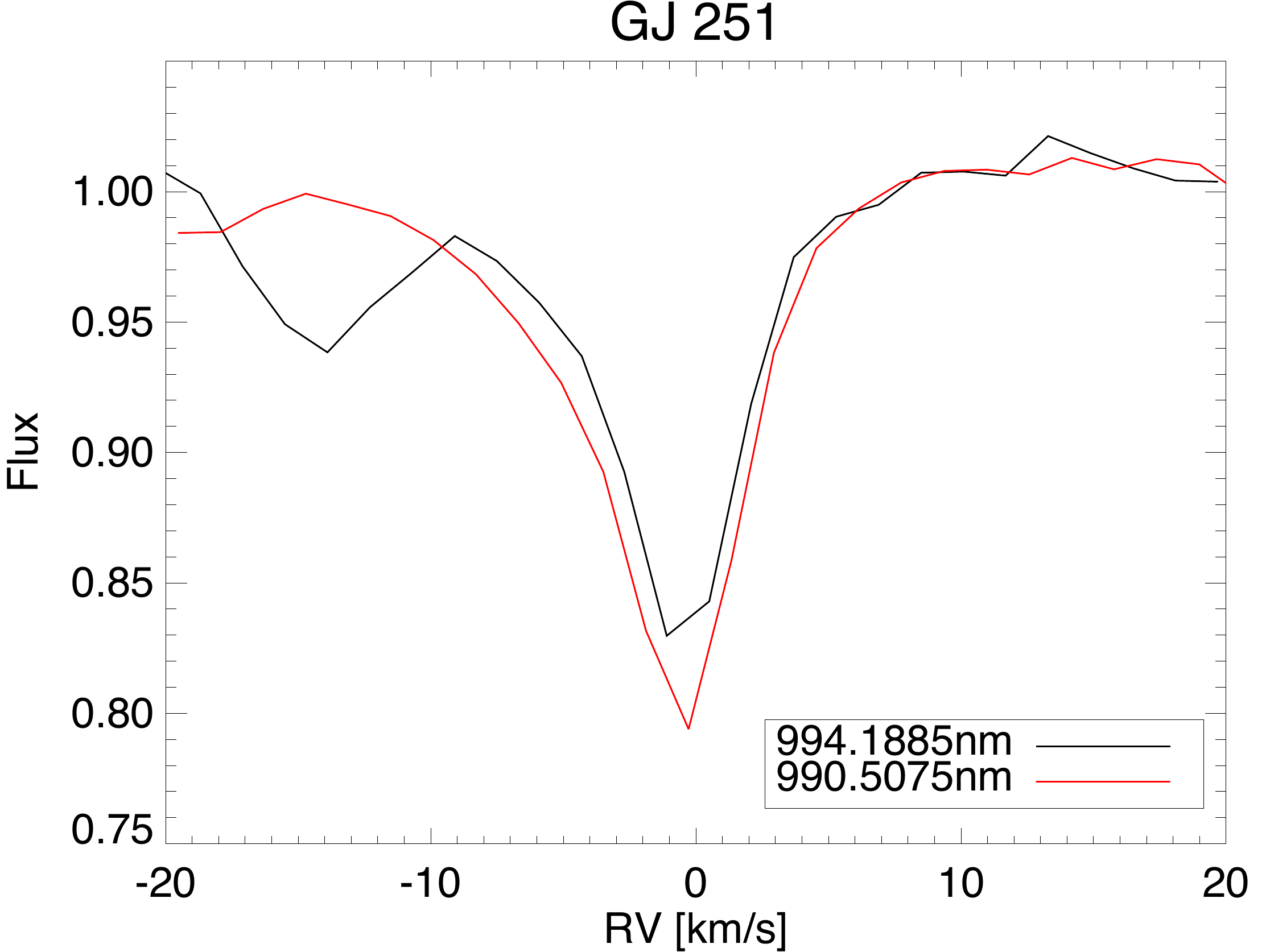}
    \caption{Average FeH lines for GJ~1289 (top left) and GJ~793 (top right), GJ~251 (bottom right) and the most broadened star GJ~388 (AD Leo, bottom left): the 995~nm line is insensitive to the magnetic field (black line) and thus used as a reference profile, while the 990~nm is magnetically sensitive (red line). Spectra with a SNR greater than 250 per CCD pixel were used to calculate the average.}
    \label{fig:feh}
\end{figure*}

\subsection{Zeeman broadening}
As shown in \citet{Saar88}, the Zeeman effect also affects the line broadening in the unpolarized light. This measurement is complementary to the detection of a polarized signature in the stellar lines, by giving access to the average surface field modulus weighted by the filling factor. The spatial scales of the polarized and unpolarized-light magnetic fields are also very different (comparable, respectively, to the full stellar sphere and the magnetic surface spots). For more details on both field diagnostics, we refer the reader to the reviews by, $e.g.$, \citet{Reiners12} and \citet{Morin13}.

We measured in the intensity spectra two unblended FeH lines of the Wind-Ford $F^4\Delta-X^4\Delta$ system near 992~nm, as their variable sensitivity to the magnetic field may be used as a proxy for the average magnetic field (called $Bf$ in the following) of the stellar surface \citep[\emph{e.g.,}][]{Reiners06,Afram08,Shulyak14}: 
\begin{itemize}
    \item The FeH line at 995.0334~nm is weakly sensitive to the magnetic field. We assume that its line width is dominated by rotation, convection, temperature, turbulence and by the intrinsic stellar profile as seen by the ESPaDOnS spectrograph.
    \item The FeH line at 990.5075~nm is magnetically sensitive \citep{Reiners07}. If there is a significant broadening of this line compared to the 995.0334~nm line, we assume it is due to Zeeman broadening in its totality.
\end{itemize}

Only spectra with passed quality criteria were kept for further analysis. In particular, we rejected all spectra with line blending, locally low signal-to-noise ratio or/and fast rotation as the lines were not properly identified and adjusted. As examples, the average FeH velocity profiles for GJ~251, GJ~793, GJ~1289, and, for comparison, GJ~388 (from the smallest to the largest measured broadening) are shown on Figure \ref{fig:feh}. The broadening is significant for all stars. For the most magnetic GJ~388 (AD Leo), the line splitting results in a significant decrease of the line amplitude, as shown in the bottom left panel of this figure.

The Zeeman broadening due to line splitting in the velocity space $\delta v_B$ (in km/s) is then related to the magnetic field modulus $Bf$ through \citep{Reiners12}:
\begin{equation}
Bf = \frac{\delta v_B}{1.4 \times 10^{-6} \lambda_{\circ} g_{\rm eff}}
\end{equation}
where $\lambda_{\circ}$ is the wavelength of the magnetically sensitive line, 990.5075~nm, $g_{\rm eff}$ is the effective Land\'e factor of the transition and $Bf$ is the average magnetic field weighted by the filling factor $f$ (in G). 
We estimated the Land\'e factor by comparing the value of the Zeeman broadening with the literature values obtained through Zeeman spectral synthesis \citep[e.g.][]{Shulyak14}. There is, however, a very small overlap of our valid measurements with the literature, since $Bf$ has been mostly derived in the past on very active, fast rotating stars. AU Mic, AD Leo, GJ~1224, GJ~9520, GJ~49 and GJ~3379 all have a valid measurement in our study and are found in the compilation of $Bf$ values in \citet{Reiners12} and \citet{Shulyak17} where they are quoted with average $Bf$ values of 2.3, 2.6-3.1, 2.7, 2.7, 0.8 and 2.3 kG, respectively. Assuming the field strength is the same in our measurements, that would induce $g_{\rm eff} = 1\pm0.24$. Note that theoretical and laboratory values of $g_{\rm eff}$, although difficult to obtain for a molecule, are quoted in a range 0.8 to 1.25 by several authors for the transitions in the $F^4\Delta-X^4\Delta$ system  \citep{Brown06,Harrison08,Afram08,Shulyak14,Crozet14}, in good agreement with our empirical estimate. We also note that, as we always take the same value of $g_{\rm eff}$ in this study, trends and behaviours intrinsic to the sample remain valid and are independent of the exact value of $g_{\rm eff}$.

\begin{figure}
\centering
\includegraphics[width=\columnwidth]{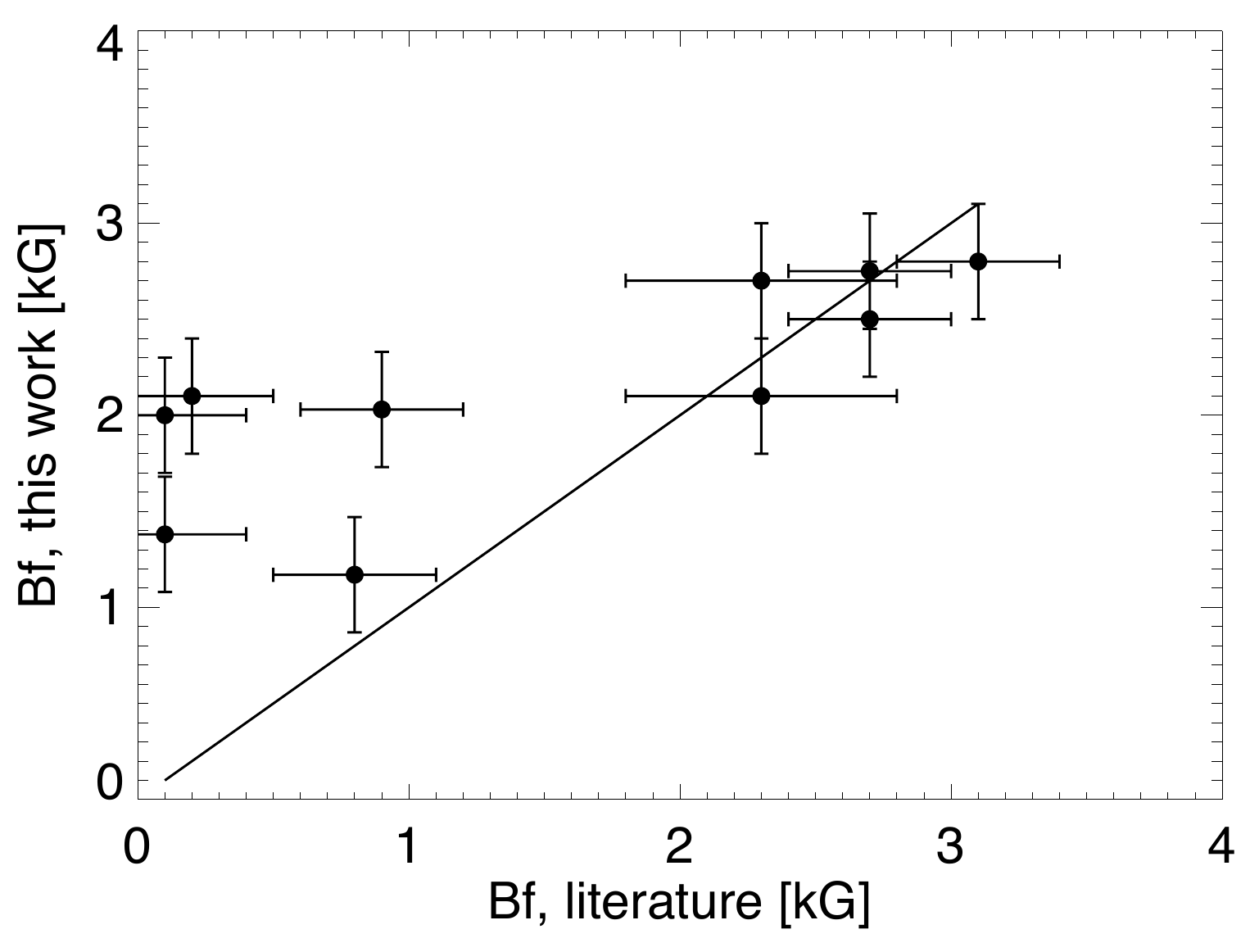}
\caption{The comparison of our $Bf$ values using $g_{\rm eff} = 1$, to different estimates available in the literature \citep{Reiners12,Shulyak17}. The line shows Y=X. While the agreement is good for stars having a large field, it is worse on low-field or slow-rotating stars. Errors on the Y axis have been set to 300~G.\label{fig:bfcomp}}
\end{figure}

The smallest $Bf$ value measured in our spectra is 0.5 kG. The typical error on the $Bf$ value is of the order of 0.3 kG. Although good agreement with spectral-synthesis analyses is found with model-fitting methods on 6 objects (see above), there are 4 stars for which the results disagree (GJ~876, GJ~410, GJ~70, and GJ~905). Figure \ref{fig:bfcomp} shows the comparison of our data with literature values. The Zeeman spectral synthesis code \citep{Shulyak17} gives much smaller values than our method for these 4 slow-rotating stars. This could be due to systematics of one of the methods at low rotation velocity, to surface heterogeneities, or to inaccurate Land\'e factors. There may thus be over-estimations of the field modulus in the following, compared to other studies of the average field, and this is yet to be understood. We note that, within the spectral-synthesis method, there may be similar discrepancies, reaching 1~kG, when using different lines \citep{Shulyak17}. It is beyond the scope of this paper to further explore other atomic or molecular lines over a wide range of spectral types, but that is an ongoing extension of this work. 

\begin{figure}
\centering
\includegraphics[width=1.1\columnwidth]{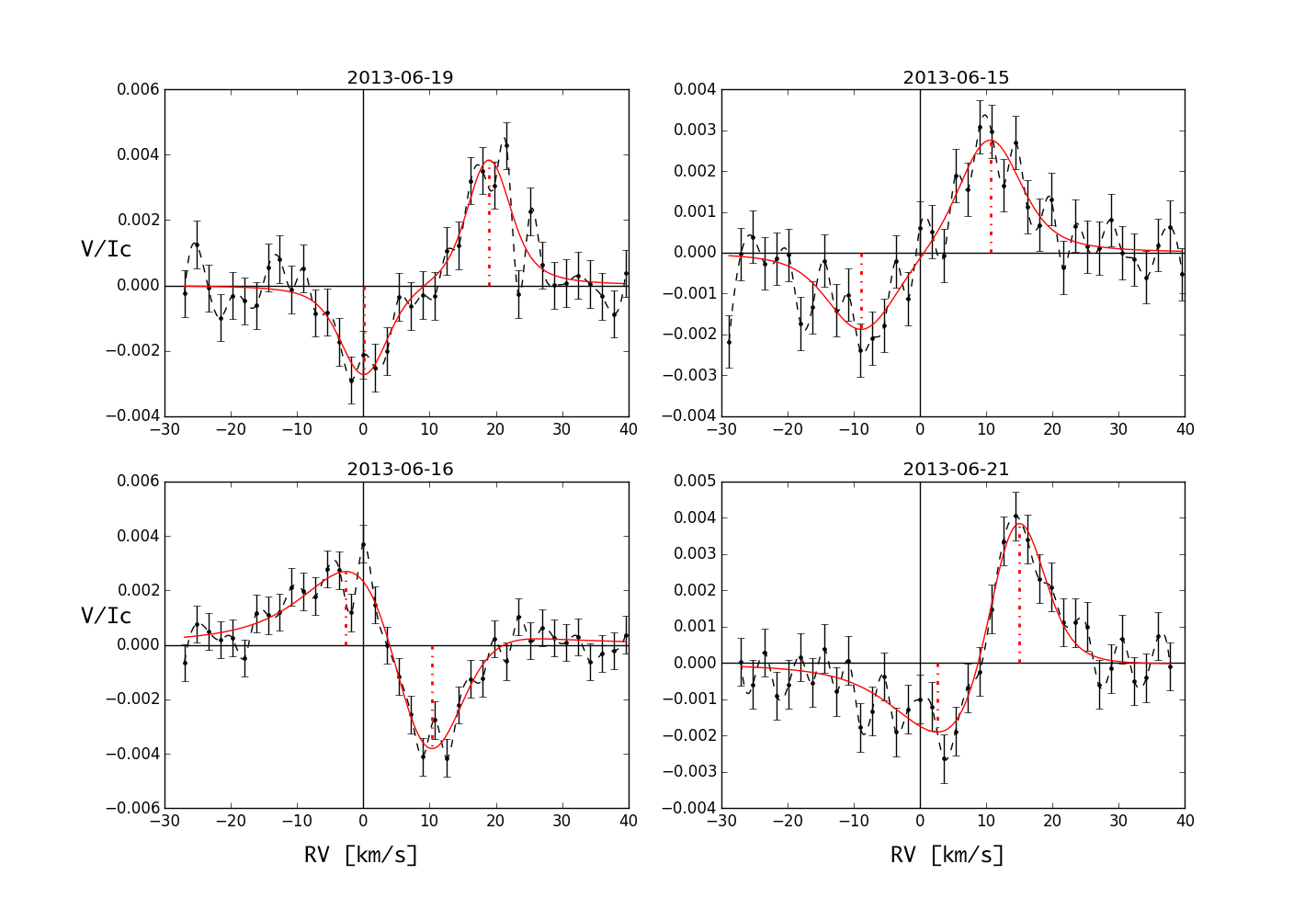}
\caption{Illustration of parametrization of Stokes $V$ profiles: the observed $V$ profiles (data points with errors), their interpolation (dashed) and the best-fit Voigt profile (red) for the star GJ~4053 at different dates. The positions of maxima have been highlighted by red dashed line.  }
\label{Pot}
\end{figure}

\subsection{From Stokes profiles to longitudinal magnetic field}
The intensity profile Stokes $I$ is derived from LSD and adjusted by a Lorentzian in all spectra. On circular-polarisation data, the Stokes $V$ profile is also calculated with LSD, as well as a null profile (labelled N). The $N$ profile results in a different combination of polarimeter positions. It allows to confirm that the detected polarization is real and not due to spurious instrumental or data reduction effects \citep{Donati97b}. It can also be used to correct the $V$ profile by removing the instrumental signature.

In order to distinguish between magnetic field detection in stellar line and noise, we use the False Alarm Probability (FAP) value, as described in \citet{Donati97b}. The $\chi^2$ function is calculated in the intensity profile and outside, both in the Stokes $V$ and $N$ spectra. A definite detection corresponds to a FAP smaller than 10$^{-5}$ and a marginal detection has a FAP between 10$^{-5}$ and 10$^{-3}$; in cases of definite and marginal detections, it is verified that the signal is detected inside the velocity range of the intensity line. If the FAP is greater than 10$^{-3}$, or if a signal is detected outside the stellar line, the detection is considered null.

The longitudinal magnetic field $B_l$ values are then determined from every observation in the polarimetric mode following the analytic method developed in \citet{Donati97b}. 

\begin{equation}
    B_l=-2.14 \times 10^{11} \frac{\int vV(v)\mathrm{d}v}{\lambda_0 \cdot g_{eff} \cdot c \cdot \int \left[ I_c - I(v) \right] \mathrm{d}v}
\end{equation}

where $v$ is the radial velocity, $V(v)$ and $I(v)$ are Stokes $V$ and Stokes $I$ profiles, and $I_c$ is the continuous value of the Stokes $I$ profile. Parameters $\lambda_0$, $g_{eff}$ and $c$ are the mean wavelength (700~nm), the effective Land\'e factor (1.25) and vacuum light speed. Errors on $B_l$ are obtained by propagating the flux errors in the polarized and unpolarized spectra.

\begin{figure}
\centering
\includegraphics[width=0.9\columnwidth]{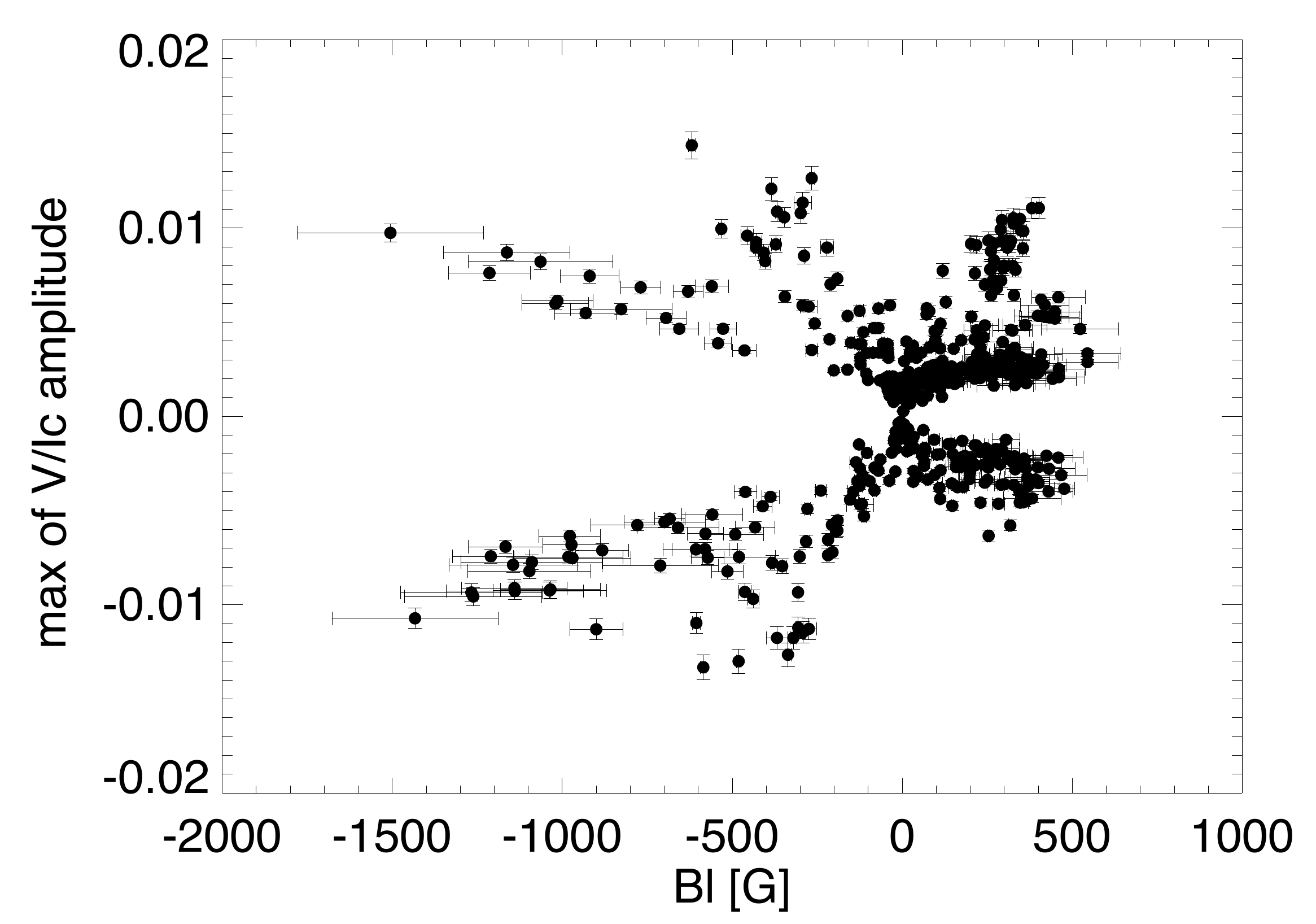}
\caption{The extreme amplitude of the V/$I_c$ profile as a function of the longitudinal field $B_l$, for all profiles where the detection of the magnetic signature in stellar lines is definite.
}
\label{amplitude}
\end{figure}

The domain between $\pm 3\sigma$ ($\sigma$ is the Lorentzian width obtained from the fit of $I(v)$) centered at the intensity profile's maximum was chosen for $B_l$ integration as it includes relevant signal while minimizing the undesired noise on the Stokes $V$ profiles. 

Stokes $V$ profiles actually offer more information than just the $B_l$ values, which only reflects the anti-symmetrical part of the profile. The complex shape of $V$ profiles was then tentatively parametrized with a combination of two Voigt models, applied to significant detections. A few examples are shown in Figure \ref{Pot} for four different profiles of star GJ~4053. The parametric model allows to explore the properties of the profile's features without depending on a complex topology model that cannot be applied in the case of scarce sampling. The amplitude and peak position of the Stokes $V$ profiles have been measured, as well as the integral of the non-anti-symmetrical component of the profiles. The Stokes-$V$ parametrization method may be useful in the future with SPIRou, to explore different activity filtering methods based on polarimetric measurements.  
Figure \ref{amplitude} shows for instance the maximum signed signal in the Stokes $V$ profiles is shown as a function of $B_l$. From experiments done on data sets having more than 20 visits, and randomly selecting pairs of spectra at different epochs, we concluded that it is hazardous to try to derive any of the physical parameters of the large-scale magnetic field from a couple of observed Stokes $V$ profiles of a given star, even with reasonable assumptions on configuration values or rotation periods.

\subsection{Zeeman Doppler imaging}
To reconstruct the magnetic map of GJ~1289, GJ~793, and GJ~251, we used the tomographic imaging technique called Zeeman Doppler imaging (ZDI). It uses the series of Stokes $V$ profiles. It may be necessary to first correct for instrumental polarization by subtracting the mean $N$ profile observed for a given star. It had a significant effect on the modeling of GJ~251 and in a lesser extent, of GJ~793.
ZDI then inverts the series of circular polarization Stokes $V$ LSD profiles into maps of the parent magnetic topology with the main assumption that profile variations are mainly due to rotational modulation. Observed Stokes $V$ are adjusted until the magnetic-field model produces profiles compatible with the data at a given reduced chi-squared $\chi^2$. In that context, longitudinal and latitudinal resolution mainly depends on the projected rotational velocity, \vsini, the star inclination with respect to the line of sight, $i$, and the phase coverage of the observations. The magnetic field is described by its radial poloidal, non-radial poloidal and toroidal components, all expressed in terms of spherical-harmonic expansions \citep{Donati06a, Donati06b}. 

The surface of the star is divided into 3000 cells of similar projected areas (at maximum visibility). Due to the low value of the rotation period of the three stars, the resolution at the surface of the star is limited.  We therefore truncate the spherical-harmonic expansions to modes with $\ell \leq $~5 \cite[\emph{e.g.},][]{Morin08b}. The synthetic Stokes $V$ LSD profiles are derived from the large-scale magnetic field map by summing up the contribution of all cells, and taking into account the Doppler broadening due to the stellar rotation, the Zeeman effect, and the continuum center-to-limb darkening. The local Stokes $V$ profile is computed using Unno-Rachkovsky's analytical solution of the transfer equations in a Milne-Eddington atmospheric model in presence of magnetic field \citep{Landi04}.
To adjust the local profile, we used the typical values of Doppler width, central wavelength and Land\'e factor of, respectively, 1.5~\kms, 700~nm and 1.25. The average line-equivalent width is adjusted to the observed value.
By iteratively comparing the synthetic profiles to the observed ones, ZDI converges to the final reconstructed map of the surface magnetic field until they match within the error bars. Since the inversion problem is ill-posed, ZDI uses the principles of Maximum-Entropy image reconstruction \citep{Skilling84} to retrieve the simplest image compatible with the data. A detailed description of ZDI and its performance can be found in \citet{Donati97b,Donati01b,Donati06a} and its previous application to low-mass slowly-rotating stars in \citet{Donati08c} and \citet{Hebrard16}.

\section{Results}

\subsection{Spectroscopic indices and rotation periods}

\begin{figure}
	\includegraphics[width=\columnwidth]{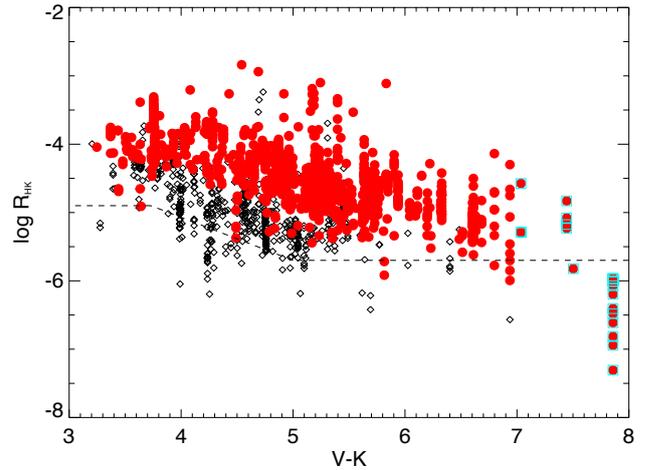}
    \caption{Variations of log($R'_{HK}$) with respect to $V-K$ colour index. In black diamonds (resp. red circles), the spectra where measured projected velocity is smaller than (resp., greater than) 3km/s. The dashed line shows the lower envelope proposed by \citet{Astudillo17a}. Very low-mass stars (cyan squares) lie beyond the log($R'_{HK}$) calibration and thus their y-value is dubious.}
    \label{fig:logR}
\end{figure}

The log($R'_{HK}$) values for our sample are shown on Figure \ref{fig:logR} as a function of $V-K$ and for different ranges of projected velocity.
It is observed that, for a given mass, faster rotators are more active than slower ones. The other trend is that log($R'_{HK}$) decreases with $V-K$, whatever the velocity. This latter observation may, however, be an indication that the calibration of the bolometric factor or the photospheric factor is not robust over a large colour range \citet{Astudillo17a}.

 For early-type and slow rotating stars, we were able to obtain estimates of the rotation period following \citet{Astudillo17a}, as discussed in the Appendix and shown on Fig. \ref{fig:prot}.  For instance, the rotation periods of GJ~1289, GJ~793 and GJ~251 derived from CaII HK are found to be 36, 27 and 85 days, in qualitative agreement with the periods derived from the ZDI analysis (54, 22 and 90 days, respectively, see section 4.4). 

For the following, and in the objective to use a global chromospheric index for stellar activity classification, we introduce the total chromospheric flux: $F_{chr}$ is the sum of the most prominent chromospheric equivalent widths found in M stars, $F_{chr} = NaD + HeI + H\alpha + S + CaIRT$. In the following we will use either the CaIRT or the $F_{chr}$ index. 

More details on chromospheric features and trends between them are presented in Appendix B.

\subsection{Large-scale magnetic field in polarized light}
\begin{figure}
\centering
\includegraphics[width=1.1\columnwidth]{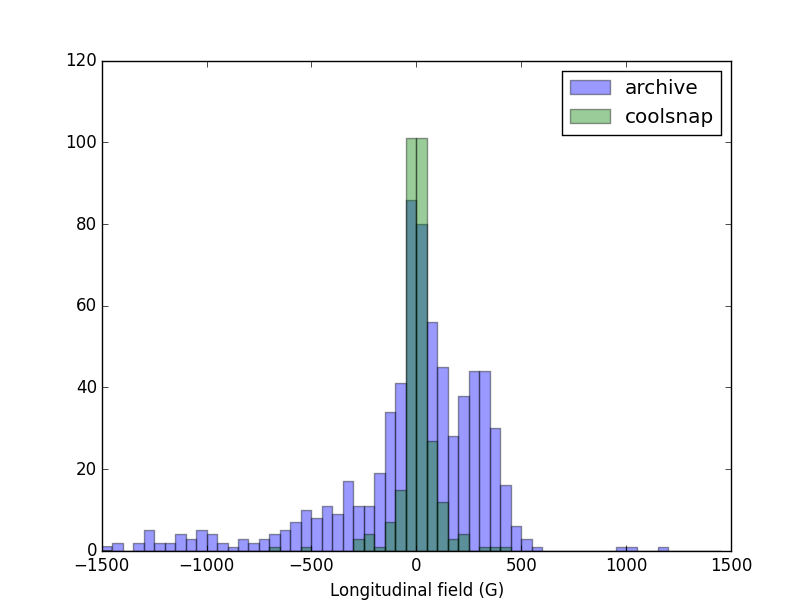}
\caption{Histogram of $B_l$ for the two polarimetric data sets.\label{histbl}}
\end{figure}

We estimated the longitudinal magnetic field in all polarized spectra of our sample. The longitudinal field values, $B_l$, are equally distributed between negative and positive values, which is expected for a random distribution. The histogram of $B_l$ value shown in Figure \ref{histbl} presents how weak the magnetic field of Coolsnap stars is while the archive stars observed in the polarimetric mode span a much wider range of the longitudinal field. This only reflects the biases of samples; the archive mostly contains fast rotators and the most active M dwarfs. Note that the tail of very high negative values of $B_l$ from the archive sample visible in Figure \ref{histbl} is mostly due to a single star, GJ 412B (WX UMa), featuring one of the strongest magnetic dipole known among M dwarfs \citep{Morin08b}.

Table \ref{tab:detstat} presents statistics on the detection of the magnetic field signatures in our data set. It shows that, while more stars have been observed in the Coolsnap program than in the full ESPaDOnS archive of M stars in the polarimetric mode, more spectra are available in the archive, and these spectra correspond to more magnetic-field detections. The percentage of null detections is greater in the Coolsnap sample while the archive contains most of the definite ones. Additional statistics show an ideal balance in the signed value of $B_l$: 50 stars have $B_l$ values always positive, while 51 stars have $B_l$ values always negative and 45 stars have $B_l$ that change sign within our data set. Finally, 300 stars have no $B_l$ estimate, either because of non detections in the polarimetric mode, or because they were observed in the spectroscopic mode.\\

\begin{figure}
\centering
\includegraphics[width=1\columnwidth]{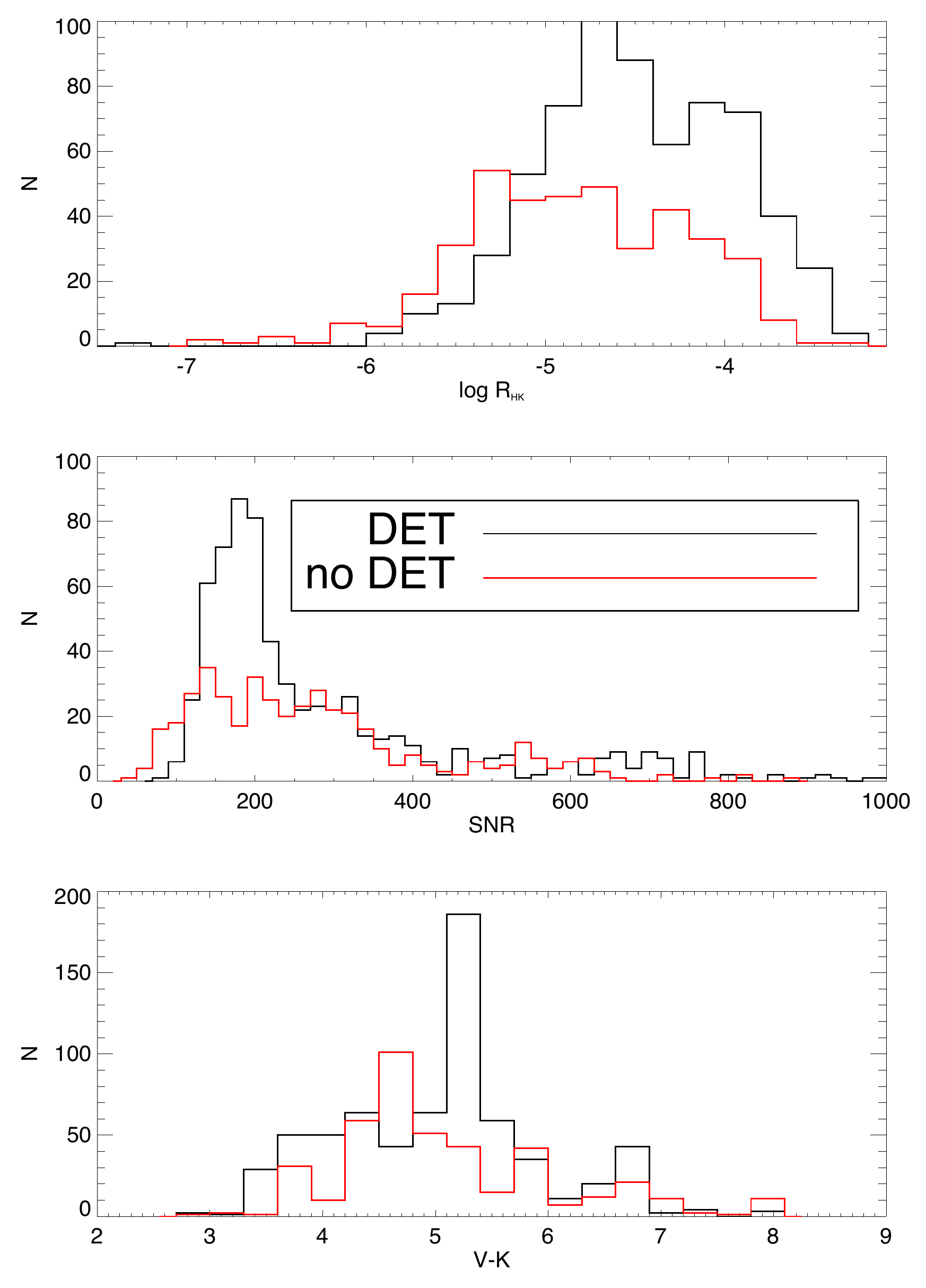}
\caption{Distribution of parameters in all spectra in the polarimetric mode, with respect to magnetic-field detections: chromospheric emission, SNR (at 809~nm), and colour index $V-K$. The red histograms show the non detections and the black histograms show the marginal and definite detections.}
\label{fig:histdet}
\end{figure}

Figure \ref{fig:histdet} shows how the detection statistics behave as a function of the spectrum SNR, the chromospheric activity, and the colour index. The black histograms correspond to marginal and definite detections while red histograms correspond to null detections. There are more detection around chromospherically active stars ( log($R'_{HK}$) greater than -5.0). The different behaviour in SNR reflects the various stellar samples (Coolsnap and archives) and shows that, beyond an SNR of 100, magnetic detection depends on stellar parameters more than on SNR.

\begin{table}
    \centering    
    \caption{Detection statistics for the polarization signal in stellar lines. The counts are given in number of polarized spectra (not in number of stars).}
    \label{tab:detstat}
    \begin{tabular}{lcc}
    \hline
 Program           & Coolsnap & Archive\\
 \hline
Total stars        & 113      &  63   \\
Total observations & 285      &  724  \\
Definite detection & 85 (30\%)      &  523 (72\%)\\ 
Marginal detection & 37 (13\%)       &  46 (6\%)\\
Null detection     & 163 (57\%)     &  155 (22\%) \\
\hline
    \end{tabular}
\end{table}

\subsection{Average surface magnetic field}
We obtained valid Zeeman broadening measurements on 396 spectra corresponding to 139 different stars (31\% of our sample), which means that the large majority of spectra and stars in our sample do not exhibit a valid or measurable Zeeman broadening. Spectroscopic binaries, stars with unknown $V-K$ and rotators with a $v \sin i$ larger than 6~km/s have been excluded. The $V-K$ of these stars span a range from 3.5 to 8, so approximately K8 to M7 types. The largest number of stellar types with broadening measurements are M3 and M4 dwarfs. 

\begin{figure}
	\includegraphics[width=\columnwidth]{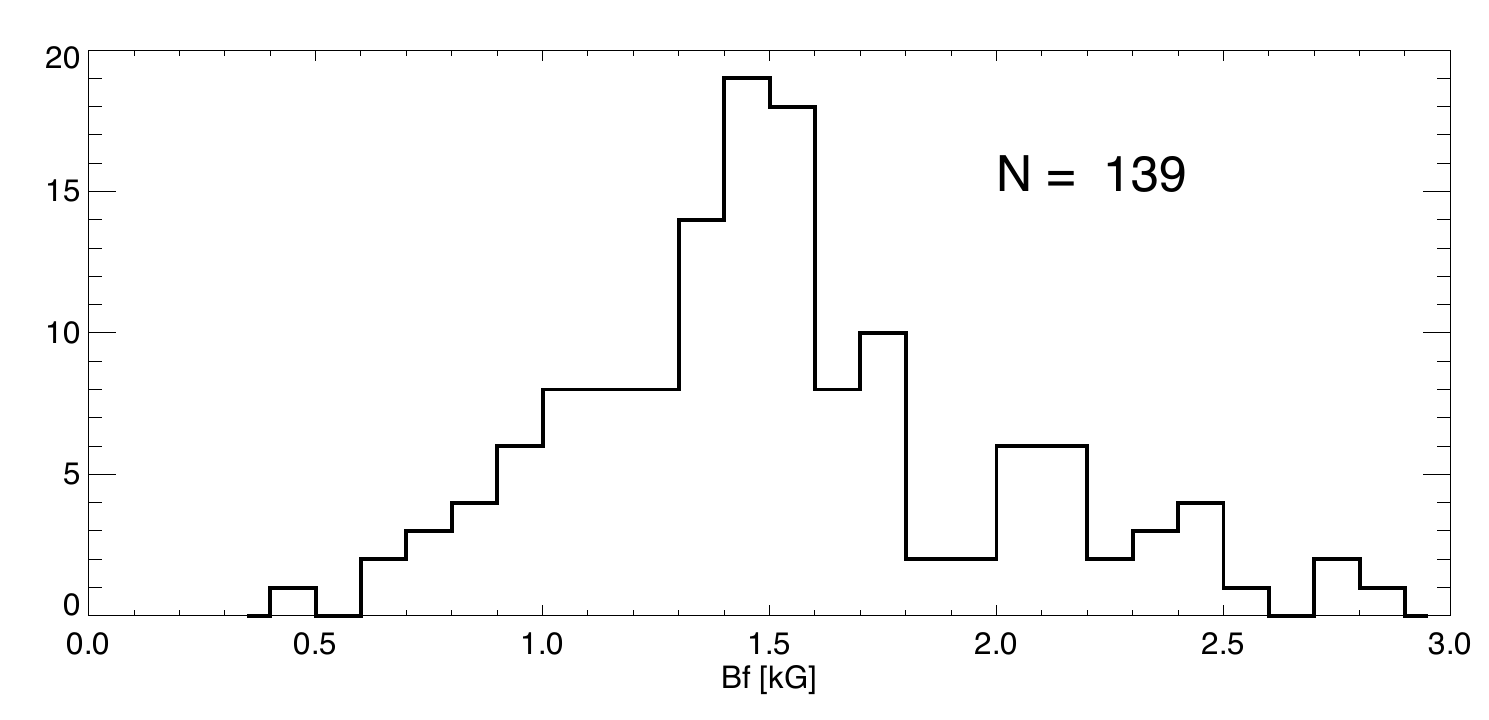}
    \caption{Histogram of $Bf$ values.}\label{fig:histobf}
\end{figure}

The histogram of $Bf$ values measured on 139 stars ranges from 0.5 to 3kG, with a marked peak at 1.5kG (Fig. \ref{fig:histobf}). The relative dispersion around the measured $Bf$ values for a given star is less than 20\% in 76\% of the sample, while only 10\% of the stars have dispersions larger than 50\%.

\begin{figure}
	\includegraphics[width=\columnwidth]{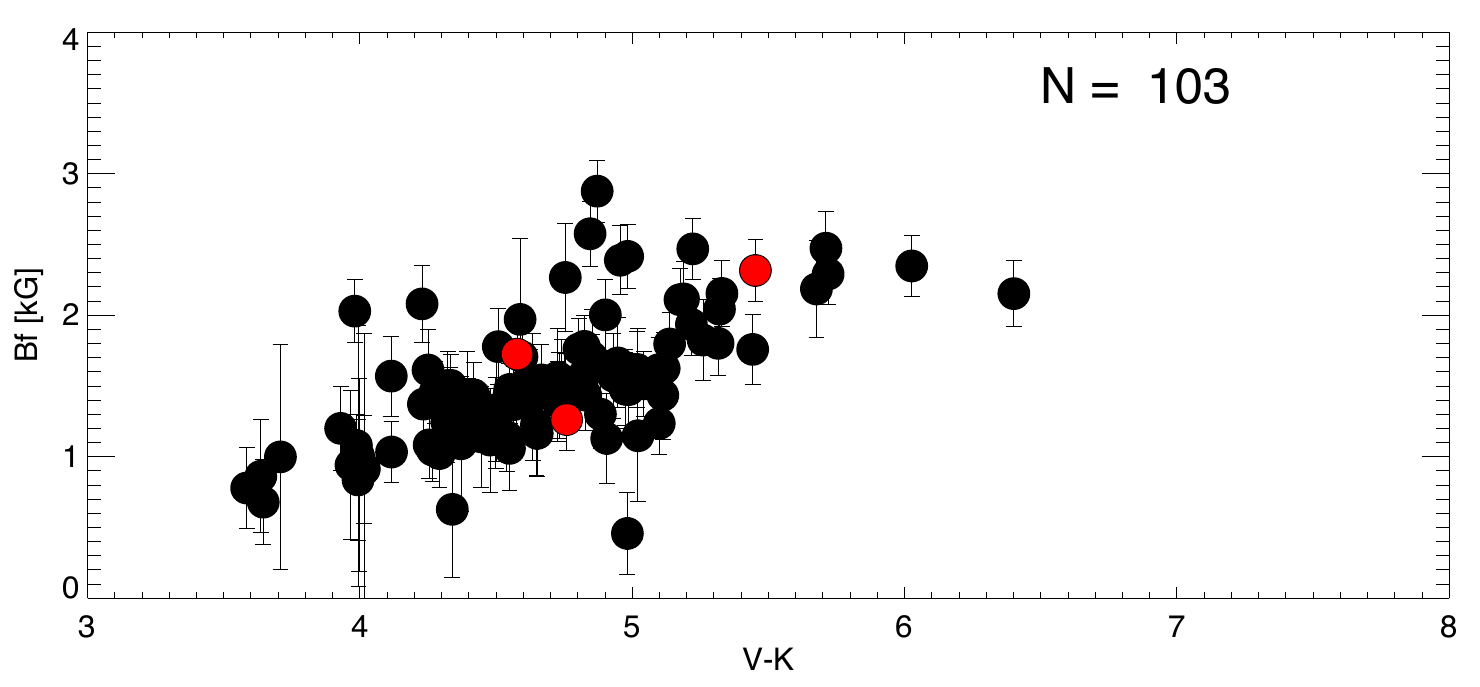}
    \caption{Relations between the small-scale field $Bf$ and $V-K$.  Each point represents a star. The error bars represent the dispersion between the different visits, when more than 2 are available (numbered N). The red points show GJ~793, GJ~251 and GJ~1289.
    }
    \label{fig:BfVK}
\end{figure}

Fig. \ref{fig:BfVK} shows the increase of the magnetic field as the star type gets redder. A slope of 0.61 kG/($V-K$)mag and a Pearson correlation coefficient of 66\% is obtained between $Bf$ and $V-K$, when spurious measurements and stars in spectroscopic binaries are excluded.
The dispersion of $Bf$ values as a function of spectral type has a similar amplitude than found on later type stars by \citet{Reiners10b}.

The measured broadening can be measured only for slow rotators. Beyond a $v \sin i$ of $\sim$6~km/s, the individual FeH lines cannot be properly measured and get blended by neighbour lines with our method. 
Faster rotators yet tend to have an average surface magnetic strength of larger amplitude. 

\subsection{The magnetic field of GJ~1289, GJ~793, and GJ~251}
\begin{figure*}
    \includegraphics[scale=0.4]{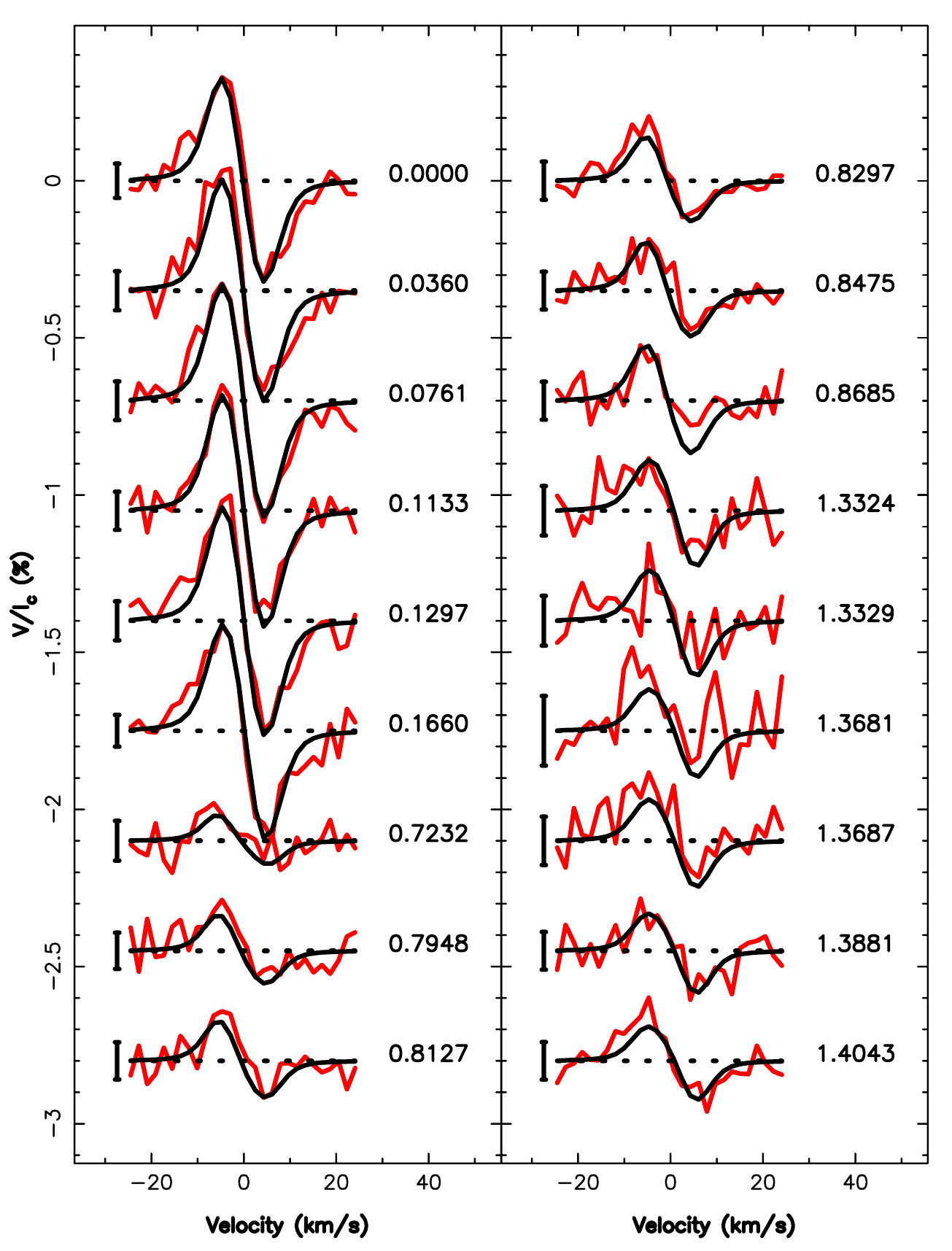}
    \includegraphics[scale=0.4]{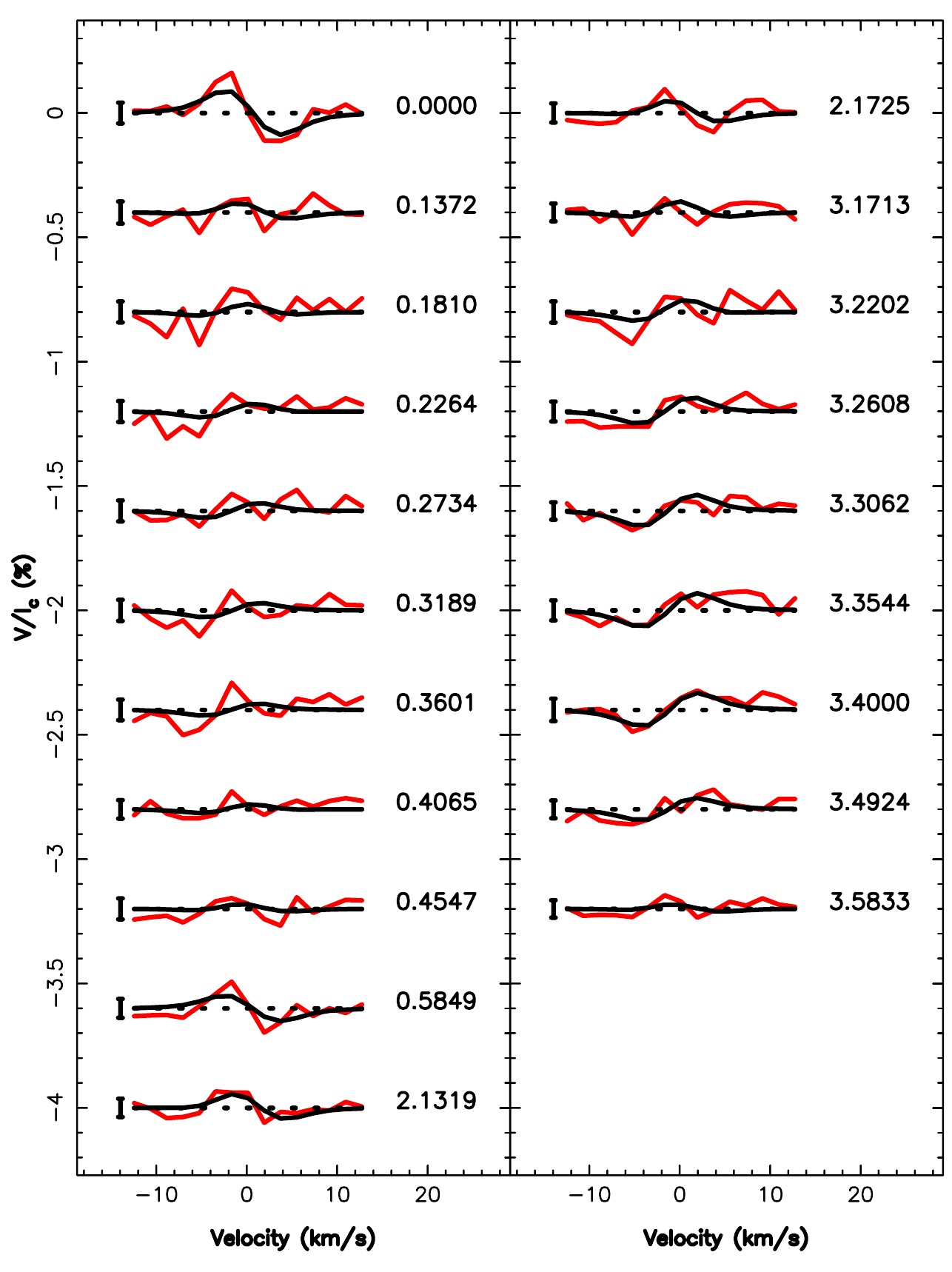}
    \includegraphics[scale=0.4]{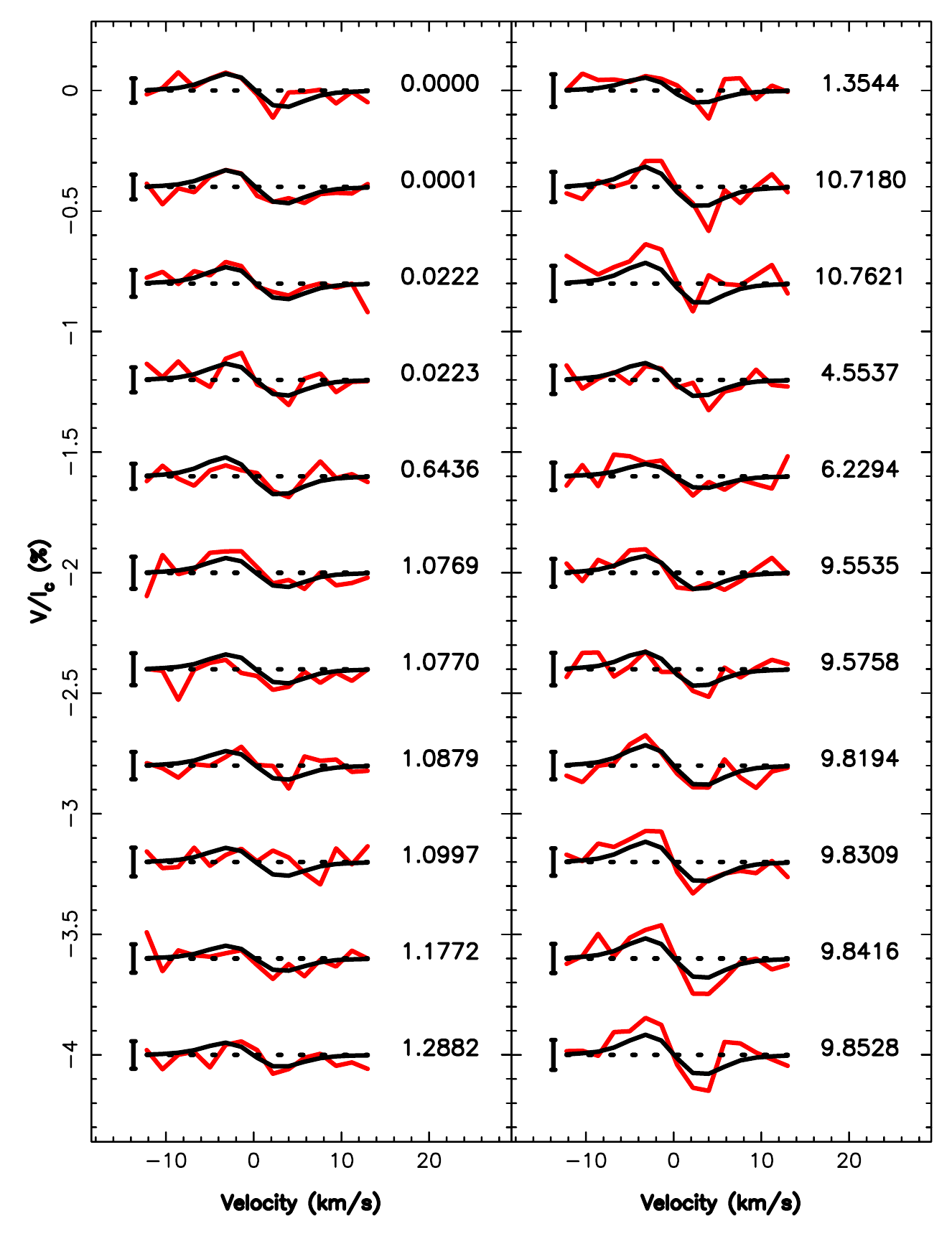}
    \caption{Maximum-entropy fit (black line) to the observed (red line) Stokes $V$ LSD photospheric profiles of GJ~1289 (left), GJ~793 (middle), and GJ~251 (right). Rotational cycles and 3$\sigma$ error bars are also shown next to each profile.}
    \label{fig:specV}
\end{figure*}

Zeeman signatures are clearly detected in Stokes $V$ LSD profiles with a maximum peak-to-peak amplitude varying from 0.1\% to 0.5\% of the unpolarized continuum level for both data sets of GJ~1289, GJ~793, and GJ~251. The temporal variations of the intensity and of the shape of the Stokes $V$ LSD profile are considered to be due to rotational modulation. Figure \ref{fig:specV} shows the observed and best-fit Stokes $V$ profiles for the three stars.  

\subsubsection{GJ 1289 = 2MASS J23430628+3632132}
GJ~1289 is a fully convective star with \mstar~=$~0.23~$\msun\ (from the absolute $K$ magnitude calibration of \citet{Benedict16}) and \rstar~$\sim~0.24~$\rsun\ (see details in Fouqu\'e et al, in prep.). 
The stellar rotation period was determined by ZDI and checked using the periodogram of various proxies ($B_l$ and several activity indices). Both methods concur to the value \pstar~=~$54~\pm~4$~d (see Fig.~\ref{fig:perio}, top panel). We thus used the following ephemeris: HJD~(d) = 2457607.01471 + 54.E, in which E is the rotational cycle and the initial Heliocentric Julian Date (HJD) is chosen arbitrarily. Also, from the convection-timescale ($\tau_c$) calibration in \citet{Kiraga07} and this rotational period, we infer a Rossby number of 0.62.\\
After several iterations, the values of stellar inclination $i$~=~$60~\pm~15\degr$ and \vsini~=~$1~\pm~1~$\kms\ are found to give the optimal field reconstruction. 

With the Zeeman Doppler Imaging technique, it is possible to adjust the Stokes $V$ profiles down to a $\chi^2_r$ of 1.5, while starting from an initial value of 9.6 for a null field map. The reconstructed large-scale magnetic field is purely poloidal (99\% of the reconstructed magnetic energy) and mainly axisymmetric (90\% of the poloidal component). The poloidal component is purely dipolar (99\%). These results are in agreement with the observed shape of the Stokes $V$ signatures which are anti-symmetric with respect to the line center with only the amplitude varying as the star rotates. The inclination of the dipole with respect to the stellar rotation axis ($\sim$~30\degr) explains these amplitude variations : the strongest profiles observed between phases 0.0 and 0.2 directly reflect the crossing of the magnetic pole in the centre of the visible hemisphere, whereas the weakest Stokes $V$ (\emph{i.e.}, phase 0.723) are associated with the magnetic equator.
The magnetic field strength averaged over the stellar surface is 275~G. Figure \ref{fig:map_field} (top) shows the reconstructed topology of the magnetic field of GJ~1289, featuring the dipole with a strong positive pole reaching $\sim$~450~G. 

The stellar inclination $i$ towards the line of sight is mildly constrained: a high inclination (>45\degr) allows a better reconstruction, \emph{i.e.}, it minimizes both $\chi^2_r$ and the large-scale field strength. 
    
In order to fit the circularly polarized profiles, we used a filling factor $f_{V}$, adjusted once for all profiles: it represents the average fraction of the flux of magnetic regions producing circular polarization at the surface of the star. ZDI reconstructs only the large-scale field, however the large-scale field can have a smaller scale structure (\emph{e.g.}, due to convection or turbulence). This parameter then allows to reconcile the discrepancy between the amplitude of Stokes $V$ signatures (constrained by the magnetic flux $B$) and the Zeeman splitting observed in Stokes $V$ profiles (constrained by the magnetic field strength $B/f_{V}$). While $f_{V}$ has no effect on the modeling of the intensity profiles of adjusting GJ~1289, it is essential to fit the width of the observed Stokes $V$ profiles, as was shown for other fully convective stars in earlier studies \citep[see, \emph{e.g.},][]{Morin08a}. The $f_{V}$ value found for GJ~1289 is 0.15.
    
We have then compared the derived field modulus obtained from this broadening $Bf$ estimated in the unpolarized light with the longitudinal field B$_l$ measured in the polarized light (its absolute value). The data is shown on Figure \ref{fig:blbf}. As expected, and consistently with the large sample, there is no strong correlation between the large-scale topology and the small-scale magnetic field; the trend, although not significant, is positive, with a Pearson coefficient of 0.4. The rotationally-modulated signal of the main dipole obeys to large-scale dynamo processes while small-scale magnetic regions may rather be induced by convective processes, especially in a fully-convective star such as GJ~1289.

\subsubsection{GJ 793 = 2MASS J20303207+6526586}

GJ~793 is a partly convective low-mass star with \mstar ~=$~0.42~$\msun\ \citep{Benedict16} and \rstar ~$\sim~0.39~$\rsun. The \vsini\ of GJ~793 is smaller than 1 km/s.
 
In the Generalised Lomb-Scargle periodogram of the activity indices and $B_l$, different peaks stand out at $\sim$~35~d, $\sim$~17~d and $\sim$~11~d with various significance levels (see Fig.~\ref{fig:perio}). The lowest false-alarm probability ($\sim$~10\%) is reached for a peak at 35.3~d in H$_\alpha$, a set of peaks associated with its first harmonic between 17 and 17.7~d in H$_\alpha$ and $B_l$, and a peak at 11.3~d in $B_l$. A peak at 22~d is also present in the periodogram of $B_l$ but with a lower significance than its first harmonic. So the rotation period cannot be unambiguously defined from the periodograms. The rotation period suggested by the CaII HK calibration is $\sim$ 27 days.
Moreover, given the poor sampling of the data and the weakness of the Stokes $V$ signatures ($\leqslant$ 0.3\% of the continuum level), the tomographic analysis is not able to precisely constrain the rotation period \pstar\ and the inclination $i$.

We then tried different values of (\pstar, $i$) to reconstruct the large-scale field, and we looked for the cases which minimise the value of $\chi^2$, from an initial $\chi^2$ of 3.1. Several minima are found corresponding to a rotation period of $\sim$~22~d or $\sim$~34~d. The best-fit associated inclination is $i$~<~40\degr\ for both cases. However a secondary minimum is found for $i\sim$~80\degr\ and \pstar~=~34~d. All maps are associated with a magnetic field strength of $\sim$210~G.
We explored the possibility that the model degeneracy could be due to differential rotation (DR).
Contrarily to a solid-body rotation hypothesis, some amount of differential rotation would allow to remove the secondary minima of the stellar inclination, and to minimise the reconstructed magnetic field strength (down to 75~G). The data sampling is, howeover, insufficient to correctly probe the differential-rotation properties.

Note that differential rotation has already been observed for faster rotating early M dwarfs (\emph{e.g.}, GJ~410 or OT-Ser in \citealt{Donati08c}). They found that the surface angular rotation shear can range 0.06 to 0.12~rad.d$^{-1}$. Such a large rotation shear for a slowly rotating star as GJ~793 is, however, rather unexpected (if 34 days were the pole rotation period, this would require a $d\Omega$ of 0.1~rad.d$^{-1}$).
A confirmation and better determination of the DR in GJ~793 would require a better sampling of the stellar rotation cycle as is available in the current data set.

Rotational cycles on Table \ref{tab:journal_obs} and Figure \ref{fig:specV} are computed from observing dates according to the following ephemeris: HJD~(d) = 2457603.95552 + 22.E.
Given the rotation period of 22d, the Rossby number of GJ~793 is 0.46, using relations in \citet{Kiraga07}.

\begin{figure*}
    \includegraphics[scale=0.34]{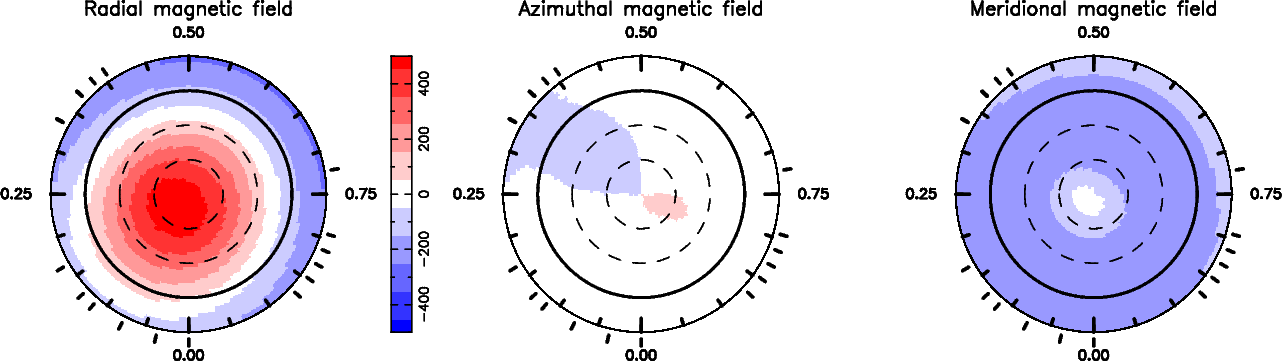}
    \includegraphics[scale=0.54]{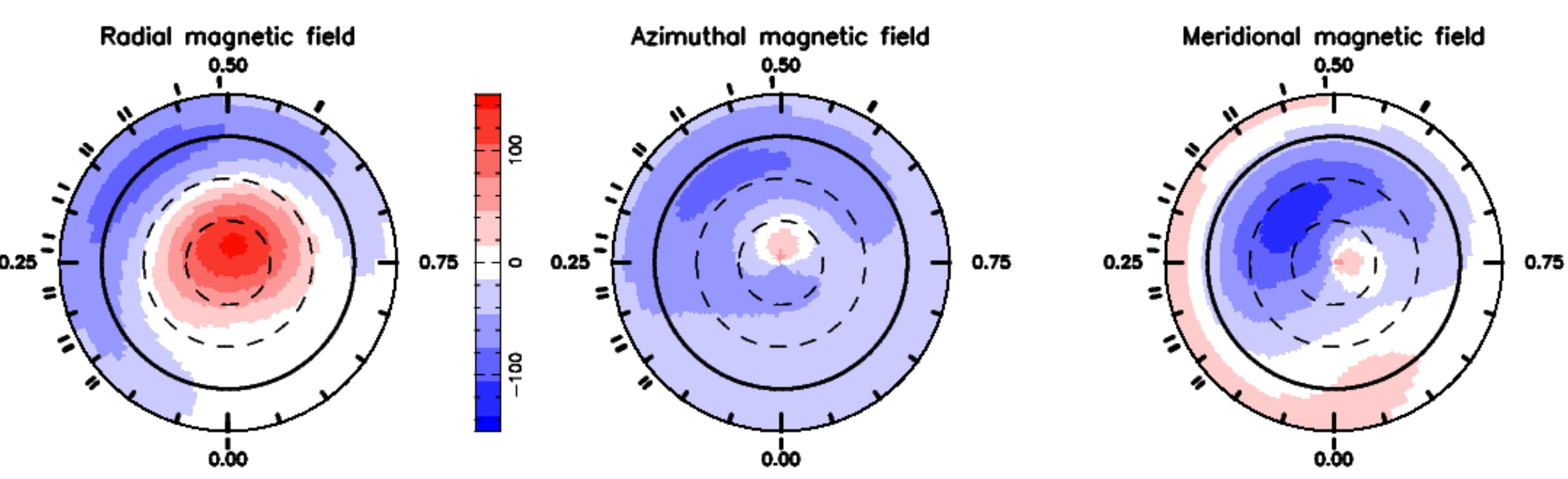}
    \includegraphics[scale=0.54]{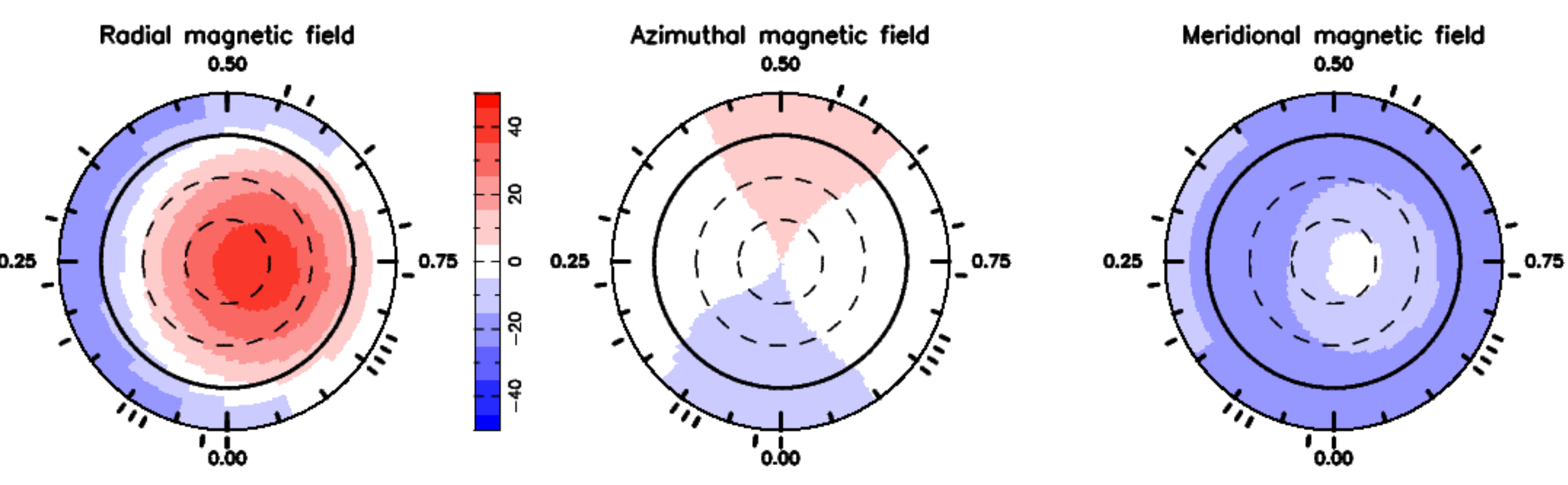}
    \caption{Surface magnetic flux as derived from our data sets of GJ~1289 (top), GJ793 (middle), and GJ~251 (bottom). The map obtained for GJ~793 is one of several equivalent solutions and more data is needed for a confirmation.
The radial (left), azimuthal (center) and meridional (right) components of the magnetic field $B$ are shown. Magnetic fluxes are labelled in G. The star is shown in a flattened polar projection down to latitude -30\degr, with the equator depicted as a bold circle and parallels as dashed circles. Radial ticks around each plot indicate phases of observations.
This figure is best viewed in colour.}
    \label{fig:map_field}
\end{figure*}

The large-scale magnetic field in the configuration shown on Fig. \ref{fig:map_field} (middle panel) is 64\% poloidal and axisymmetric (82\% of the poloidal component). The poloidal component would be mainly quadrupolar (>~66\%). Due to the low signal and the uncertainties in the rotational properties, however, this field reconstruction is the least robust of the three and would benefit additional data; several equivalent solutions differ in topology and field strength (Table \ref{tab:zdi}). In this case, the filling factor $f_{V}$ was not needed to fit the Stokes $V$ profiles, which is expected for stars with a shorter rotation period or not fully convective \citep{Morin08a}.

\begin{figure}
	\includegraphics[width=\columnwidth]{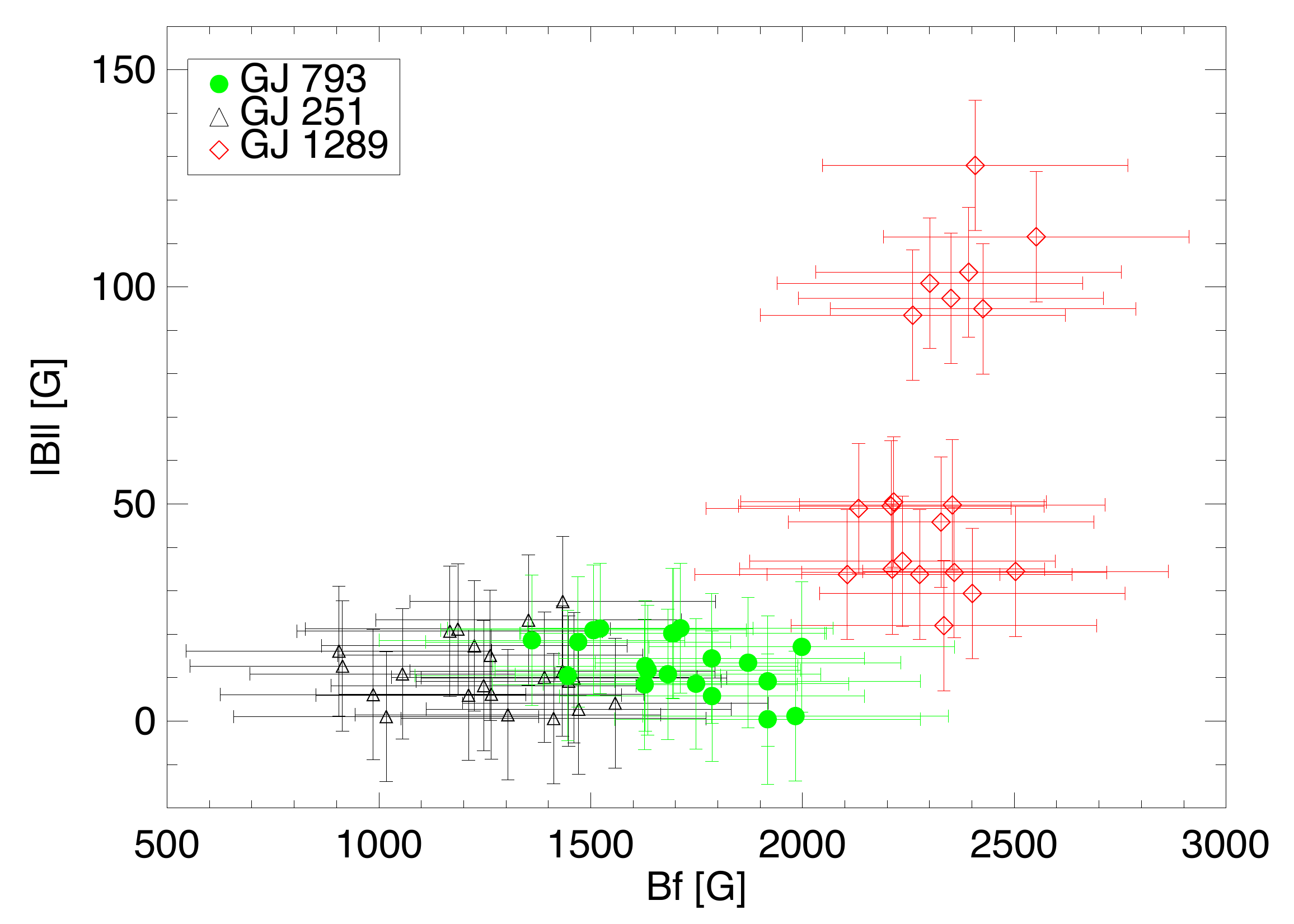}
    \caption{$Bf$ as a function of the absolute value of the longitudinal field for GJ 1289 (red diamonds), GJ 251 (black triangles) and GJ 793 (green circles).}
    \label{fig:blbf}
\end{figure}

The average small-scale field strength ranges from 1400 to 2000 G. The average $Bf$ of GJ~793 is significantly lower than for the lower-mass GJ~1289. 
In Figure \ref{fig:blbf}, compared to GJ~1289, the location of the data points of GJ~793 is more confined in $|B_l|$, and more spread out in $Bf$. A slight negative trend between both measurements is visible, with a negative Pearson coefficient -0.5.

\subsubsection{GJ~251 = 2MASS J06544902+3316058 }
GJ~251 is another partly convective low-mass star with \mstar ~=$~0.39~$\msun\ \citep{Benedict16} and \rstar ~$\sim~0.37~$\rsun\ but a longer rotation period than GJ~793.
A rotation period estimate of $\sim$85~days from the activity index seems to correspond to the observed profile variability and is refined to 90 days by the ZDI analysis. The ephemeris of HJD~(d) = 24576914.643 + 90.E has been used. A projected velocity \vsini\ smaller than 1 km/s is used for this star.
The Rossby number of GJ~251, considering a rotation period of 90 days, is 1.72 \citep{Kiraga07}.

The Stokes $V$ profiles of GJ~251, after correction of the mean residual $N$ signature, show a low level of variability over almost 10 stellar rotation cycles. The amplitude of the circularly polarised signatures never exceeds 0.2\% peak-to-peak. It is difficult with such a weak signal to strongly constrain the rotational period and the stellar inclination. The initial $\chi_r^2$ of the data is 2.2. A minimum $\chi_r^2$ of 1.1 is found for a period of 90$\pm$10 days and an inclination of 30$\pm$10\degr. However, a secondary
minimum of 40 d is found for the rotational period, due to scarce data sampling. 

The reconstructed ZDI map is shown on Figure \ref{fig:map_field} (using these values of \pstar and $i$) and features a topology with a strong poloidal component encompassing 99\% of the magnetic energy, a pure dipole. This poloidal component is also mostly axisymmetric (88\%), with an average field of only 27.5 G.

GJ~251 has a lower magnetic field than the other two stars, as shown on Fig. \ref{fig:blbf}. For this star, the trend between $Bf$ and $|B_l|$ values is insignificant and lower than for both other stars (Pearson coefficient is -0.1). 

The sparse spectropolarimetric sampling and very long rotation period make it difficult to get a robust reconstruction, and it is the first time that data spanning such a long period of time are used in ZDI, so the results for this star have to be taken with caution, although the field reconstruction seems robust. It is possible, for instance, that the hypothesis of the signal modulation being due to rotation is wrong, as the topology itself could vary over several years.

Table \ref{tab:zdi} summarizes the ZDI parameters and fundamental parameters for the three stars.

\begin{table*}
\begin{center}
\caption{Summary of magnetic field parameters for GJ~1289, GJ~793 and GJ~251. $\tau_c$ is the convection timescale \citep{Kiraga07}, used to calculate the Rossby number $Ro$, d$\Omega$ is the differential rotation and $i$ is the inclination of the rotation axis with respect to the line of sight. The topology is characterized by the mean large-scale magnetic flux B, the percentage of magnetic energy in the poloidal component (\% pol) and the percentage of energy in the axisymmetric component of the poloidal field (\% sym). For GJ~793, we report a wide range of values, since several configurations are compatible with the data. \label{tab:zdi}}
\begin{tabular}{lccccccccccccc}
\hline
Name& Mass & Radius& SpT& $\tau_c$& $Ro$ & $v \sin i$&$P_{rot}$& d$\Omega$& $i$& B & \% pol& \% sym& $<Bf>$\\
    & \msun & \rsun&    & d &      &km/s& d & rad.d$^{-1}$ &\degr& G&&& kG \\
    \hline
GJ~1289&0.23 &0.24 & M4.5V & 82.4 & 0.66 & $<$1 & 54 & 0    & 60 & 275 & 99 & 90 & 2.32 \\
GJ~793 &0.42 &0.39 & M3V & 48.3 & 0.46 & $<$1 & 22-34 &0-0.1 & 40-80  &  75-200  & 64-98 & 82-97 & 1.72\\ 
GJ~251 &0.39 &0.37 & M3.5V & 52.3 & 1.72 & $<$1 & 90 & 0    & 30  &27.5 & 99 & 88 & 1.26\\
\hline
\end{tabular}
\end{center}
\end{table*}

\subsection{Magnetic topology of M dwarfs}

\begin{figure*}
    \includegraphics[scale=0.7]{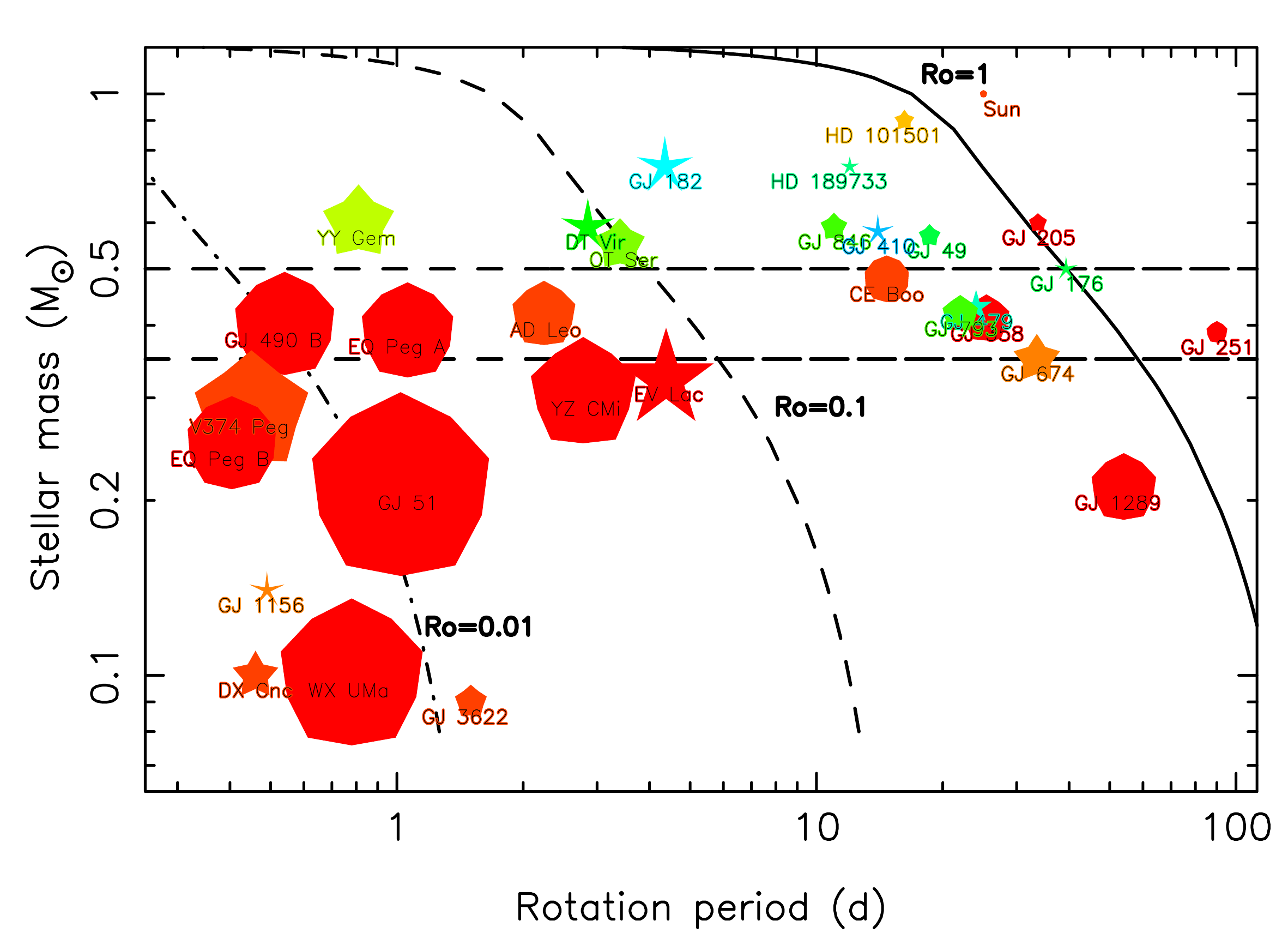}
    \caption{Properties of the magnetic topologies of mainly M dwarfs as a function of rotation period and stellar mass. Larger symbols indicate larger magnetic fields while symbol shapes depict the different degrees of axisymmetry of the reconstructed magnetic field (from decagons for purely axisymmetric fields to sharp stars for purely non-axisymmetric fields). colours illustrate the field configuration (dark blue for purely toroidal fields, dark red for purely poloidal fields and intermediate colours for intermediate configurations). The solid line represents the contour of constant Rossby number $R_o$ = 1 (from a smoothed interpolation of \citet{Kiraga07}). The dotted line correspond to the 0.5 and 0.35 \msun thresholds. HD~101501 and HD~189733 \citep{Fares10} and the Sun are shown for comparison, as well as the other M dwarfs coming from \citet{Donati08c, Morin08b, Morin10} and \citet{Hebrard16}. While GJ~1289 and GJ~251 stand in unexplored regions of the diagram, GJ~793 sits on top of other stars GJ~358 and GJ~479 analysed in \citet{Hebrard16}, but its symbol requires confirmation.}
    \label{fig:config_magn}
\end{figure*}

The study of the large-scale stellar magnetic field is interesting for the exoplanetary study as it allows to better explore conditions for habitability \citep{Vidotto14}. However, understanding its origin remains challenging, and more particularly for fully convective stars.
The large-scale magnetic field is generated in the stellar interior. In solar and partially convective stars, a shearing is expected to take place in a boundary layer located between the inner radiative core and the outer convective envelope. Part of the magnetic field generation comes from the convective envelope itself. Stars less massive than 0.35\msun\ are fully-convective and therefore their convective envelope is fully responsible for the magnetic field generation.
To study and compare the magnetic field of low-mass stars, we fill up the \mstar\  - \pstar\ diagram with characteristics of the topology, as initiated in \citet{Donati09}. 
In order to solve questions about the dynamo, it is crucial to detect and have access to the geometry of the field and explore the space of parameters (stellar internal structure and rotation properties). 
In that perspective, results we obtained for fully convective slowly rotating stars like GJ~1289 are very interesting and bring new observational constraints to models in which the dynamo originates throughout the convection zone. 

The 3 stars presented in this paper, with their long rotation periods, cover a poorly explored domain so far (see Figure \ref{fig:config_magn}). 
The magnetic topology of GJ~793 and GJ~251 resembles the diverse topologies and weak fields found so far for the partly convective slowly rotation stars, like GJ~479 or GJ~358.
On the other hand, we find that the magnetic field detected for GJ~1289 exhibits a strength of a few hundreds of Gauss, as AD Leo. While much larger than the field of GJ~793 and GJ~251, the field of GJ~1289 is lower by a factor 3 than those of more active and rapidly rotating mid-M dwarfs \citep{Morin08b}. Its large-scale magnetic field is dipole dominated and therefore is similar to the topology of more rapidly rotating low-mass stars, rather than to the field of slowly rotating Sun-like stars. Both GJ~1289 and GJ~251 have a large rotation period ($\geqslant$ 50~d), but a different internal structure. 
Our results shows that slowly rotating stars without tachocline (as GJ~1289) tend to have a relatively strong dipolar field rather than the weaker field of slow and partly convective stars. Therefore the maps we obtained tend to confirm the key role of the stellar structure. This is also supported by the earlier observations of large-scale magnetic fields of fully convective stars, although on faster rotators \citep{Donati08c}.

A recent X-ray study carried out using Chandra \citep{Wright16} showed that slowly rotating fully convective M dwarfs can behave like partly convective stars in terms of X-ray luminosity - rotation relation. X-ray luminosity is a tracer of the surface magnetic activity and is believed to be driven by the stellar magnetic dynamo. Their result may thus give another observational evidence that a tachocline is not necessarily critical for the generation of a large-scale magnetic field, and that both stars with and without a tachocline appear to be operating similar magnetic dynamo mechanisms.

\section{Discussion}
\subsection{Relations between magnetism and activity}

\begin{figure}
	\includegraphics[width=\columnwidth]{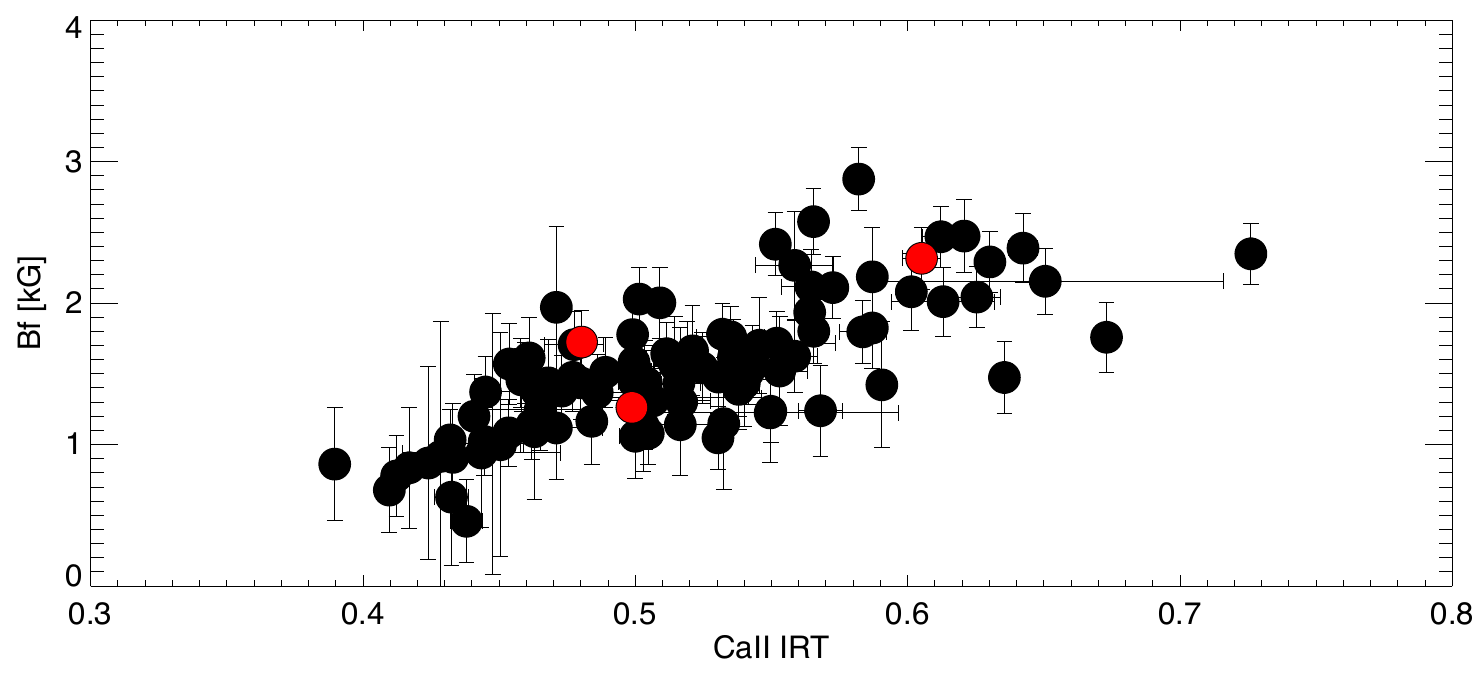}
    \caption{Relations between the small-scale field $Bf$and the CaII IRT index.  Each point represents a star. The error bars represent the dispersion between the different visits, when more than 2 are available (numbered N). The red points show GJ~793, GJ~251 and GJ~1289.
    }
    \label{fig:act5}
\end{figure}

In Figure \ref{fig:act5}, the collected data has been combined so that each data point is a star of the sample rather than a spectrum. The median (average) value of each plotted quantity has been calculated when more than three (resp., only two) spectra are available. The error bars represent the dispersion between the measurements for a single star, when it was possible to calculate the standard deviation. When spectroscopic mode and polarimetric mode spectra were available, the data were all combined together, since the spectral resolution of both modes is similar. The red symbols show results for GJ~1289, GJ~793 and GJ~251. 

We observe the trend that more active stars (\emph{i.e.}, stars with a larger average CaII IRT index) have a stronger small-scale magnetic field.
There is a Pearson correlation coefficient of 70\% between the CaII IRT index and the $Bf$ in our sample of 139 stars where these quantities are measured. For instance, GJ~793 and GJ~251 have a similar average CaII IRT index but marginally different $Bf$ values of 1.72$\pm$0.22 and 1.26$\pm$0.22 kG, respectively.
Similar behaviour of chromospheric indices as a function of field modulus were found by \citet{Reiners10b}. 

\begin{figure}
	\includegraphics[width=\columnwidth]{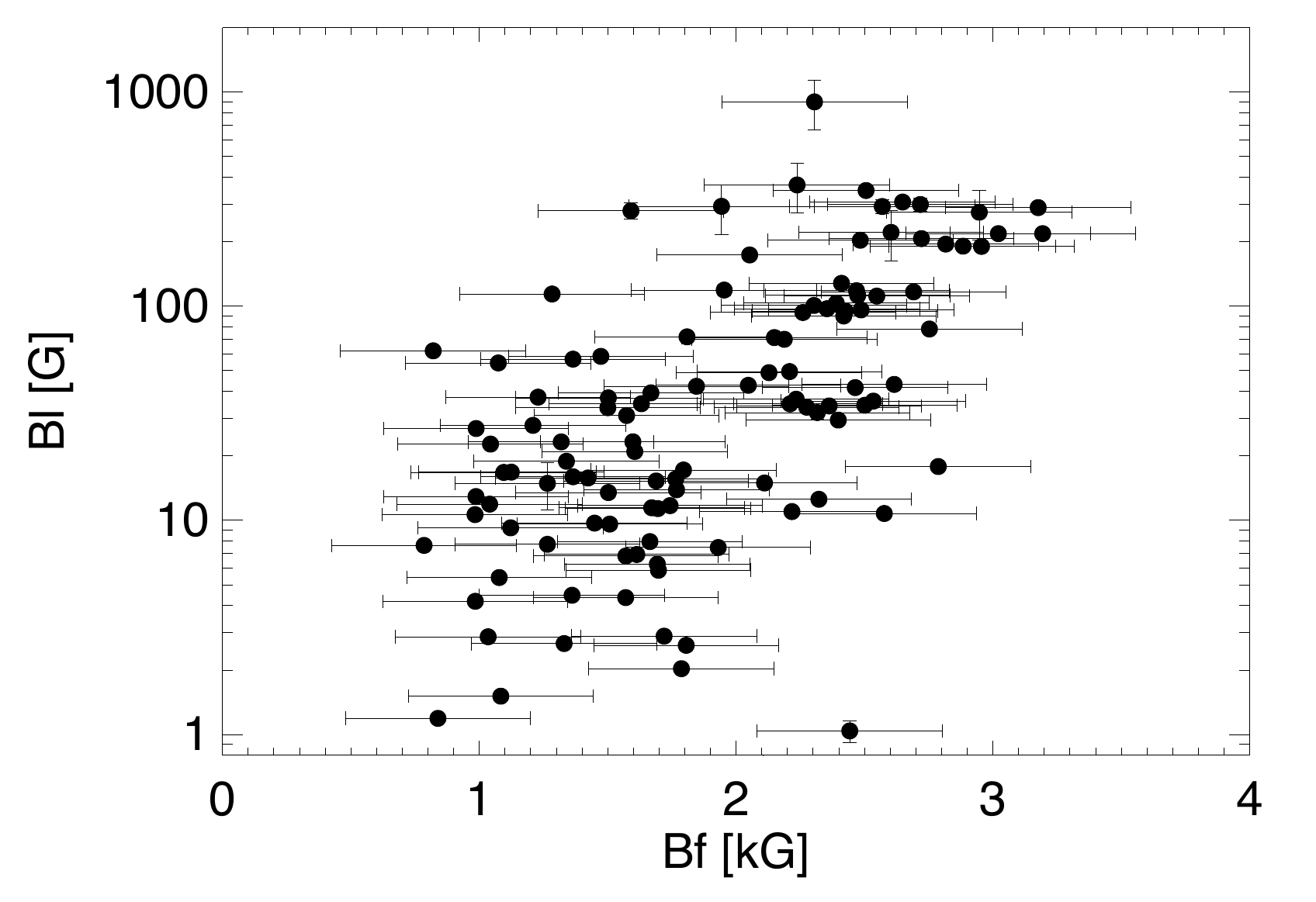}
    \caption{The large-scale longitudinal field as a function of the small-scale average field. Note the linear scale of the x-axis and logarithmic scale of the y-axis; most $B_l$ errors are within the symbol size. 
    }
    \label{fig:blzee}
\end{figure}

Finally, in Figure \ref{fig:blzee}, we show how the measurement of the unsigned value of the longitudinal large-scale field compares to the small-scale field measured through the Zeeman broadening of the FeH magnetically sensitive line. Both parameters have been conjointly measured on a total of 151 spectra and 59 different stars. The figure illustrates how the large-scale field can span several orders of magnitude (y-axis is in logarithmic scale) for a given small-scale field value (x-axis in linear scale). There is a 47\% correlation coefficient between both quantities. While inclination and topology impact the way one pictures the large-scale field, the small-scale field accounted for in $Bf$ is concentrated on active regions that can be seen at a wider range of inclinations and, for active stars, at most rotational phases.  A possible retroaction of one scale to another may also be due to physical processes (related convection and dynamo) which differ from a star to another. So a moderate correlation between these quantities of the global sample may result from a mix of stars where the correlation may vary widely due to differences in the field topology. The percentage of the maximum longitudinal field with respect to the total $Bf$ field is less than 5\% in most of the sample and rarely beyond 10\%. For GJ~1289, GJ~793, and GJ~251, it is, respectively, 5.5, 1.9, and 1.6\%.

\subsection{Prospects for stellar jitter and planet search}
Ultimately, the study of activity diagnostics on stars on which radial-velocity planet search will be conducted needs to assess how each diagnostic contributes to the stellar RV jitter.
In this study, we do not use the stellar radial-velocity jitter measured by ESPaDOnS, because this spectrograph is not optimized for RV precision better than $\sim$ 20m/s (instrumental floor) \citep{Moutou07}. 
For instance, our 22 RV measurements of GJ~793 have an $rms$ of 17~m/s. 

\begin{figure}
	\includegraphics[width=\columnwidth]{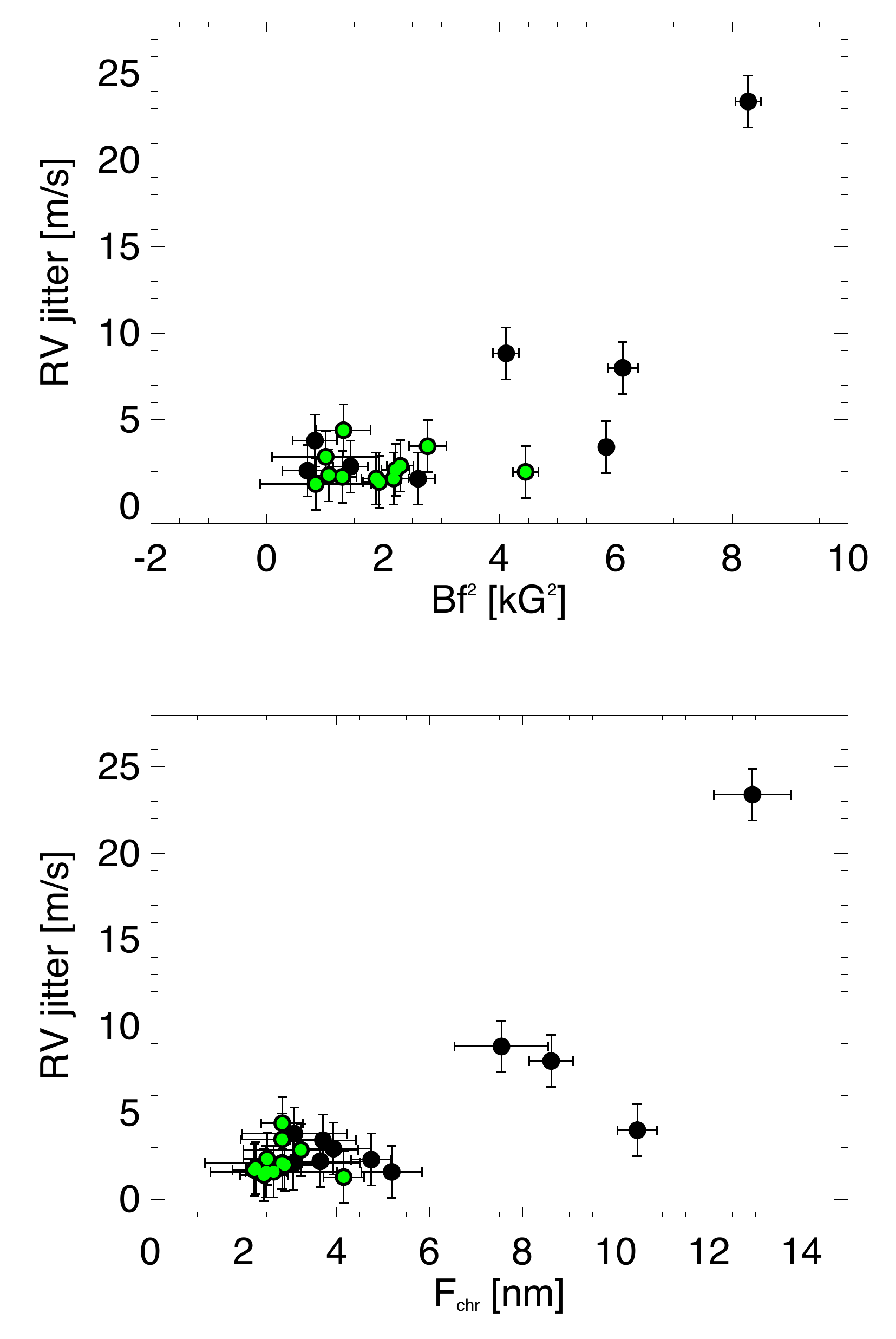}
    \caption{Literature RV jitter values as a function of the small-scale magnetic field energy $(Bf)^2$ (top) and the chromospheric index $F_{chr}$ (bottom)}. Green symbols indicate systems where one or several planets have been identified, and their signal removed from the RV variations. 
    \label{fig:rvj}
\end{figure}

We have thus searched the literature for all published RV jitter values due to rotational activity and cross-matched these values with our sample, focusing on the stars for which a small-scale magnetic field is measured (using the value averaged over all different spectra for a given star). We found 20 stars for which both measurements are available. It must be noted, however, that $Bf$ and RV jitter measurements are not contemporaneous, since $Bf$ measurements come from the ESPaDOnS spectra and RV jitter measurements come from ESO/HARPS or Keck/HIRES instruments. All data roughly come from the last decade, but this span may be very large compared to characteristic activity timescales of some of these stars. This time discrepancy is expected to increase the dispersion since these activity proxies are naturally expected to evolve with time. Also, there may be signals of yet-undetected planets in some of these stars (signals for some, and noise for others!), which would artificially increase the jitter value. Finally, the RV jitter is expected to (slowly) vary with the wavelength and we did not take this into account since all spectrographs are in the optical. As these RV jitters have been measured with HARPS or HIRES, which have very similar spectral ranges, and on M stars, we expect that the effective wavelength of this jitter is toward the red end of the instrumental bandpasses, at about 650~nm. The most sensitive wavelength of M-star spectra with ESPaDOnS is around 730~nm. An arbitrary error of 1.5 m/s was applied to all RV jitter values, in excess to the values quoted in the respective papers; it aims at accounting for the uncertainty due to the reasons described above (non-contemporaneity primarily, then chromaticity). The possible presence of planet signals could, evidently, account for a larger excess.

In Table \ref{tab:rvjitter}, the stars with a know RV jitter 
are listed and references to the RV jitter values are provided. The quantities are plotted on Figure \ref{fig:rvj} as a function of $(Bf)^2$ and of the total chromospheric flux $F_{chr}$ (see section 4.1).
Apart from AD Leo (GJ~388), all stars have a reported RV jitter smaller than 10m/s and also lie at the lower end of the magnetic field/activity scale. 
The stars shown with an insert green circle are those where exoplanets have already been found and characterized (see references in Table \ref{tab:rvjitter}) and their signal(s) have been removed from the RV variations shown here. It is clear that the RV jitter of these stars is closer to the instrumental threshold than most of the others, except for GJ~179 \citep{Montet14} for which the jitter is relatively high while the small-scale magnetic field is weak. From the original discovery paper, however, it seems likely that instrumental jitter may partly account for the excess jitter \citep{Howard10}. On the other hand, it is surprising that GJ~876 shows a low RV jitter and a large Zeeman broadening. There are, however, only two ESPaDOnS measurements of GJ~876, taken 20 days apart (the rotational period of this star is estimated to be 91 days from \citet{Suarez17b}) while RV measurements span a period of more than 8 years. 
Also, as discussed previously, the average magnetic strength of GJ~876 was undetected by the spectral-synthesis approach and an upper limit of 0.2 kG had been estimated \citep{Reiners07}.

Assuming that the RV jitter is mainly due to the Zeeman broadening -which can be significant in these stars-, we searched for a quadratic behaviour of the RV jitter, as advocated in \citet{Reiners13} and found no significant correlation. In the same way, jitter and chromospheric activity are only mildly related for stars where these quantities are small; using log($R'_{HK}$) rather than $F_{chr}$ does not make the trend stronger. The latter conclusion does not support the findings of \citet{Suarez17c}, in which a linear relationship could be found between the RV jitter and the log($R'_{HK}$), per spectral type. This disagreement may come from the choice of targets, or it may due to the contemporaneity of data.

\begin{table*}
\begin{center}
\caption{For all stars in our sample that have radial-velocity activity-jitter values in the literature, are also listed: the photometric index $V-K$, the measured average field modulus, the adopted projected rotational velocity \vsini, the total chromospheric emission $F_{chr}$, the maximum of the absolute value of the longitudinal field, and the activity merit function $AMF$ (see text). Note that RV jitter are not contemporaneous to the other measurements. The "p" in col. 3 is a flag for known exoplanet systems. }
\label{tab:rvjitter}
\begin{tabular}{llccccccccc}
\hline
2MASS  & other  & Planet & $V-K$ & $Bf$  & \vsini & $F_{chr}$ & max(|$B_l$|) & $AMF$ & RV jitter & Ref \\ 
name   & name   & flag   &       & [kG]  & [km/s]     &  nm    &    [G] &     &  [m/s] &  \\
\hline
 J00182256+4401222&  GJ~15A& p &4.12&  1.0 & 2.6    &  2.30 &  5.4 &     94.4&  1.80&  \citet{Howard14} \\
 J04520573+0628356&  GJ~179&p  & 5.02& 1.1 & $<$ 2.0&  2.83 &   -  &     41.2&  4.40&   \citet{Howard10}\\
 J05312734-0340356&  GJ~205&   &3.93&  1.2 & $<$ 2.0&  4.74 &   -  &     33.6&  2.30&  \citet{Hebrard16}\\
 J06103462-2151521&  GJ~229A&p &4.02&  0.9 & $<$ 2.0&  4.15 &  5.7 &     39.0&  1.29&  \citet{Suarez17b}\\
 J07272450+0513329&  GJ~273 &p & 4.97& 1.4 & $<$ 2.0&  2.65 &  9.3 &     38.1&  1.60&  \citet{Astudillo17b}\\
 J08405923-2327232&  GJ~317& p &4.95&  1.6 & $<$ 2.0&  2.83 & 66.4 &     39.1&  3.47&  \citet{Butler17}\\
 J10121768-0344441&  GJ~382  & & 4.25& 1.6 & $<$ 2.0&  5.18 & 12.7 &     33.9&  1.59&  \citet{Suarez17b}\\
 J10193634+1952122&  GJ~388&   &4.87&  2.9 & 4.1    & 12.94 &218.4 &     12.2& 23.40&  \citet{Reiners13}\\
 J11023832+2158017&  GJ~410&   &3.98&  2.0 & 3.0    &  7.54 & 27.2 &     45.9&  8.84&  \citet{Hebrard16}\\
 J11421096+2642251&  GJ~436& p &4.54&  1.1 & $<$ 2.0&  2.23 & 24.5 &     49.1&  1.70&  \citet{Lanotte14}\\
 J13295979+1022376&  GJ~514&   &3.99&  0.8 & 2.0    &  3.07 & 11.3 &     72.7&  2.06& \citet{Suarez17b}\\
 J13454354+1453317&  GJ~526&   &4.01&  0.9 & $<$ 2.0&  3.09 & 12.7 &     70.1&  3.80& \citet{Suarez17b}\\
 J14010324-0239180&  GJ~536& p &3.99&  1.0 & $<$ 2.0&  3.23 & 47.5 &     70.3&  2.86& \citet{Suarez536}\\
 J15192689-0743200&  GJ~581& p &4.73&  1.4 & $<$ 2.0&  2.44 & 16.7 &     35.2&  1.41&\citet{Suarez17b}\\
 J16252459+5418148&  GJ~625& p &4.23&  1.4 & 2.2    &  2.49 & 26.9 &     29.6&  1.60& \citet{Suarez17c}\\
 J16301808-1239434&  GJ~628 & p&5.04&  1.5 & $<$ 2.0&  2.51 & 37.6 &     40.1&  2.34& \citet{Astudillo17b}\\
 J18073292-1557464&  GJ~1224&  &5.71&  2.5 & 4.3    &  8.61 & 316.0&     10.6&  8.00&  \citet{Bonfils13}\\
 J22021026+0124006&  GJ~846&   &3.92&  2.4 & $<$ 2.0&  3.70 & 13.2 &     37.8&  3.42& \citet{Suarez17b}\\
 J22094029-0438267&  GJ~849& p &4.75&  1.5 & $<$ 2.0&  2.83 &   -  &     70.3&  2.10&  \citet{Bonfils13}\\
 J22531672-1415489&  GJ~876& p &5.18&  2.1 & $<$ 2.0&  2.88 &  2.7 &     44.4&  1.99& \citet{Suarez17b}\\
\hline
\end{tabular}
\end{center}
\end{table*}

Although most RV programs have leaned towards removing most active stars from their input sample in the past, it is not the only way to handle the problem of activity induced jitter. For M stars especially, eliminating active stars may be a strong limiting factor, especially for later types where activity is more pronounced. \citet{Hebrard16} has shown that the temporal behaviour of RV jitter in M stars was twofold: 1) a rotational component with signatures modulated at the rotational period and its harmonics and 2) a random component. This study also demonstrated that characterizing the rotational properties of stars (rotation period and differential rotation) from Doppler Imaging turned out to be a powerful asset in modeling the rotationally-modulated component of this jitter. The use of ZDI also proved powerful in filtering out the activity for the T Tauri stars V830 Tau \citep{Donati16} and TaP 26 \citep{Yu17}, allowing the detection of the hot-Jupiter planets orbiting these extremely active stars. Thus, as dealing with activity of M dwarfs in a planet-search RV survey is inevitable, it actually has mitigating solutions when the activity signature can be understood, measured and (at least partially) filtered out.

\subsection{A merit function for activity?}
We then attempt a classification of stars by considering the multiple measurements that are indirectly related to activity: the non-thermal radiation from the chromosphere in different lines, the average field modulus, the properties of the large-scale field, and the rotational velocity. The degree of correlation between these diagnostics is variable and not very high, as seen earlier, but we can still combine various indicators to build up a quantitative merit function that is relevant to the level of jitter amplitude and thus, compare the relevance of stars for exoplanet search. While chromospheric emission and rotational velocity can be estimated in all spectra, the other diagnostics are not always measurable, which results in some inhomogeneities in building up this merit function. Our attempt to estimate the activity merit function of a given star is given below:
\begin{equation}
AMF = w_c\frac{N_c}{F_{chr}} +w_{B}\frac{N_{B}}{max|B_l|}+w_v\frac{N_v}{v \sin i}+w_{Z}\frac{N_{Z}}{\delta v_B}
\end{equation}
where $F_{chr}$ is the total chromospheric emission, $max|B_l|$ the absolute value of the longitudinal field, and $\delta v_B$ the Zeeman broadening. The different $w$ and $N$ are weight factors and normalization factors, respectively. Normalization factors are chosen as the median of each parameter. Weight factors are more arbitrary as they depend upon the objective of the ranking. 

\begin{figure}
	\includegraphics[width=\columnwidth]{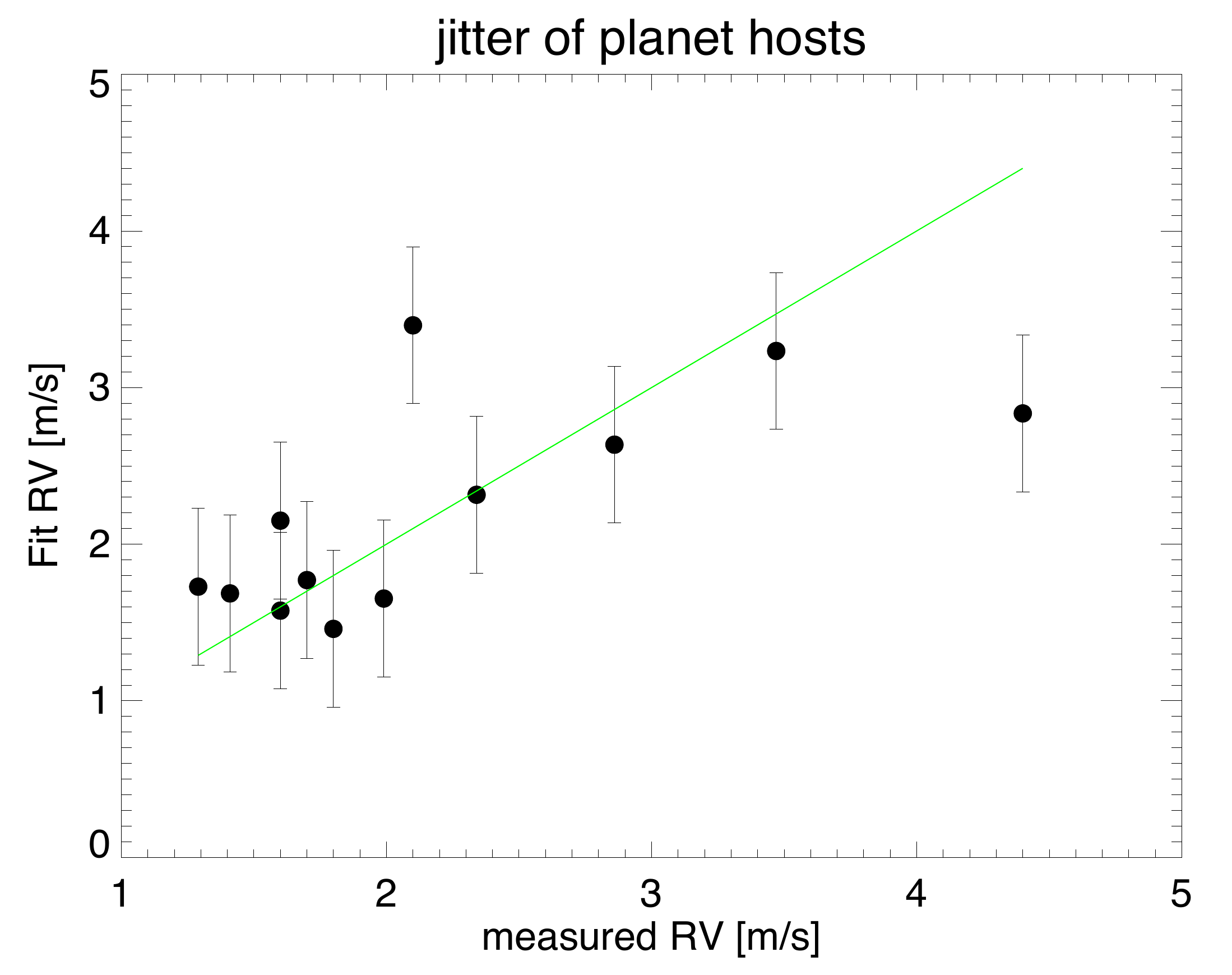}
    \caption{The adjusted RV jitter for known planet host stars in our sample, as a function of the measured one. The line shows Y=X. The RV fit is done from the measured or assumed values of max(|$B_l$|), $Bf$, \vsini\ and $F_{chr}$.\label{rvjfit}}
\end{figure}

In order to find the best weighting factors, we used a multi-variable fit with the four diagnostic parameters and adjust the RV jitter of the planet-host sample. This sub-sample is preferred to the whole sample with RV jitters shown earlier, as planet signals should mostly be removed. The best-fit is shown on Figure \ref{rvjfit}. The system with a large measured RV jitter and small predicted jitter is still GJ~179 \citep{Montet14}, where the jitter may be over-estimated, as discussed earlier. On the other hand, GJ~876 predicted jitter is 1.7 m/s, very close to the 1.99 m/s measured value, despite the large field modulus. The coefficients derived from this fit tell us that 39, 18, 35 and 8\% of the jitter contribution come from, respectively, max(|$B_l$|), $Bf$, $F_{chr}$ and \vsini. It thus gives a larger weight to the longitudinal field and chromospheric emission. We then used this weighting factors to derive the activity merit function and a predicted jitter value for the 442 stars in our sample. 

When no longitudinal field is available because the spectroscopic mode is used, we impose a median value to this parameter based on the histogram of $B_l$, in order to have a neutral effect on the ranking. When the spectrum is in polarimetric mode and the detection is null, we adopt a high value. Finally, when the Zeeman broadening is not detected while the star is a slow rotator, we also adopt a high value. 
The ranking is measured for each of the 1878 spectra and then averaged out per star. The final $AMF$ ranges from 0 to 103. The individual values for the sub-sample of stars on which the Zeeman broadening is detected are listed in the Appendix Table \ref{tab:jitterderived}.  Figure \ref{fig:meritf} shows the values obtained as a function of the $V-K$ colour index. There is no visible colour effect between $V-K$ values of 3 to 6, while most late-type stars tend to have a low activity merit function. In order to select the quietest stars in each bin of spectral type, one should set a simple horizontal threshold, such as the line shown in Figure \ref{fig:meritf}. Most of the slowest rotators in our sample lie above the line (and thus are deemed relevant targets for planet search). Interestingly, all slow rotators below the line, except one (GJ~406) are actually part of a spectroscopic binary system (denoted with a circling cyan diamond for clarity); this is not clear if the activity of those stars is actually enhanced, or if, in some cases, the measurements are impacted by the binarity. The predicted median jitter for stars with $AMF$ greater than 20 is 2.3~m/s, with most stars having a jitter less than 4~m/s. On the other hand, the median is 4.2~m/s for stars having an $AMF$ smaller than 20 and their distribution has a long tail towards large jitter values. 
The threshold of $AMF=20$, as shown on the figure selects $\sim$40\% of the stars, with some distribution in colour. The mean $V-K$ colour of stars above (below) the line is 4.57 (resp., 4.93), so there is a definite tendency for late-type M in our sample to be less quiet than earlier type M stars. Finally, it is interesting to note that the merit function shows a bimodal distribution, in the same way as rotation periods of M stars show \citep[\emph{e.g.,}][]{McQuillan13,Newton16}.\\

Stars with known exoplanets are featured in Figure \ref{fig:meritf} as blue stars, and they all lie at high values of the activity merit function. It could be expected since they have been used to derive the coefficients or the $AMF$, but it is also due to the fact that past RV surveys have barely observed active M stars, and even less found planets around them. The stars GJ~251, GJ~793, and GJ~1289 (red squares) also get a relatively high ranking of, respectively, 44, 39 and 23 and predicted jitter values of 2.9, 2.6 and 6.0 m/s. Stars with measured RV jitter values are depicted as green circles, with a size that is proportional to the jitter. Apart from GJ~388 and GJ~1224, all stars with measured jitter have a high activity merit function.\\ 

\begin{figure*}
	\includegraphics[width=2\columnwidth]{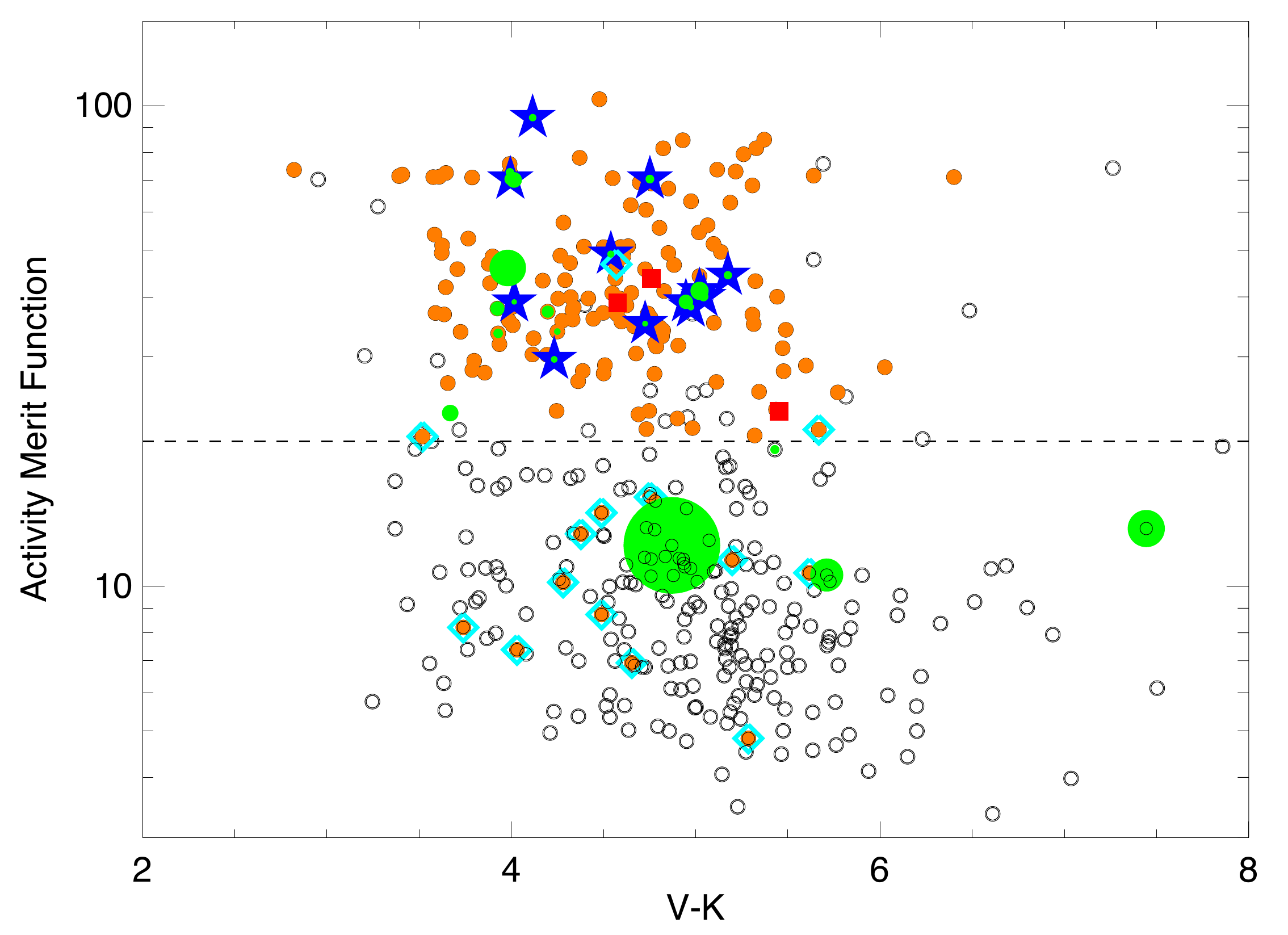}
    \caption{The activity "merit function" (see text) as a function of the $V-K$ colour index. Symbols with an inner orange circle show the slow rotators (unresolved profiles), green symbols show the stars with known activity jitter measurements with a size that is proportional to the jitter, and red squares show the three stars GJ~251, GJ~793 and GJ~1289. Cyan diamonds indicate the stars that are a component of spectroscopic binary systems. The blue star symbols show hosts of known exoplanet systems. The dash line is an arbitrary threshold proposed to select stars most favorable to planet searches.}\label{fig:meritf}
\end{figure*}

Further improvements of the activity merit function could still be obtained with a more thorough investigation of the stars with a measured jitter value, and a larger sample of these stars with a wider range of jitter values. It would be important, for instance, to precisely quantify the part of the RV $rms$ due to rotation modulated activity from other noise sources, as planets and instrumental systematics. It would then be possible to derive the best merit function from the non-rotationally modulated jitter itself.

\section{Summary and Conclusion}
In this study, we have collected and homogeneously analyzed all CFHT/ESPaDOnS data taken on the large sample of 442 M dwarfs, with a focus on their activity properties. Stellar activity takes many different faces, and we are interested in any and all diagnostics that relate to the radial-velocity stellar jitter, with the objectives of selecting proper targets and preparing efficient activity filtering techniques in planet-search programs. As ESPaDOnS cannot measure radial velocity jitter itself, we have cross-matched our data with HARPS or HIRES published data and shown that the amplitude of the RV jitter is somehow predictable: large Zeeman broadening, strong magnetic strength in the large scale, fast rotation and/or large chromospheric emission are all prone to higher activity jitter, with identified relative contributions: the maximum longitudinal field proves to be a quantity as important as the non-thermal emission, which shows the importance of measuring circular polarisation of stars. 

As commented in \citet{Hebrard16}, this RV jitter usually has a rotational component and a non-periodic component. If the first one can reasonably be filtered out from RV time series \citep[\emph{e.g.},][]{Boisse09,Boisse11,Petit15,Donati16,Yu17}, the second component is more evasive and would benefit from additional contemporaneous spectroscopic indicators, as the Zeeman broadening variation and chromospheric (flaring) emission. The mild correlation between the emission of several chromospheric emission lines in the optical demonstrates that these activity tracers are not straightforward and deserve a cautious analysis. On the other hand, the role of the rotation period to improve the activity filtering efficiency is critical. If rotation periods can be measured photometrically \citep{Newton16}, or are usually clearly found in the line circular polarization signal \citep{Hebrard16}, it can also be inferred from the level of the non-thermal emission observed in the CaII or H$\alpha$ lines \citep{Astudillo17a,Newton17}. These latter relations are only valid for the slow rotators out of the saturation regime. It is a primordial characterization of the system, as it modulates the activity and may interfere with planetary signals. 

In the nIR domain where the next-generation spectropolarimeter CFHT/SPIRou will operate, spectroscopic diagnostics of activity are still to be explored, in particular their effect on the radial-velocity jitter. SPIRou spectra will, however, allow a more general measurement of the Zeeman broadening since this effect is larger in the nIR for a given magnetic field modulus, and it is expected that this measurement on several atomic and biatomic lines will allow to trace the jitter due to localized magnetic regions, as simulated in \citet{Reiners13} and convincingly shown for the Sun \citep{Haywood16}.

In addition to the global description of the 442 star sample, we have reconstructed the magnetic topology at the surface of three stars that had not yet been scrutinized: the partly convective stars GJ~793 and GJ~251 and the fully convective star GJ~1289. All three stars have long rotation periods (22, 90, and 54 days, respectively) and are relatively quiet. With a mass lower than 0.45 \msun, they represent the type of stars that SPIRou could monitor in search for exoplanets. Their surface magnetic field is similar to the field of most M stars with much shorter rotation periods or Rossby numbers smaller than 0.1, with a predominent poloidal topology. Further work will focus on the implications for the dynamo processes in M stars, especially in the transition zone from partly to fully convective stars, as in \citet{Morin10,Morin11} and \citet{Gastine13}.

The SPIRou input catalog generation will use the inputs from this work as well as the distribution of fundamental parameters (temperature, gravity, mass, metalicity and projected velocity, see Fouqu\'e et al, in prep.) to characterize and select targets of the SPIRou survey. In addition to the archive of M stars observed with ESPaDOnS, other catalogs and other archives are explored to complete the list of low-mass stars with relevant properties for planet search. This information gathering and method of target selection, including activity characterization, will be published in a subsequent work (Malo et al, in prep.).

\section*{Acknowledgements}
The authors thank the referee Prof. Basri for his critical reading and highly appreciate the comments which significantly contributed to improving the quality of the publication. The authors made use of CFHT/ESPaDOnS data. CFHT is operated by the National Research Council (NRC) of Canada, the Institut National des Science de l'Univers of the Centre National de la Recherche Scientifique (CNRS) of France, and the University of Hawaii. This work is based in part on data products available at the Canadian Astronomy Data Centre (CADC) as part of the CFHT Data Archive. CADC is operated by the National Research Council of Canada with the support of the Canadian Space Agency. This work has been partially supported by the Labex OSUG2020. X.D. acknowledges the support of the CNRS/INSU PNP and PNPS (Programme National de Plan\'etologie and Physique Stellaire).

\bibliographystyle{mnras}
\bibliography{coolsnap3} 

\begin{thebibliography}{}
\makeatletter
\relax
\def\mn@urlcharsother{\let\do\@makeother \do\$\do\&\do\#\do\^\do\_\do\%\do\~}
\def\mn@doi{\begingroup\mn@urlcharsother \@ifnextchar [ {\mn@doi@}
  {\mn@doi@[]}}
\def\mn@doi@[#1]#2{\def\@tempa{#1}\ifx\@tempa\@empty \href
  {http://dx.doi.org/#2} {doi:#2}\else \href {http://dx.doi.org/#2} {#1}\fi
  \endgroup}
\def\mn@eprint#1#2{\mn@eprint@#1:#2::\@nil}
\def\mn@eprint@arXiv#1{\href {http://arxiv.org/abs/#1} {{\tt arXiv:#1}}}
\def\mn@eprint@dblp#1{\href {http://dblp.uni-trier.de/rec/bibtex/#1.xml}
  {dblp:#1}}
\def\mn@eprint@#1:#2:#3:#4\@nil{\def\@tempa {#1}\def\@tempb {#2}\def\@tempc
  {#3}\ifx \@tempc \@empty \let \@tempc \@tempb \let \@tempb \@tempa \fi \ifx
  \@tempb \@empty \def\@tempb {arXiv}\fi \@ifundefined
  {mn@eprint@\@tempb}{\@tempb:\@tempc}{\expandafter \expandafter \csname
  mn@eprint@\@tempb\endcsname \expandafter{\@tempc}}}

\bibitem[\protect\citeauthoryear{{Afram}, {Berdyugina}, {Fluri}, {Solanki}  \&
  {Lagg}}{{Afram} et~al.}{2008}]{Afram08}
{Afram} N.,  {Berdyugina} S.~V.,  {Fluri} D.~M.,  {Solanki} S.~K.,   {Lagg} A.,
   2008, \mn@doi [\aap] {10.1051/0004-6361:20079300}, \href
  {http://adsabs.harvard.edu/abs/2008A%26A...482..387A} {482, 387}

\bibitem[\protect\citeauthoryear{{Andersen} \& {Korhonen}}{{Andersen} \&
  {Korhonen}}{2015}]{Andersen15}
{Andersen} J.~M.,  {Korhonen} H.,  2015, \mn@doi [\mnras]
  {10.1093/mnras/stu2731}, \href
  {http://adsabs.harvard.edu/abs/2015MNRAS.448.3053A} {448, 3053}

\bibitem[\protect\citeauthoryear{{Artigau}, {Donati}  \& {Delfosse}}{{Artigau}
  et~al.}{2011}]{Artigau11}
{Artigau} {\'E}.,  {Donati} J.-F.,   {Delfosse} X.,  2011, in {Johns-Krull} C.,
   {Browning} M.~K.,   {West} A.~A.,  eds,  Astronomical Society of the Pacific
  Conference Series Vol. 448, 16th Cambridge Workshop on Cool Stars, Stellar
  Systems, and the Sun. p.~771

\bibitem[\protect\citeauthoryear{{Astudillo-Defru} et~al.,}{{Astudillo-Defru}
  et~al.}{2017a}]{Astudillo17b}
{Astudillo-Defru} N.,  et~al., 2017a, preprint, \href
  {http://esoads.eso.org/abs/2017arXiv170305386A} {} (\mn@eprint {arXiv}
  {1703.05386})

\bibitem[\protect\citeauthoryear{{Astudillo-Defru}, {Delfosse}, {Bonfils},
  {Forveille}, {Lovis}  \& {Rameau}}{{Astudillo-Defru}
  et~al.}{2017b}]{Astudillo17a}
{Astudillo-Defru} N.,  {Delfosse} X.,  {Bonfils} X.,  {Forveille} T.,  {Lovis}
  C.,   {Rameau} J.,  2017b, \mn@doi [\aap] {10.1051/0004-6361/201527078},
  \href {http://adsabs.harvard.edu/abs/2017A%26A...600A..13A} {600, A13}

\bibitem[\protect\citeauthoryear{{Baraffe}, {Chabrier}, {Allard}  \&
  {Hauschildt}}{{Baraffe} et~al.}{1998}]{Baraffe98}
{Baraffe} I.,  {Chabrier} G.,  {Allard} F.,   {Hauschildt} P.~H.,  1998, \aap,
  \href {http://adsabs.harvard.edu/abs/1998A\%26A...337..403B} {337, 403}

\bibitem[\protect\citeauthoryear{{Benedict} et~al.,}{{Benedict}
  et~al.}{2016}]{Benedict16}
{Benedict} G.~F.,  et~al., 2016, \mn@doi [\aj] {10.3847/0004-6256/152/5/141},
  \href {http://esoads.eso.org/abs/2016AJ....152..141B} {152, 141}

\bibitem[\protect\citeauthoryear{{Boisse} et~al.,}{{Boisse}
  et~al.}{2009}]{Boisse09}
{Boisse} I.,  et~al., 2009, \mn@doi [\aap] {10.1051/0004-6361:200810648}, \href
  {http://cdsads.u-strasbg.fr/abs/2009A\%26A...495..959B} {495, 959}

\bibitem[\protect\citeauthoryear{{Boisse}, {Bouchy}, {H{\'e}brard}, {Bonfils},
  {Santos}  \& {Vauclair}}{{Boisse} et~al.}{2011}]{Boisse11}
{Boisse} I.,  {Bouchy} F.,  {H{\'e}brard} G.,  {Bonfils} X.,  {Santos} N.,
  {Vauclair} S.,  2011, \aap, 528

\bibitem[\protect\citeauthoryear{{Bonfils} et~al.,}{{Bonfils}
  et~al.}{2007}]{Bonfils07}
{Bonfils} X.,  et~al., 2007, \mn@doi [\aap] {10.1051/0004-6361:20077068}, \href
  {http://adsabs.harvard.edu/abs/2007A%26A...474..293B} {474, 293}

\bibitem[\protect\citeauthoryear{{Bonfils}, {Delfosse}, {Udry}, {Forveille},
  {Mayor}, {Perrier}  \& {Bouchy}}{{Bonfils} et~al.}{2013}]{Bonfils13}
{Bonfils} X.,  {Delfosse} X.,  {Udry} S.,  {Forveille} T.,  {Mayor} M.,
  {Perrier} C.,   {Bouchy} F.,  2013, \aap, 549, A109

\bibitem[\protect\citeauthoryear{{Brown}, {K{\"o}rsgen}, {Beaton}  \&
  {Evenson}}{{Brown} et~al.}{2006}]{Brown06}
{Brown} J.~M.,  {K{\"o}rsgen} H.,  {Beaton} S.~P.,   {Evenson} K.~M.,  2006,
  \mn@doi [\jcp] {10.1063/1.2198843}, \href
  {http://esoads.eso.org/abs/2006JChPh.124w4309B} {124, 234309}

\bibitem[\protect\citeauthoryear{{Butler} et~al.,}{{Butler}
  et~al.}{2017}]{Butler17}
{Butler} R.~P.,  et~al., 2017, preprint, \href
  {http://esoads.eso.org/abs/2017arXiv170203571B} {} (\mn@eprint {arXiv}
  {1702.03571})

\bibitem[\protect\citeauthoryear{{Chabrier} \& {Baraffe}}{{Chabrier} \&
  {Baraffe}}{2000}]{Chabrier00}
{Chabrier} G.,  {Baraffe} I.,  2000, \mn@doi [\araa]
  {10.1146/annurev.astro.38.1.337}, \href
  {http://cdsads.u-strasbg.fr/abs/2000ARA\%26A..38..337C} {38, 337}

\bibitem[\protect\citeauthoryear{{Crozet}, {Dobrev}, {Richard}  \&
  {Ross}}{{Crozet} et~al.}{2014}]{Crozet14}
{Crozet} P.,  {Dobrev} G.,  {Richard} C.,   {Ross} A.~J.,  2014, \mn@doi
  [Journal of Molecular Spectroscopy] {10.1016/j.jms.2014.07.005}, \href
  {http://esoads.eso.org/abs/2014JMoSp.303...46C} {303, 46}

\bibitem[\protect\citeauthoryear{{Delfosse}, {Forveille}, {Perrier}  \&
  {Mayor}}{{Delfosse} et~al.}{1998}]{Delfosse98}
{Delfosse} X.,  {Forveille} T.,  {Perrier} C.,   {Mayor} M.,  1998, \aap, \href
  {http://adsabs.harvard.edu/abs/1998A%26A...331..581D} {331, 581}

\bibitem[\protect\citeauthoryear{{Donati}}{{Donati}}{2003}]{Donati03c}
{Donati} J.-F.,  2003. p.~41

\bibitem[\protect\citeauthoryear{{Donati} \& {Landstreet}}{{Donati} \&
  {Landstreet}}{2009}]{Donati09}
{Donati} J.,  {Landstreet} J.~D.,  2009, \mn@doi [\araa]
  {10.1146/annurev-astro-082708-101833}, \href
  {http://adsabs.harvard.edu/abs/2009ARA%26A..47..333D} {47, 333}

\bibitem[\protect\citeauthoryear{{Donati}, {Semel}, {Carter}, {Rees}  \&
  {Collier Cameron}}{{Donati} et~al.}{1997}]{Donati97b}
{Donati} J.-F.,  {Semel} M.,  {Carter} B.~D.,  {Rees} D.~E.,   {Collier
  Cameron} A.,  1997, \mnras, \href
  {http://adsabs.harvard.edu/abs/1997MNRAS.291..658D} {291, 658}

\bibitem[\protect\citeauthoryear{{Donati}, {Wade}, {Babel}, {Henrichs}, {de
  Jong}  \& {Harries}}{{Donati} et~al.}{2001}]{Donati01b}
{Donati} J.-F.,  {Wade} G.~A.,  {Babel} J.,  {Henrichs} H.~f.,  {de Jong}
  J.~A.,   {Harries} T.~J.,  2001, \mn@doi [\mnras]
  {10.1046/j.1365-8711.2001.04713.x}, \href
  {http://adsabs.harvard.edu/abs/2001MNRAS.326.1265D} {326, 1265}

\bibitem[\protect\citeauthoryear{{Donati}, {Forveille}, {Cameron}, {Barnes},
  {Delfosse}, {Jardine}  \& {Valenti}}{{Donati} et~al.}{2006a}]{Donati06a}
{Donati} J.-F.,  {Forveille} T.,  {Cameron} A.~C.,  {Barnes} J.~R.,  {Delfosse}
  X.,  {Jardine} M.~M.,   {Valenti} J.~A.,  2006a, \mn@doi [Science]
  {10.1126/science.1121102}, \href
  {http://adsabs.harvard.edu/abs/2006Sci...311..633D} {311, 633}

\bibitem[\protect\citeauthoryear{{Donati} et~al.,}{{Donati}
  et~al.}{2006b}]{Donati06b}
{Donati} J.-F.,  et~al., 2006b, \mn@doi [\mnras]
  {10.1111/j.1365-2966.2006.10558.x}, \href
  {http://adsabs.harvard.edu/abs/2006MNRAS.370..629D} {370, 629}

\bibitem[\protect\citeauthoryear{{Donati} et~al.,}{{Donati}
  et~al.}{2008}]{Donati08c}
{Donati} J.-F.,  et~al., 2008, \mn@doi [\mnras]
  {10.1111/j.1365-2966.2008.13799.x}, \href
  {http://adsabs.harvard.edu/abs/2008MNRAS.390..545D} {390, 545}

\bibitem[\protect\citeauthoryear{{Donati} et~al.,}{{Donati}
  et~al.}{2016}]{Donati16}
{Donati} J.~F.,  et~al., 2016, \mn@doi [\nat] {10.1038/nature18305}, \href
  {http://cdsads.u-strasbg.fr/abs/2016Natur.534..662D} {534, 662}

\bibitem[\protect\citeauthoryear{{Dressing} \& {Charbonneau}}{{Dressing} \&
  {Charbonneau}}{2015}]{Dressing15}
{Dressing} C.~D.,  {Charbonneau} D.,  2015, \mn@doi [\apj]
  {10.1088/0004-637X/807/1/45}, \href
  {http://adsabs.harvard.edu/abs/2015ApJ...807...45D} {807, 45}

\bibitem[\protect\citeauthoryear{{Dumusque} et~al.,}{{Dumusque}
  et~al.}{2017}]{Dumusque17}
{Dumusque} X.,  et~al., 2017, \mn@doi [\aap] {10.1051/0004-6361/201628671},
  \href {http://esoads.eso.org/abs/2017A%26A...598A.133D} {598, A133}

\bibitem[\protect\citeauthoryear{{Fares} et~al.,}{{Fares}
  et~al.}{2010}]{Fares10}
{Fares} R.,  et~al., 2010, \mn@doi [\mnras] {10.1111/j.1365-2966.2010.16715.x},
  \href {http://cdsads.u-strasbg.fr/abs/2010MNRAS.406..409F} {406, 409}

\bibitem[\protect\citeauthoryear{{Gastine}, {Morin}, {Duarte}, {Reiners},
  {Christensen}  \& {Wicht}}{{Gastine} et~al.}{2013}]{Gastine13}
{Gastine} T.,  {Morin} J.,  {Duarte} L.,  {Reiners} A.,  {Christensen} U.~R.,
  {Wicht} J.,  2013, \mn@doi [\aap] {10.1051/0004-6361/201220317}, \href
  {http://cdsads.u-strasbg.fr/abs/2013A%26A...549L...5G} {549, L5}

\bibitem[\protect\citeauthoryear{{Gomes Da Silva}, {Santos}, {Bonfils}  \&
  {Delfosse}}{{Gomes Da Silva} et~al.}{2011}]{DaSilva11}
{Gomes Da Silva} J.,  {Santos} N.,  {Bonfils} X.,   {Delfosse} X.,  2011, \aap,
  534, A30

\bibitem[\protect\citeauthoryear{{Harrison}, {Brown}, {Chen}, {Steimle}  \&
  {Sears}}{{Harrison} et~al.}{2008}]{Harrison08}
{Harrison} J.~J.,  {Brown} J.~M.,  {Chen} J.,  {Steimle} T.~C.,   {Sears}
  T.~J.,  2008, \mn@doi [\apj] {10.1086/587169}, \href
  {http://esoads.eso.org/abs/2008ApJ...679..854H} {679, 854}

\bibitem[\protect\citeauthoryear{{Haywood} et~al.,}{{Haywood}
  et~al.}{2016}]{Haywood16}
{Haywood} R.~D.,  et~al., 2016, \mn@doi [\mnras] {10.1093/mnras/stw187}, \href
  {http://esoads.eso.org/abs/2016MNRAS.457.3637H} {457, 3637}

\bibitem[\protect\citeauthoryear{{H{\'e}brard}, {Donati}, {Delfosse}, {Morin},
  {Moutou}  \& {Boisse}}{{H{\'e}brard} et~al.}{2016}]{Hebrard16}
{H{\'e}brard} {\'E}.~M.,  {Donati} J.-F.,  {Delfosse} X.,  {Morin} J.,
  {Moutou} C.,   {Boisse} I.,  2016, \mn@doi [\mnras] {10.1093/mnras/stw1346},
  \href {http://esoads.eso.org/abs/2016MNRAS.461.1465H} {461, 1465}

\bibitem[\protect\citeauthoryear{{Houdebine}}{{Houdebine}}{2010}]{Houdebine10}
{Houdebine} E.~R.,  2010, \mn@doi [\mnras] {10.1111/j.1365-2966.2010.16827.x},
  \href {http://esoads.eso.org/abs/2010MNRAS.407.1657H} {407, 1657}

\bibitem[\protect\citeauthoryear{{Houdebine}}{{Houdebine}}{2012}]{Houdebine12}
{Houdebine} E.~R.,  2012, \mn@doi [\mnras] {10.1111/j.1365-2966.2012.20543.x},
  \href {http://esoads.eso.org/abs/2012MNRAS.421.3180H} {421, 3180}

\bibitem[\protect\citeauthoryear{{Houdebine} \& {Mullan}}{{Houdebine} \&
  {Mullan}}{2015}]{Houdebine15}
{Houdebine} E.~R.,  {Mullan} D.~J.,  2015, \mn@doi [\apj]
  {10.1088/0004-637X/801/2/106}, \href
  {http://esoads.eso.org/abs/2015ApJ...801..106H} {801, 106}

\bibitem[\protect\citeauthoryear{{Howard} et~al.,}{{Howard}
  et~al.}{2010}]{Howard10}
{Howard} A.~W.,  et~al., 2010, \mn@doi [\apj] {10.1088/0004-637X/721/2/1467},
  \href {http://esoads.eso.org/abs/2010ApJ...721.1467H} {721, 1467}

\bibitem[\protect\citeauthoryear{{Howard} et~al.,}{{Howard}
  et~al.}{2014}]{Howard14}
{Howard} A.~W.,  et~al., 2014, \mn@doi [\apj] {10.1088/0004-637X/794/1/51},
  \href {http://esoads.eso.org/abs/2014ApJ...794...51H} {794, 51}

\bibitem[\protect\citeauthoryear{{Kiraga}}{{Kiraga}}{2012}]{Kiraga12}
{Kiraga} M.,  2012, \actaa, \href
  {http://esoads.eso.org/abs/2012AcA....62...67K} {62, 67}

\bibitem[\protect\citeauthoryear{{Kiraga} \& {Stepien}}{{Kiraga} \&
  {Stepien}}{2007}]{Kiraga07}
{Kiraga} M.,  {Stepien} K.,  2007, Acta Astronomica, \href
  {http://adsabs.harvard.edu/abs/2007AcA....57..149K} {57, 149}

\bibitem[\protect\citeauthoryear{{Kurucz}}{{Kurucz}}{1993}]{Kurucz93}
{Kurucz} R.,  1993, CDROM \#~13 (ATLAS9 atmospheric models) and \#~18 (ATLAS9
  and SYNTHE routines, spectral line database).
Smithsonian Astrophysical Observatory, Washington D.C.

\bibitem[\protect\citeauthoryear{{Landi degl'Innocenti} \& {Landolfi}}{{Landi
  degl'Innocenti} \& {Landolfi}}{2004}]{Landi04}
{Landi degl'Innocenti} E.,  {Landolfi} M.,  2004, {Polarisation in spectral
  lines}.
Dordrecht/Boston/London: Kluwer Academic Publishers

\bibitem[\protect\citeauthoryear{{Lanotte} et~al.,}{{Lanotte}
  et~al.}{2014}]{Lanotte14}
{Lanotte} A.~A.,  et~al., 2014, \mn@doi [\aap] {10.1051/0004-6361/201424373},
  \href {http://esoads.eso.org/abs/2014A%26A...572A..73L} {572, A73}

\bibitem[\protect\citeauthoryear{{Maldonado} et~al.,}{{Maldonado}
  et~al.}{2017}]{Maldonado17}
{Maldonado} J.,  et~al., 2017, \mn@doi [\aap] {10.1051/0004-6361/201629223},
  \href {http://esoads.eso.org/abs/2017A%26A...598A..27M} {598, A27}

\bibitem[\protect\citeauthoryear{{Mart{\'{\i}}nez-Arn{\'a}iz},
  {L{\'o}pez-Santiago}, {Crespo-Chac{\'o}n}  \&
  {Montes}}{{Mart{\'{\i}}nez-Arn{\'a}iz} et~al.}{2011}]{Martinez11}
{Mart{\'{\i}}nez-Arn{\'a}iz} R.,  {L{\'o}pez-Santiago} J.,  {Crespo-Chac{\'o}n}
  I.,   {Montes} D.,  2011, \mn@doi [\mnras]
  {10.1111/j.1365-2966.2011.18584.x}, \href
  {http://adsabs.harvard.edu/abs/2011MNRAS.414.2629M} {414, 2629}

\bibitem[\protect\citeauthoryear{{McQuillan}, {Aigrain}  \&
  {Mazeh}}{{McQuillan} et~al.}{2013}]{McQuillan13}
{McQuillan} A.,  {Aigrain} S.,   {Mazeh} T.,  2013, \mn@doi [\mnras]
  {10.1093/mnras/stt536}, \href {http://esoads.eso.org/abs/2013MNRAS.432.1203M}
  {432, 1203}

\bibitem[\protect\citeauthoryear{{Montet}, {Crepp}, {Johnson}, {Howard}  \&
  {Marcy}}{{Montet} et~al.}{2014}]{Montet14}
{Montet} B.~T.,  {Crepp} J.~R.,  {Johnson} J.~A.,  {Howard} A.~W.,   {Marcy}
  G.~W.,  2014, \mn@doi [\apj] {10.1088/0004-637X/781/1/28}, \href
  {http://esoads.eso.org/abs/2014ApJ...781...28M} {781, 28}

\bibitem[\protect\citeauthoryear{{Morin} et~al.,}{{Morin}
  et~al.}{2008a}]{Morin08a}
{Morin} J.,  et~al., 2008a, \mnras, 384, 77

\bibitem[\protect\citeauthoryear{{Morin} et~al.,}{{Morin}
  et~al.}{2008b}]{Morin08b}
{Morin} J.,  et~al., 2008b, \mn@doi [\mnras]
  {10.1111/j.1365-2966.2008.13809.x}, \href
  {http://adsabs.harvard.edu/abs/2008MNRAS.390..567M} {390, 567}

\bibitem[\protect\citeauthoryear{{Morin}, {Donati}, {Petit}, {Delfosse},
  {Forveille}  \& {Jardine}}{{Morin} et~al.}{2010}]{Morin10}
{Morin} J.,  {Donati} J.,  {Petit} P.,  {Delfosse} X.,  {Forveille} T.,
  {Jardine} M.~M.,  2010, \mn@doi [\mnras] {10.1111/j.1365-2966.2010.17101.x},
  \href {http://adsabs.harvard.edu/abs/2010MNRAS.407.2269M} {407, 2269}

\bibitem[\protect\citeauthoryear{{Morin}, {Dormy}, {Schrinner}  \&
  {Donati}}{{Morin} et~al.}{2011}]{Morin11}
{Morin} J.,  {Dormy} E.,  {Schrinner} M.,   {Donati} J.-F.,  2011, \mn@doi
  [\mnras] {10.1111/j.1745-3933.2011.01159.x}, \href
  {http://cdsads.u-strasbg.fr/abs/2011MNRAS.418L.133M} {418, L133}

\bibitem[\protect\citeauthoryear{{Morin} et~al.,}{{Morin}
  et~al.}{2013}]{Morin13}
{Morin} J.,  et~al., 2013, \mn@doi [Astronomische Nachrichten]
  {10.1002/asna.201211771}, \href
  {http://esoads.eso.org/abs/2013AN....334...48M} {334, 48}

\bibitem[\protect\citeauthoryear{{Moutou} et~al.,}{{Moutou}
  et~al.}{2007}]{Moutou07}
{Moutou} C.,  et~al., 2007, \mn@doi [\aap] {10.1051/0004-6361:20077795}, \href
  {http://adsabs.harvard.edu/abs/2007A\%26A...473..651M} {473, 651}

\bibitem[\protect\citeauthoryear{{Newton}, {Irwin}, {Charbonneau},
  {Berta-Thompson}, {Dittmann}  \& {West}}{{Newton} et~al.}{2016}]{Newton16}
{Newton} E.~R.,  {Irwin} J.,  {Charbonneau} D.,  {Berta-Thompson} Z.~K.,
  {Dittmann} J.~A.,   {West} A.~A.,  2016, \mn@doi [\apj]
  {10.3847/0004-637X/821/2/93}, \href
  {http://cdsads.u-strasbg.fr/abs/2016ApJ...821...93N} {821, 93}

\bibitem[\protect\citeauthoryear{{Newton}, {Irwin}, {Charbonneau}, {Berlind},
  {Calkins}  \& {Mink}}{{Newton} et~al.}{2017}]{Newton17}
{Newton} E.~R.,  {Irwin} J.,  {Charbonneau} D.,  {Berlind} P.,  {Calkins}
  M.~L.,   {Mink} J.,  2017, \mn@doi [\apj] {10.3847/1538-4357/834/1/85}, \href
  {http://adsabs.harvard.edu/abs/2017ApJ...834...85N} {834, 85}

\bibitem[\protect\citeauthoryear{{Pecaut} \& {Mamajek}}{{Pecaut} \&
  {Mamajek}}{2013}]{Pecaut13}
{Pecaut} M.~J.,  {Mamajek} E.~E.,  2013, \mn@doi [\apjs]
  {10.1088/0067-0049/208/1/9}, \href
  {http://adsabs.harvard.edu/abs/2013ApJS..208....9P} {208, 9}

\bibitem[\protect\citeauthoryear{{Petit} et~al.,}{{Petit}
  et~al.}{2015}]{Petit15}
{Petit} P.,  et~al., 2015, preprint, \href
  {http://cdsads.u-strasbg.fr/abs/2015arXiv150300180P} {} (\mn@eprint {arXiv}
  {1503.00180})

\bibitem[\protect\citeauthoryear{{Reiners}}{{Reiners}}{2012a}]{Reiners2012a}
{Reiners} A.,  2012a, \mn@doi [Living Reviews in Solar Physics]
  {10.12942/lrsp-2012-1}, \href {http://esoads.eso.org/abs/2012LRSP....9....1R}
  {9, 1}

\bibitem[\protect\citeauthoryear{{Reiners}}{{Reiners}}{2012b}]{Reiners12}
{Reiners} A.,  2012b, \mn@doi [Living Reviews in Solar Physics]
  {10.12942/lrsp-2012-1}, \href
  {http://adsabs.harvard.edu/abs/2012LRSP....9....1R} {9, 1}

\bibitem[\protect\citeauthoryear{{Reiners} \& {Basri}}{{Reiners} \&
  {Basri}}{2006}]{Reiners06}
{Reiners} A.,  {Basri} G.,  2006, \mn@doi [\apj] {10.1086/503324}, \href
  {http://adsabs.harvard.edu/abs/2006ApJ...644..497R} {644, 497}

\bibitem[\protect\citeauthoryear{{Reiners} \& {Basri}}{{Reiners} \&
  {Basri}}{2007}]{Reiners07}
{Reiners} A.,  {Basri} G.,  2007, \mn@doi [\apj] {10.1086/510304}, \href
  {http://cdsads.u-strasbg.fr/abs/2007ApJ...656.1121R} {656, 1121}

\bibitem[\protect\citeauthoryear{{Reiners} \& {Basri}}{{Reiners} \&
  {Basri}}{2010}]{Reiners10b}
{Reiners} A.,  {Basri} G.,  2010, \mn@doi [\apj] {10.1088/0004-637X/710/2/924},
  \href {http://adsabs.harvard.edu/abs/2010ApJ...710..924R} {710, 924}

\bibitem[\protect\citeauthoryear{{Reiners}, {Shulyak}, {Anglada-Escud{\'e}},
  {Jeffers}, {Morin}, {Zechmeister}, {Kochuklov}  \& {Piskunov}}{{Reiners}
  et~al.}{2013}]{Reiners13}
{Reiners} A.,  {Shulyak} D.,  {Anglada-Escud{\'e}} G.,  {Jeffers} S.,  {Morin}
  J.,  {Zechmeister} M.,  {Kochuklov} O.,   {Piskunov} N.,  2013, \aap, 552

\bibitem[\protect\citeauthoryear{{Robertson}, {Mahadevan}, {Endl}  \&
  {Roy}}{{Robertson} et~al.}{2014}]{Robertson14}
{Robertson} P.,  {Mahadevan} S.,  {Endl} M.,   {Roy} A.,  2014, \mn@doi
  [Science] {10.1126/science.1253253}, \href
  {http://adsabs.harvard.edu/abs/2014Sci...345..440R} {345, 440}

\bibitem[\protect\citeauthoryear{{Robertson}, {Endl}, {Henry}, {Cochran},
  {MacQueen}  \& {Williamson}}{{Robertson} et~al.}{2015}]{Robertson15}
{Robertson} P.,  {Endl} M.,  {Henry} G.~W.,  {Cochran} W.~D.,  {MacQueen}
  P.~J.,   {Williamson} M.~H.,  2015, \mn@doi [\apj]
  {10.1088/0004-637X/801/2/79}, \href
  {http://adsabs.harvard.edu/abs/2015ApJ...801...79R} {801, 79}

\bibitem[\protect\citeauthoryear{{Saar}}{{Saar}}{1988}]{Saar88}
{Saar} S.~H.,  1988, \mn@doi [\apj] {10.1086/165907}, \href
  {http://adsabs.harvard.edu/abs/1988ApJ...324..441S} {324, 441}

\bibitem[\protect\citeauthoryear{{Scandariato} et~al.,}{{Scandariato}
  et~al.}{2017}]{Scandariato17}
{Scandariato} G.,  et~al., 2017, \mn@doi [\aap] {10.1051/0004-6361/201629382},
  \href {http://esoads.eso.org/abs/2017A%26A...598A..28S} {598, A28}

\bibitem[\protect\citeauthoryear{{Shkolnik} \& {Barman}}{{Shkolnik} \&
  {Barman}}{2014}]{Shkolnik14}
{Shkolnik} E.~L.,  {Barman} T.~S.,  2014, \mn@doi [\aj]
  {10.1088/0004-6256/148/4/64}, \href
  {http://cdsads.u-strasbg.fr/abs/2014AJ....148...64S} {148, 64}

\bibitem[\protect\citeauthoryear{{Shulyak}, {Reiners}, {Seemann}, {Kochukhov}
  \& {Piskunov}}{{Shulyak} et~al.}{2014}]{Shulyak14}
{Shulyak} D.,  {Reiners} A.,  {Seemann} U.,  {Kochukhov} O.,   {Piskunov} N.,
  2014, \mn@doi [\aap] {10.1051/0004-6361/201322136}, \href
  {http://adsabs.harvard.edu/abs/2014A%26A...563A..35S} {563, A35}

\bibitem[\protect\citeauthoryear{{Shulyak}, {Reiners}, {Engeln}, {Malo},
  {Yadav}, {Morin}  \& {Kochukhov}}{{Shulyak} et~al.}{2017}]{Shulyak17}
{Shulyak} D.,  {Reiners} A.,  {Engeln} A.,  {Malo} L.,  {Yadav} R.,  {Morin}
  J.,   {Kochukhov} O.,  2017, \mn@doi [Nature Astronomy]
  {10.1038/s41550-017-0184}, \href
  {http://esoads.eso.org/abs/2017NatAs...1E.184S} {1, 0184}

\bibitem[\protect\citeauthoryear{{Skilling} \& {Bryan}}{{Skilling} \&
  {Bryan}}{1984}]{Skilling84}
{Skilling} J.,  {Bryan} R.~K.,  1984, \mnras, \href
  {http://cdsads.u-strasbg.fr/abs/1984MNRAS.211..111S} {211, 111}

\bibitem[\protect\citeauthoryear{{Skrutskie} et~al.,}{{Skrutskie}
  et~al.}{2006}]{Skrutskie06}
{Skrutskie} M.~F.,  et~al., 2006, \mn@doi [\aj] {10.1086/498708}, \href
  {http://esoads.eso.org/abs/2006AJ....131.1163S} {131, 1163}

\bibitem[\protect\citeauthoryear{{Stelzer}, {Marino}, {Micela},
  {L{\'o}pez-Santiago}  \& {Liefke}}{{Stelzer} et~al.}{2013}]{Stelzer13}
{Stelzer} B.,  {Marino} A.,  {Micela} G.,  {L{\'o}pez-Santiago} J.,   {Liefke}
  C.,  2013, \mn@doi [\mnras] {10.1093/mnras/stt225}, \href
  {http://adsabs.harvard.edu/abs/2013MNRAS.431.2063S} {431, 2063}

\bibitem[\protect\citeauthoryear{{Su{\'a}rez Mascare{\~n}o}, {Rebolo},
  {Gonz{\'a}lez Hern{\'a}ndez}  \& {Esposito}}{{Su{\'a}rez Mascare{\~n}o}
  et~al.}{2015}]{Suarez15}
{Su{\'a}rez Mascare{\~n}o} A.,  {Rebolo} R.,  {Gonz{\'a}lez Hern{\'a}ndez}
  J.~I.,   {Esposito} M.,  2015, \mn@doi [\mnras] {10.1093/mnras/stv1441},
  \href {http://esoads.eso.org/abs/2015MNRAS.452.2745S} {452, 2745}

\bibitem[\protect\citeauthoryear{{Su{\'a}rez Mascare{\~n}o}
  et~al.,}{{Su{\'a}rez Mascare{\~n}o} et~al.}{2017a}]{Suarez17c}
{Su{\'a}rez Mascare{\~n}o} A.,  et~al., 2017a, preprint, \href
  {http://adsabs.harvard.edu/abs/2017arXiv170506537S} {} (\mn@eprint {arXiv}
  {1705.06537})

\bibitem[\protect\citeauthoryear{{Su{\'a}rez Mascare{\~n}o}, {Rebolo},
  {Gonz{\'a}lez Hern{\'a}ndez}  \& {Esposito}}{{Su{\'a}rez Mascare{\~n}o}
  et~al.}{2017b}]{Suarez17b}
{Su{\'a}rez Mascare{\~n}o} A.,  {Rebolo} R.,  {Gonz{\'a}lez Hern{\'a}ndez}
  J.~I.,   {Esposito} M.,  2017b, \mn@doi [\mnras] {10.1093/mnras/stx771},
  \href {http://esoads.eso.org/abs/2017MNRAS.468.4772S} {468, 4772}

\bibitem[\protect\citeauthoryear{{Su{\'a}rez Mascare{\~n}o}
  et~al.,}{{Su{\'a}rez Mascare{\~n}o} et~al.}{2017c}]{Suarez536}
{Su{\'a}rez Mascare{\~n}o} A.,  et~al., 2017c, \mn@doi [\aap]
  {10.1051/0004-6361/201629291}, \href
  {http://esoads.eso.org/abs/2017A%26A...597A.108S} {597, A108}

\bibitem[\protect\citeauthoryear{{Vidotto}, {Jardine}, {Morin}, {Donati}  \&
  {Opher}}{{Vidotto} et~al.}{2014}]{Vidotto14}
{Vidotto} A.,  {Jardine} M.,  {Morin} J.,  {Donati} J.,   {Opher} M.,  2014,
  \mnras, 438, 1162

\bibitem[\protect\citeauthoryear{{Wright} \& {Drake}}{{Wright} \&
  {Drake}}{2016}]{Wright16}
{Wright} N.~J.,  {Drake} J.~J.,  2016, \mn@doi [\nat] {10.1038/nature18638},
  \href {http://cdsads.u-strasbg.fr/abs/2016Natur.535..526W} {535, 526}

\bibitem[\protect\citeauthoryear{{Yu} et~al.,}{{Yu} et~al.}{2017}]{Yu17}
{Yu} L.,  et~al., 2017, \mn@doi [\mnras] {10.1093/mnras/stx009}, \href
  {http://cdsads.u-strasbg.fr/abs/2017MNRAS.tmp...26Y} {}

\makeatother
\end{thebibliography}


\appendix

\section{Observing log of GJ~1289, GJ~793, and GJ~251 and period searches}

Table \ref{tab:journal_obs} gives the log of ESPaDOnS observations for GJ~1289 and GJ~793 in 2016 that were used to reconstruct the magnetic field maps shown in the article.\\

\begin{table*}
\begin{center}
\caption{Journal of observations for GJ~1289 (top), GJ~793 (middle), and GJ~251 (bottom). Columns 1 to 4, respectively, list the rotational cycle according to ephemeris given in the text, the date of the beginning of the night, the Julian Date, the peak S/N in Stokes $V$ spectra (per 2.6~\kms\ velocity bin at 871~nm). Column 5-6 respectively give \bl\ and RV values.}
\label{tab:journal_obs}
\begin{tabular}{cccccccc}
\hline
Cycle  & Date   &   HJD          & S/N     &  \bl & RV\\
Rot    & 2016   &  (+ 2 457 000) &         &(G)   & (\kms)\\
\hline
0.0001 & 06Aug16 & 607.0190 &  243 & 106.67 $\pm$ 10.74 & -2.45 \\
0.0360 & 08Aug16 & 608.9610 &  215 & 129.08 $\pm$ 12.29 & -2.48 \\
0.0762 & 10Aug16 & 611.1290 &  215 & 115.44 $\pm$ 12.24 & -2.44 \\
0.1134 & 12Aug16 & 613.1390 &  218 & 83.66 $\pm$ 12.08 & -2.47 \\
0.1298 & 13Aug16 & 614.0260 &  214 & 122.10 $\pm$ 12.19 & -2.46 \\
0.1661 & 15Aug16 & 615.9830 &  260 & 124.90 $\pm$ 9.94 & -2.45 \\
0.7233 & 14Sep16 & 646.0710 &  213 & -22.55 $\pm$ 12.58 & -2.56 \\
0.7949 & 18Sep16 & 649.9420 &  226 & 24.13 $\pm$ 11.52 & -2.56 \\
0.8128 & 19Sep16 & 650.9080 &  220 & -4.70 $\pm$ 11.73 & -2.60 \\
0.8298 & 20Sep16 & 651.8260 &  217 & 23.25 $\pm$ 11.88 & -2.59 \\
0.8476 & 21Sep16 & 652.7880 &  220 & 20.92 $\pm$ 11.74 & -2.54 \\
0.8687 & 22Sep16 & 653.9220 &  217 & -5.37 $\pm$ 11.67 & -2.69 \\
1.3325 & 17Oct16 & 678.9670 &  175 & 44.63 $\pm$ 15.55 & -2.50 \\
1.3330 & 17Oct16 & 678.9960 &  180 & 5.17 $\pm$ 15.62 & -2.53 \\
1.3682 & 19Oct16 & 680.8950 &  131 & 15.56 $\pm$ 21.70 & -2.47 \\
1.3688 & 19Oct16 & 680.9310 &  176 & 33.65 $\pm$ 14.88 & -2.51 \\
1.3882 & 20Oct16 & 681.9790 &  223 & 2.56 $\pm$ 11.75 & -2.53 \\
1.4044 & 21Oct16 & 682.8520 &  221 & 30.34 $\pm$ 11.82 & -2.51 \\
\hline
\hline
0.0002 & 03Aug16 & 603.9590 &  267 & 10.47  $\pm$ 7.69 & 10.68 \\
0.1373 & 06Aug16 & 606.9770 &  260 & -8.62  $\pm$ 8.09 & 10.72 \\
0.1811 & 07Aug16 & 607.9410 &  269 & -18.62 $\pm$ 7.72 & 10.70 \\
0.2266 & 08Aug16 & 608.9410 &  270 & -21.39 $\pm$ 7.72 & 10.69 \\
0.2736 & 09Aug16 & 609.9740 &  267 & -13.45 $\pm$ 7.73 & 10.70 \\
0.3190 & 10Aug16 & 610.9740 &  266 & -14.44 $\pm$ 7.75 & 10.70 \\
0.3603 & 11Aug16 & 611.8810 &  269 & -20.23 $\pm$ 7.63 & 10.71 \\
0.4067 & 12Aug16 & 612.9030 &  297 & -10.80 $\pm$ 6.88 & 10.70 \\
0.4548 & 13Aug16 & 613.9620 &  267 & -11.72 $\pm$ 7.67 & 10.72 \\
0.5851 & 16Aug16 & 616.8280 &  295 & -1.17  $\pm$ 6.92 & 10.67 \\
2.1321 & 19Sep16 & 650.8620 &  301 & -0.44  $\pm$ 6.72 & 10.69 \\
2.1727 & 20Sep16 & 651.7550 &  291 & -9.170 $\pm$ 7.06 & 10.65 \\
3.1714 & 12Oct16 & 673.7270 &  312 & -5.78  $\pm$ 6.50 & 10.69 \\
3.2203 & 13Oct16 & 674.8020 &  274 & -20.22 $\pm$ 7.65 & 10.72 \\
3.2609 & 14Oct16 & 675.6960 &  285 & -21.33 $\pm$ 7.34 & 10.69 \\
3.3064 & 15Oct16 & 676.6950 &  317 & -12.67 $\pm$ 6.45 & 10.69 \\
3.3545 & 16Oct16 & 677.7540 &  284 & -20.95 $\pm$ 7.28 & 10.72 \\
3.4002 & 17Oct16 & 678.7590 &  255 & -17.10 $\pm$ 8.19 & 10.71 \\
3.4925 & 19Oct16 & 680.7910 &  311 & -18.24 $\pm$ 6.52 & 10.69 \\
3.5833 & 21Oct16 & 682.7890 &  318 & -8.431 $\pm$ 6.37 & 10.70 \\
\hline
\hline
0.0000 & 13sep14 & 914.6430 & 237   & 15.15  $\pm$ 5.03 & 23.11 \\
0.0001 & 13sep14 & 914.6480 & 236 & 5.983 $\pm$ 5.03 & 23.13 \\
0.0222 & 15sep14 & 916.6410 & 218   & 21.22 $\pm$ 5.53 & 23.09 \\
0.0223 & 15sep14 & 916.6460 & 231 & 16.13 $\pm$ 5.18 & 23.09 \\
0.6436 & 10nov14 & 972.5710 & 239   & 2.746  $\pm$ 5.09 & 22.99 \\
1.0769 & 19dec14 & 1011.5700 & 203  & 12.66  $\pm$ 6.58 & 23.07 \\
1.0770 & 19dec14 & 1011.5700 & 203& 4.125  $\pm$ 6.39 & 23.06 \\
1.0879 & 20dec14 & 1012.5500 & 228  & 1.033  $\pm$ 5.53 & 23.05 \\
1.0997 & 21dec14 & 1013.6200 & 220  & 0.599  $\pm$ 5.73 & 23.05 \\
1.1772 & 28dec14 & 1020.5900 & 222  & 11.47 $\pm$ 5.68 & 22.99 \\
1.2882 & 07jan15 & 1030.5800 & 231  & 9.20   $\pm$ 5.54 & 22.95 \\
1.3544 & 13jan15 & 1036.5400 & 202  & 10.86 $\pm$ 6.52 & 22.99 \\
10.718 & 28oct15 & 1324.4700 & 221  & 11.50 $\pm$ 5.70 & 23.02 \\
10.762 & 27mar16 & 1475.2900 & 239  & 8.173  $\pm$ 5.62 & 23.09 \\
4.5537  & 20jan17 & 1774.4600 & 255 & 9.973 $\pm$ 6.59 & 23.02 \\
6.2294  & 22jan17 & 1776.4600 & 228 & 6.080  $\pm$ 6.34 & 23.09 \\
9.5535  & 13feb17 & 1798.3900 & 263 & 10.10 $\pm$ 6.54 & 22.93 \\
9.5758  & 14feb17 & 1799.4300 & 265 & 27.56  $\pm$ 7.59 & 22.96 \\
9.8194  & 15feb17 & 1800.3900 & 252 & 23.27 $\pm$ 6.49 & 22.99 \\
9.8310  & 16feb17 & 1801.4000 & 246 & 17.35  $\pm$ 6.49 & 23.01 \\
9.8416  & 05may17 & 1879.2600 & 256 & 6.203  $\pm$ 6.93 & 23.02 \\
9.8528  & 09may17 & 1883.2300 & 227 & 20.73 $\pm$ 7.04 & 23.02 \\
\hline
\end{tabular}
\end{center}
\end{table*}

\begin{figure*}
\includegraphics[width=18cm, height=10cm,angle=0]{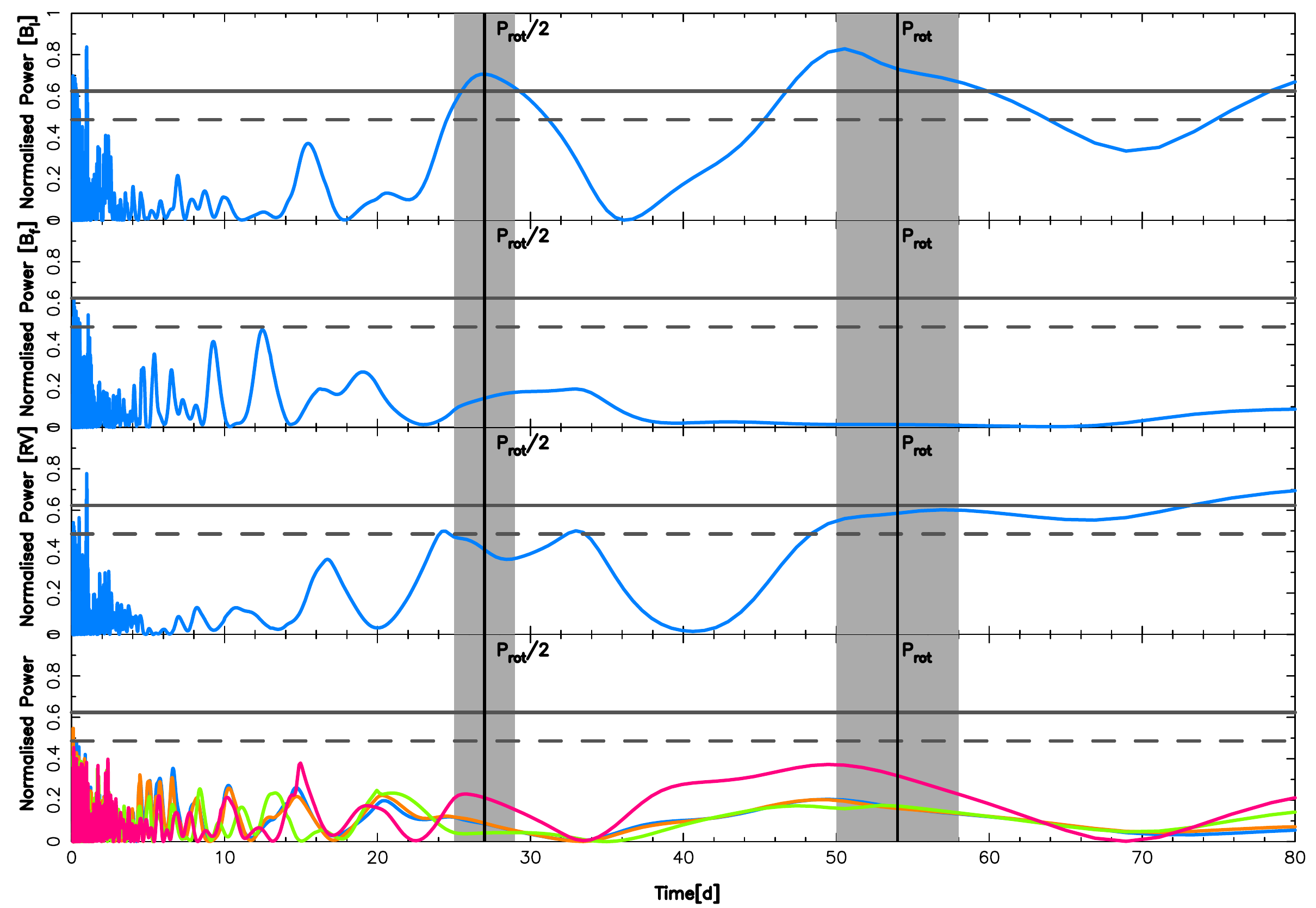}
\includegraphics[width=18cm, height=10cm,angle=0]{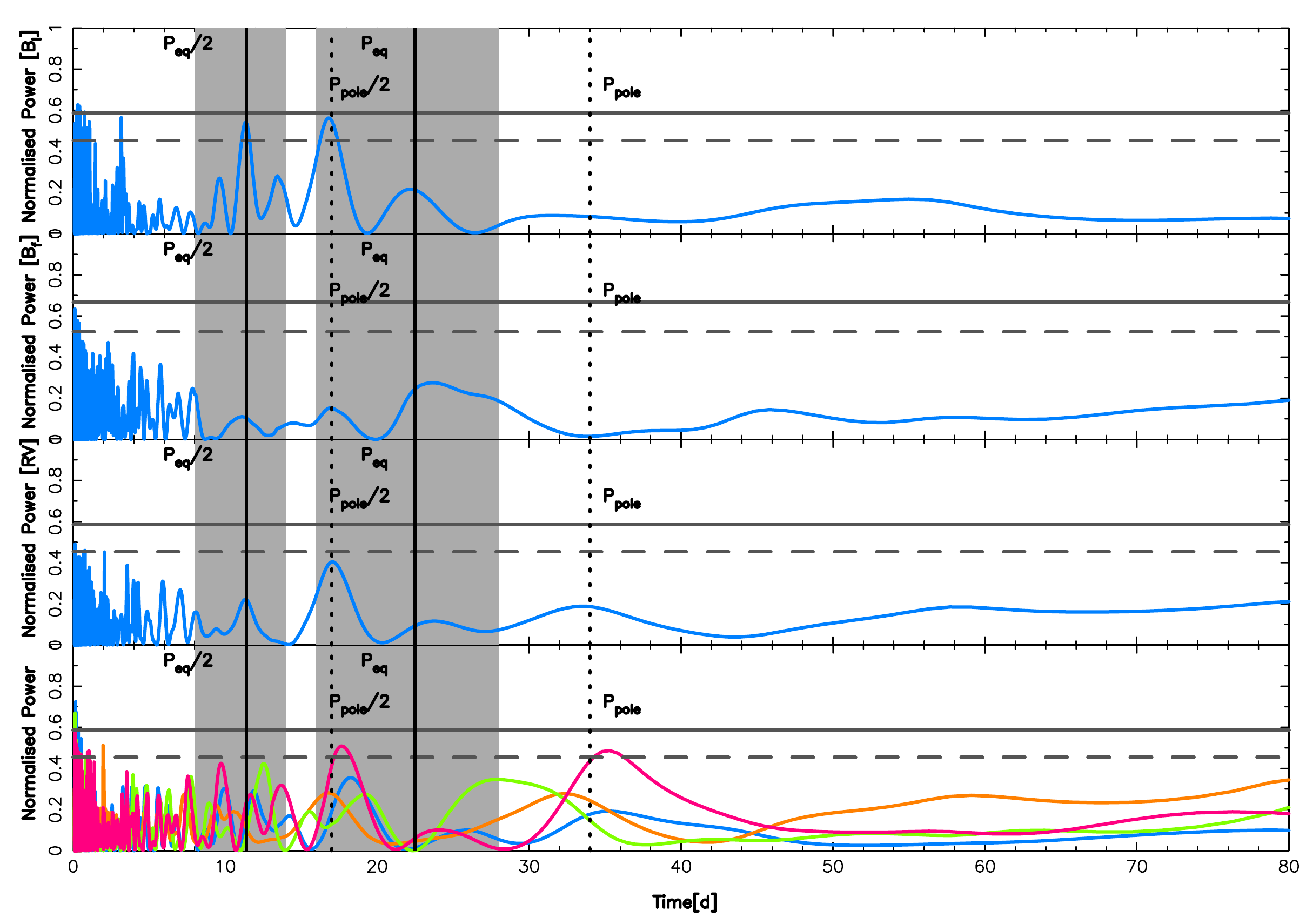}
\caption{\textit{Top :}  Generalised Lomb-Scargle periodogram of GJ~1289 longitudinal magnetic field (top panel), B$f$ (second panel), raw RVs (third panel), and $S$, H$\alpha$, HeI and NaD activity indices (bottom panel, resp. in pink, blue, orange and green). On each periodogram the False Alarm Probability at 1\% (solid line) and at 10\% (dashed line) are represented in black. The rotation period and its first harmonics are shown in vertical solid black line. 
\textit{Bottom:} Same for GJ~793. 
Here, the vertical red line represents the best-fit equator rotation period ($\sim$22 days) and its first harmonics. The grey bands include the rotation at high latitudes on this star where differential rotation seem to be detected.}
\label{fig:perio}
\end{figure*}

Periodograms used in the analysis of GJ~1289 and GJ~793 are shown on Figure \ref{fig:perio}, for $B_l$, $Bf$, RV and some of the activity indices. The adopted rotation period is shown as the grey area, together with the first harmonics. For GJ~793, a range of periods appears in the periodogram of the longitudinal field, chromospheric indicators, and also the radial velocities. It is indicative of differential rotation (see Section 4.4.2). For GJ~1289, the $Bf$ shows a modulation at about 12.5 days at 10\% FAP, while for GJ~793, there is no significant peak in the periodogram of $Bf$. This quantity does not seem to show modulation with the rotation of the star, for the few examples shown here.

\section{Additional activity data}
\subsection{S$_{HK}$ and rotation-period calibrations  }
\begin{figure}
	\includegraphics[width=\columnwidth]{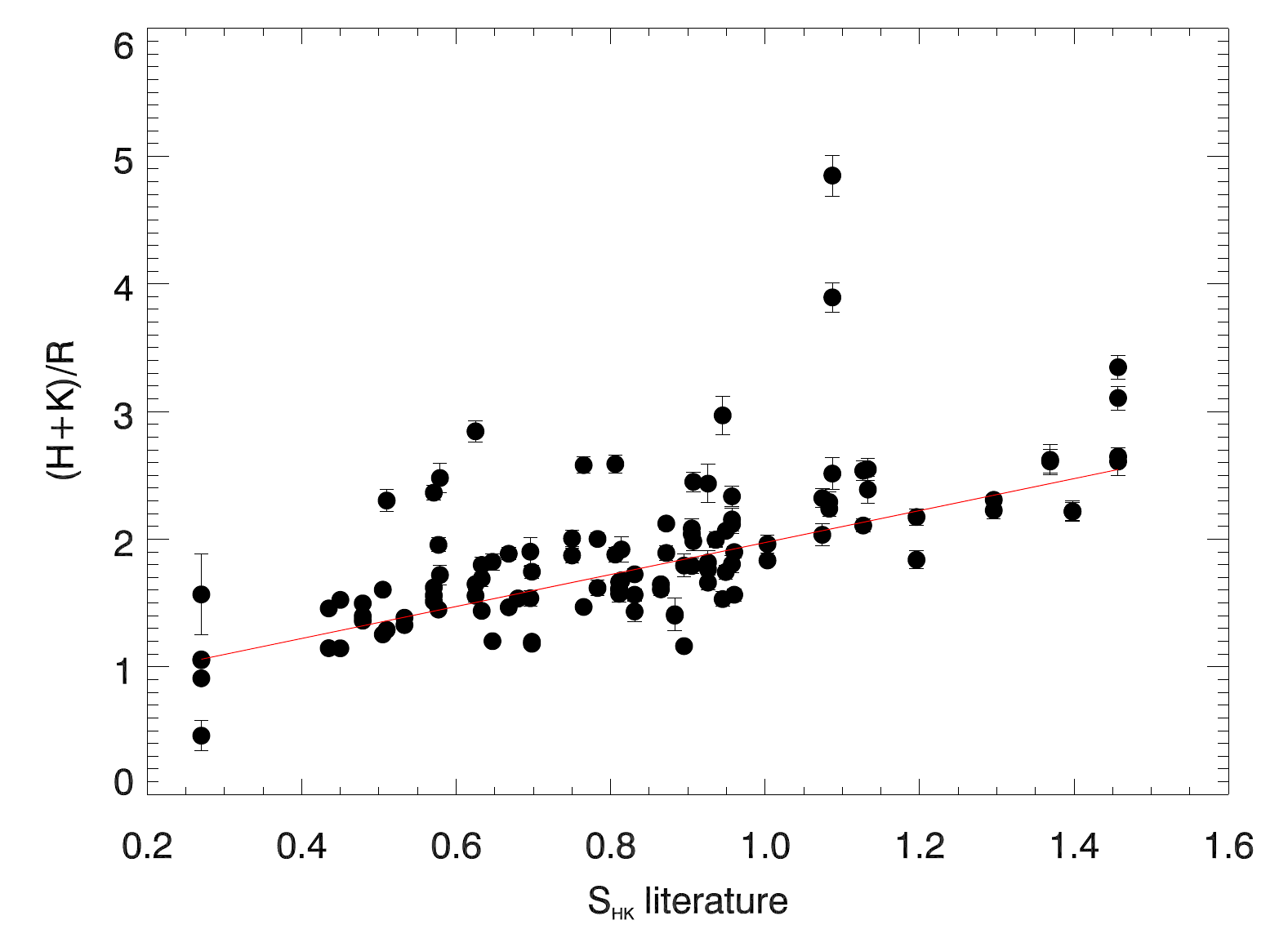}
    \caption{Calibration of S$_{HK}$ with literature values}
    \label{fig:calib}
\end{figure}

The calibration of the S$_{HK}$ index with literature values has been made using 114 spectra and 54 different stars, using the windows defined in Table \ref{tab:index}. As it is expected that S$_{HK}$ varies with time, using all available spectra for each given star allows to include some natural error in the calibration law. Reference values from the literature were taken from \citet{Astudillo17a}. Figure \ref{fig:calib} shows the calibration.

We then used the following equation to derive the $S$ index:
\begin{equation}
S=((F_H+F_K)/F_R-0.606)/1.443
\end{equation}
where $F_H$, $F_K$ and $F_R$ are, respectively, the flux integrated over the CaII H and K band, and the red continuum.

\begin{figure}
	\includegraphics[width=\columnwidth]{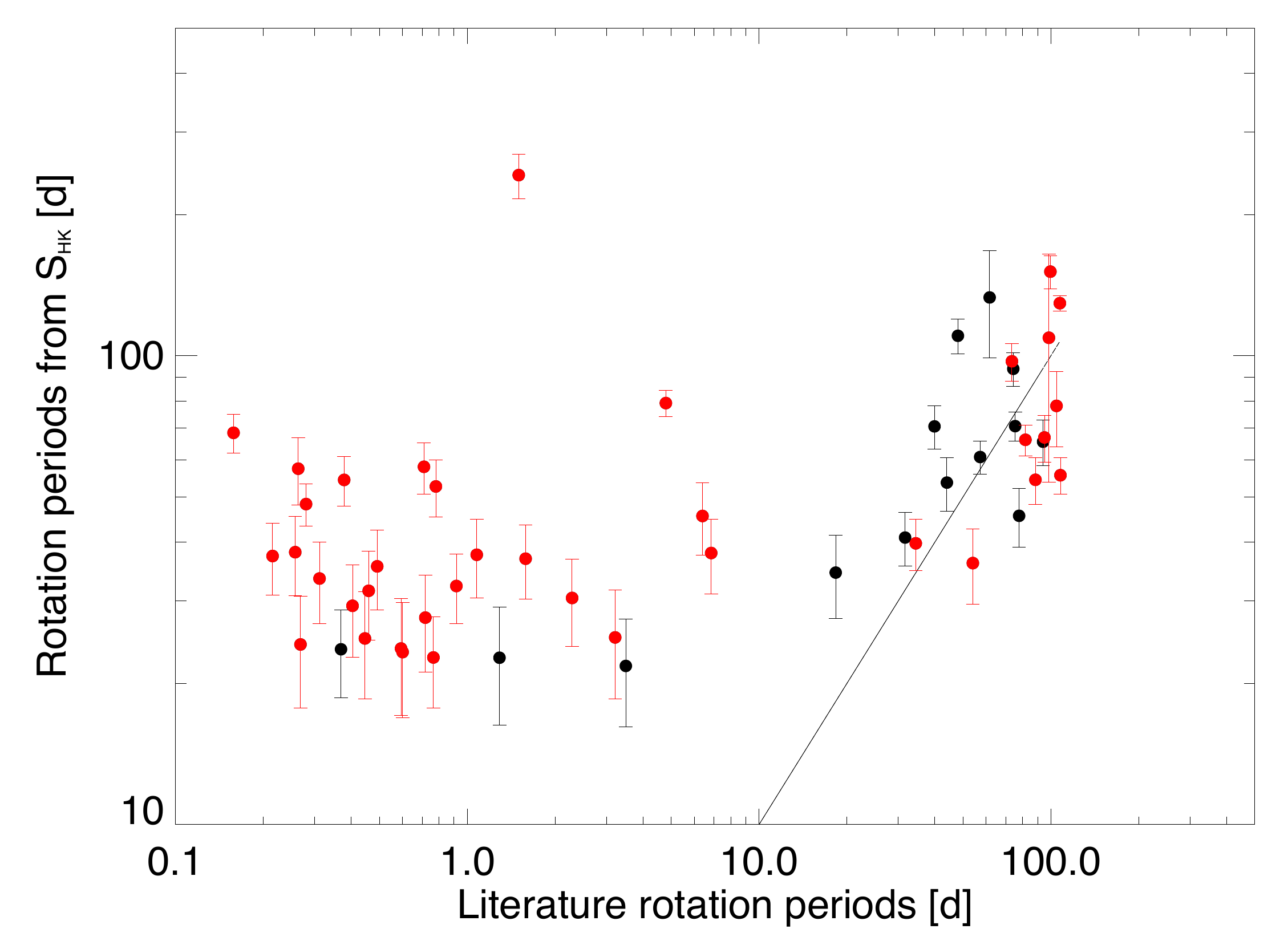}
    \caption{Comparison of rotation periods obtained from the log($R'_{HK}$) with literature values. Red symbols indicate stars with $V-K$ greater than 5.0. The identity line is shown. The plot illustrates how unreliable the calibration of rotational periods from log($R'_{HK}$) is for fast rotators, especially for late-type M dwarfs.  \label{fig:prot}}
\end{figure}

As another quality control for our measurements, we have compared the rotation periods obtained from the calibration of the log($R'_{HK}$) values adapted from \citet{Astudillo17a} with available literature values. The latter includes periods obtained from ZDI analysis \citep{Donati08c,Morin10,Hebrard16}, from photometric monitoring \citep{Kiraga07,Kiraga12,Newton16}, from statistically-significant averages of CaII HK values \citep{Astudillo17a,Suarez15,Suarez17b}, or from $v \sin i$ \citep{Houdebine10,Houdebine12,Houdebine15}, the first two methods being the most robust and the most scarce. The comparison, shown in Figure \ref{fig:prot}, shows how the log($R'_{HK}$)/rotation period calibration adapted from \citet{Astudillo17a} fails at detecting fast rotators, especially for the later type M stars (red dots). For periods larger than 10 days, the match is better although the scatter is large. 

\subsection{Trends, correlations and data table}

\begin{figure}
	\includegraphics[width=\columnwidth]{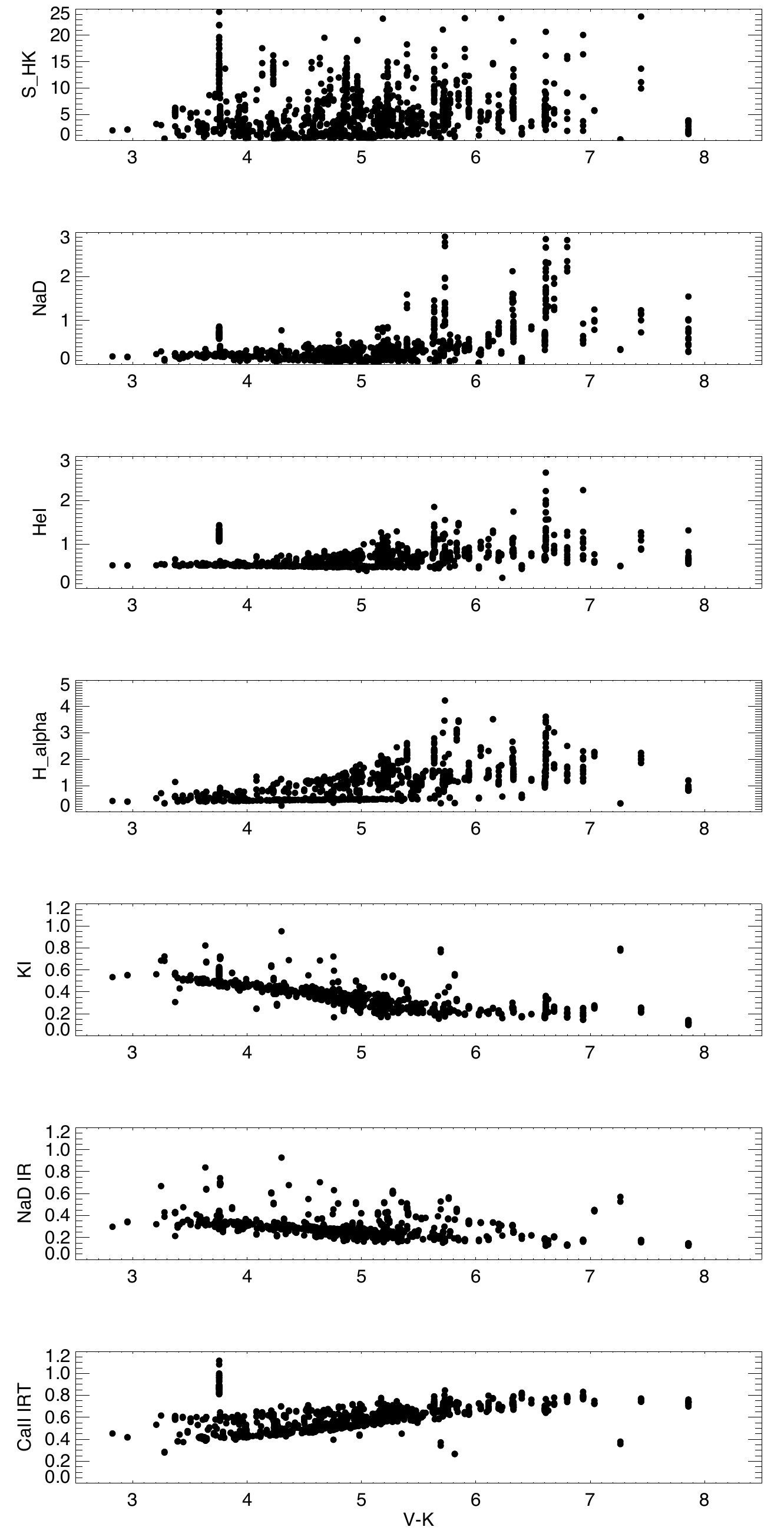}
    \caption{Variations of all activity indices with respect to $V-K$ colour index.}
    \label{fig:act11}
\end{figure}

We show the trend of all measured spectroscopic indices with the colour index (taken as a proxy for spectral type) on Figure \ref{fig:act11}. Each symbol on these plots represents a spectrum rather than a star. As some stars in the sample are extremely variable, it is then possible to consider each individual spectrum as a different  configuration of both the stellar surfaces and chromospheres. Specific stars with many visits (as V374 Peg with 110 individual spectra, a star originally studied in \citet{Donati06a}) appear with vertical lines, as their $V-K$ is constant.

\begin{figure}
	\includegraphics[width=\columnwidth]{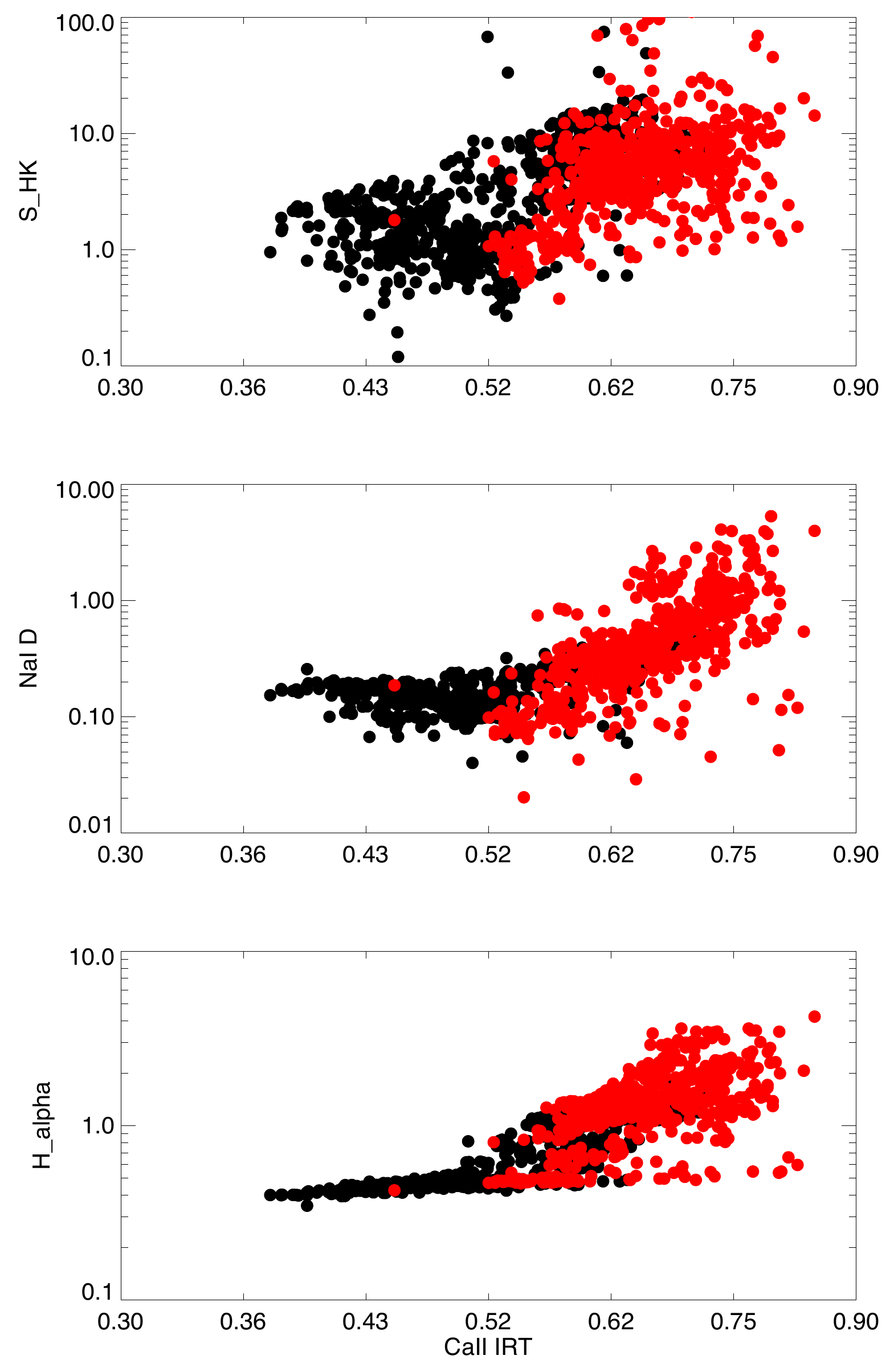}
    \caption{Variations of CaII HK, NaI D and H$\alpha$ activity indices with respect to the CaII IRT index. Black (red) symbols show stars bluer (respectively, redder) than $V-K$ of 5.0. }\label{fig:act10}
\end{figure}

As shown on Figure \ref{fig:act11}, some activity indices  have a positive trend with $V-K$ while others have a negative trend or no trend. The KI Doublet near 767nm shows the most pronounced negative trend as the CaII IRT shows the clearest positive trend. All other indices have much larger dispersions: CaII HK $S$ index shows strong dispersions for all $V-K$ while NaD, HeI, H$\alpha$ and NaI IR mostly show dispersion for the redder stars.  On Figure \ref{fig:act10} we show how activity indices NaD, H$_\alpha$, and CaII HK vary against the redder CaII IRT index. Here, the sample has been divided in $V-K$, to enhance the fact that the coolest M stars (red symbols, later than a spectral type of $\sim$M4) in our sample systematically have higher activity index ranges than early M dwarfs (black symbols). Table \ref{tab:pearson} summarizes the correlation coefficients between the various features. The correlation coefficients are also calculated separately for stars bluer and redder than $V-K=5$ (spectral types earlier/later than $\sim$ M4). Some discrepancies are notable: the correlation between the CaII IRT and HeI, NaD and H$\alpha$ is always significantly higher for early M dwarfs than for stars of later types.\\

\begin{figure*}
\centering
\includegraphics[width=2\columnwidth]{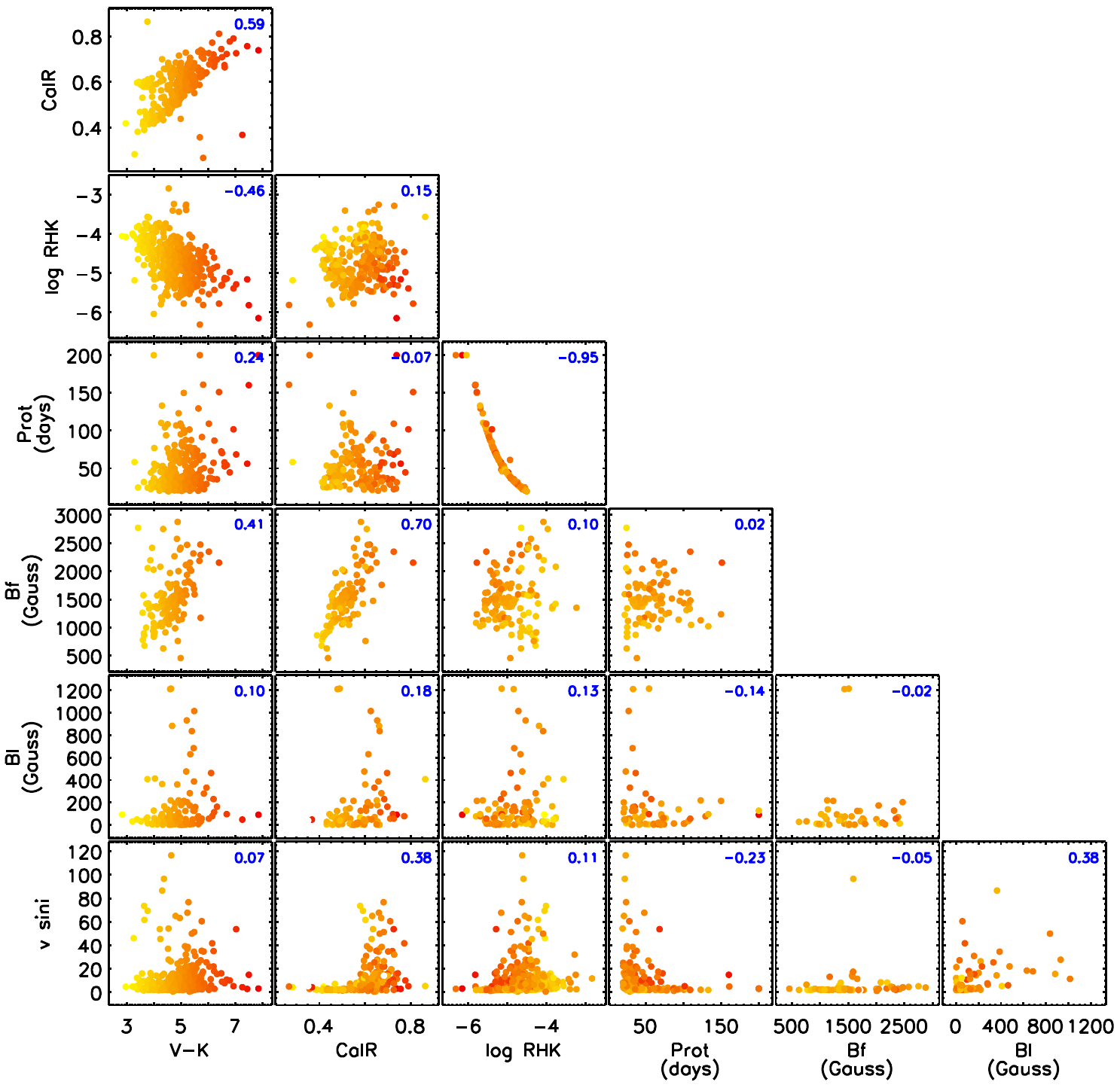}
\caption{The relationships between the main parameters of our data catalog are shown, including: $V-K$, CaII IRT, log($R'_{HK}$), Prot (from CaII HK imperfect calibration, see main text), $<Bf>$, the maximum of |$B_l$|, and \vsini. The correlation coefficient for each pair of parameters is given in the corner. The colour coding indicates the stellar colour (the reddest, the coolest).}
\label{tri}
\end{figure*}

Finally, the triangular plot Fig. \ref{tri} shows the relations between most parameters in our data set, with one average measurement per star. The colour coding shows the $V-K$ colour of the stars (the reddest the symbol, the coolest the star) and the correlation coefficient is shown in the corner of each plot. These coefficients include all stars, with no cut in SNR or spectroscopic binaries; hence some values differ from those quoted in the text, where these cuts have been applied.

Online table \ref{tab:jitterderived} summarizes the activity parameters for the stars where the Zeeman broadening is measured and there is no value for the RV jitter in the literature. The activity merit function $AMF$ as described in the text is also listed. \\


\begin{table}
\begin{center}
\caption{Pearson coefficients of correlation between various activity indices. These coefficients are calculated on individual spectra of all stars, after excluding low-SNR values (SNR$<$30) spectra and spectroscopic binary systems.}
\label{tab:pearson}
\begin{tabular}{lccc}
\hline
Line pair     & all stars  & $V-K < 5.0$ & $V-K \geq 5.0$ \\
\hline
CaIRT/CaII HK   &    0.27    &    0.07    &  0.03     \\
CaIRT/HeI       &    0.66    &    0.74    &  0.50     \\
CaIRT/NaD       &    0.47    &    0.73    &  0.43     \\
CaIRT/H$\alpha$ &    0.62    &    0.82    &  0.58     \\
CaIRT/KI       &   -0.33    &   -0.45    & -0.54     \\
CaIRT/NaI IR    &   -0.08    &   -0.18    & -0.30     \\
CaII HK/H$\alpha$&   0.20    &   0.09    &   0.06    \\
CaII HK/NaD     &    0.10    &   0.09    &   0.04    \\
H$\alpha$/NaD   &    0.21    &   0.82    &   0.37    \\
H$\alpha$/HeI   &    0.54    &   0.96    &   0.79    \\
KI/NaI IR      &    0.77    &    0.81    & 0.72      \\
\hline
\end{tabular}
\end{center}
\end{table}

\onecolumn
\begin{longtable}{lccccccc}
\caption{Calculated $Bf$ values and measured or adopted values for the photometric index, projected rotational velocity, chromospheric emission $F_{chr}$, maximum of the absolute value of the longitudinal field, and Activity Merit Function.}
\label{tab:jitterderived}\\
\hline
2MASS  & Other & $V-K$ & $Bf$  & $v \sin i$& $F_{chr}$ & max(|$B_l$|) & $AMF$ \\
name   & name  &       & [kG]  &  [km/s]   &   nm      & [G]          &      \\
\hline
 J00182549+4401376&           &  5.11&  1.4&  2.3    &  2.40& 11.3& 26.61  \\
 J00210932+4456560&           &  4.97&  1.6& 14.3    & 11.60&    -&  6.98  \\
 J00294322+0112384&           &  3.97&  0.9&  $<$ 2.0&  3.61&114.0& 47.58  \\
 J00385879+3036583& Wolf 1056 &  4.39&  1.4&  $<$ 2.0&  2.70& 10.4& 50.89  \\
 J00570261+4505099&           &  4.73&  1.4&  2.1    &  3.17&  4.5& 60.70  \\
 J00582789-2751251&           &  4.88&  1.3&  2.5    &  2.52& 53.8& 46.61  \\
 J01012006+6121560& GJ~47     &  4.32&  1.5&  2.7    &  3.11& 15.78&  40.0  \\
 J01023213+7140475& GJ~48     &  4.59&  1.7&  $<$ 2.0&  2.72& 13.90&  50.0  \\
 J01023895+6220422&  GJ~49    &  4.19&  1.2&  $<$ 2.0&  5.05& -   & 30.0  \\
 J01123052-1659570&           &  5.68&  2.2&  3.4    &  4.74&119.0& 16.70  \\
 J01432015+0419172&   GJ~70   &  4.33&  1.4&  $<$ 2.0&  2.84& 73.9& 37.54  \\
 J01515108+6426060&           &  4.48&  1.1&  2.2    &  3.86&  5.9&103.14  \\
 J01591239+0331092&           &  3.88&  1.2&  $<$ 2.0&  3.20&    -& 46.84  \\
 J01592349+5831162&           &  5.19&  1.6& 13.4    &  9.33&    -&  7.52  \\
 J02001278+1303112&   TZ Ari  &  5.72&  2.3&  4.4    &  8.65& 52.6& 17.49  \\
 J02013533+6346118&  GJ~3126  &  4.63&  1.4&  2.2    &  2.89& 75.9& 51.00  \\
 J02441537+2531249&  GJ~109   & 4.62&   1.5 & $<$ 2.0&  2.56 & 56.3 & 38.4 \\ 
 J03143273+5926160&           &  4.60&  1.4&  $<$ 2.0&  3.35& 26.2& 35.55  \\
 J03392972+2458028&   KP Tau  &  4.90&  2.0&  3.0    &  5.00& 52.0& 22.34  \\
 J03531041+6234081&           &  4.50&  1.2&  $<$ 2.0&  2.30&210.5& 37.05  \\
 J04311147+5858375&           &  5.26&  1.8&  $<$ 2.0&  2.81& 58.0& 79.24  \\
 J04425586+2128230&           &  4.82&  1.8&  $<$ 2.0&  3.58& 17.2& 81.52  \\
 J04535004+1549156&           &  4.59&  1.9&  $<$ 2.0&  2.33&    -& 39.67  \\
 J05032009-1722245&           &  4.82&  1.6&  $<$ 2.0&  2.57& 30.9& 34.05  \\
 J05124223+1939566&           &  4.32&  1.2&  $<$ 2.0&  2.41& 20.8& 47.06  \\
 J05280015+0938382&           &  4.91&  1.1&  2.2    &  2.10&219.1& 31.68  \\
 J05345212+1352471&   GJ~3356 &  4.93&  1.6&  2.0    &  2.68& 38.7& 84.73  \\
 J05420897+1229252&           &  5.10&  1.2&  $<$ 2.0&  2.16&214.7& 35.32  \\
 J06000351+0242236&   GJ~3379 &  5.27&  1.9&  5.9    & 10.94& -   & 11.09 \\
 J06011106+5935508&   GJ~3378 &  5.10&  1.6&  $<$ 2.0&  2.88&  8.0& 51.56  \\
 J06101978+8206256&  GJ~226   &  4.37&  1.0 & $<$ 2.0&  2.59 &  7.6 &77.9\\
 J06544902+3316058&  GJ~251   &4.76&  1.2   & $<$ 2.0&  2.31 & 65.7 & 43.6\\
 J06573891+4951540&           &  3.86&  2.2&  8.4    &  6.90&    -& 10.91  \\
 J07284541-0317524& LHS 1920  &  4.73&  1.6&  2.0    &  2.43&  3.4& 45.67  \\
 J07315735+3613477&           &  4.98&  2.4&  2.7    &  10.0& 12.6& 21.34  \\
 J07320291+1719103& HIP 36637 &  3.62&  0.9&  3.0    &  4.08&    -& 51.21  \\
 J07345632+1445544&           &  4.75&  2.3&  4.8    &  9.88& 96.2& 15.59  \\
 J07384089-2113276&  LHS 1935 &  4.67&  1.5&  $<$ 2.0&  2.40& 55.2& 34.75  \\
 J07581269+4118134&           &  5.14&  1.8&  $<$ 2.0&  2.63& 15.3& 49.61  \\
 J08160798+0118091&           &  4.33&  1.5&  $<$ 2.0&  2.86& 73.1& 35.90  \\
 J09142298+5241125&  GJ~338A  &  3.64&  0.7 & $<$ 2.0&  4.07 &  4.8 & 41.9\\
 J09142485+5241118&  GJ~338B  &  3.58&  0.8 & 2.1    &  4.06 &   -  &     53.9\\
 J09304457+0019214& GJ~1125   &  4.85&  1.6&  $<$ 2.0&  2.69& 39.4& 49.36  \\
 J09360161-2139371&           &  4.45&  1.1&  2.5    &  2.11&    -& 36.02  \\
 J09423493+7002024&   GJ~360  &  4.51&  1.8&  2.1    &  6.08& 13.0& 28.84  \\
 J09560868+6247185&  GJ~373   &  3.79&  1.4&  2.5    &  4.87&    -& 28.18  \\
 J10112218+4927153&   GJ~380  &  3.64&  0.8&  2.3    &  3.66& 48.2& 36.78  \\
 J10123481+5703495&           &  4.83&  1.4&  2.8    &  2.49& 18.9& 41.13  \\
 J10141918+2104297& HIP 50156 &  3.82&  1.3&  5.5    &  9.39&    -& 16.21  \\
 J10285555+0050275&   GJ~393  &  4.28&  1.4&  $<$ 2.0&  2.86& 13.5& 35.68  \\
 J11032023+3558117&  GJ~411   &  4.27&  1.0 & $<$ 2.0&  2.03 & 61.9 & 48.7 \\
 J11000432+2249592&  GJ~408   &  4.57&  1.3&  $<$ 2.0&  3.19& 27.8& 36.83  \\
 J11110245+3026415&    GJ~414B&  4.98&  0.4&  3.1    &  3.56& 27.8& 36.89  \\
 J11115176+3332111&    GJ~3647&  4.84&  2.6&  4.6    &  7.85&619.0&  9.30  \\
 J11200526+6550470&     SZ UMa&  3.77&  2.0&  $<$ 2.0&  2.90& 15.3& 52.90  \\
 J11414471+4245072&    GJ~1148&  5.12&  1.6&  $<$ 2.0&  2.41&  3.6& 73.59  \\
 J11474143+7841283&     GJ~445&  4.82&  1.4&  $<$ 2.0&  2.26& 35.6& 33.13  \\
 J11474440+0048164&     GJ~447&  5.44&  1.7&  2.1    &  3.20& 42.3& 40.03  \\
 J11510737+3516188&   GJ~450  &  4.11&  1.6&  2.2    &  4.53& 19.9& 30.36  \\
 J12100559-1504156&  GJ~3707  &  5.22&  1.9&  $<$ 2.0&  2.70& 48.7& 72.93  \\
 J12385241+1141461&  GJ~480   &  4.80&  1.7 & $<$ 2.0&  3.13 & 11.7 & 55.7\\
 J12475664+0945050&     GJ~486&  5.02&  1.6&  $<$ 2.0&  2.28& 10.5& 54.50  \\
 J12574030+3513306&  GJ~490A  &  3.97&  1.4&  8.2    & 10.97&    -& 10.02  \\
 J13085124-0131075&           &  4.68&  1.7&  $<$ 2.0&  2.63&    -& 30.50  \\
 J13282106-0221365&           &  4.65&  1.2&  2.2    &  2.67& 45.2& 62.08  \\
 J13315057+2323203&           &  4.42&  1.4&  $<$ 2.0&  2.59& 12.6& 39.70  \\
 J13424328+3317255& GJ~3801   &     -&  1.7&  2.0    &  2.71& 15.6& 15.50  \\
 J13455074-1758047&           &  4.97&  1.5&  $<$ 2.0&  2.26& 10.0& 63.25  \\
 J13455096-1209502&           &  3.71&  1.0&  $<$ 2.0&  3.52& 34.9& 45.68  \\
 J14154197+5927274&           &  3.41&  2.8&  $<$ 2.0&  2.54&    -& 71.90  \\
 J14170294+3142472& GJ~3839   &  5.50&  1.6&  17.6   &  7.46&    -&  6.78  \\
 J14172437+4526401& GJ~541.2  &  3.61&  1.6&  $<$ 2.0&  3.14&    -& 71.12  \\
 J14341683-1231106&  GJ~555   &  5.37&  2.0&  $<$ 2.0&  2.79& 15.0& 84.96  \\
 J15215291+2058394&  GJ~9520  &  4.34&  2.7&  5.2    & 12.54& 78.3& 12.88  \\
 J15323737+4653048&           &  4.09&  1.4&  3.4    & 11.76&    -& 17.06  \\
 J15581883+3524236&           &  4.81&  1.20& $<$ 2.0&  2.53&    -& 34.55  \\
 J16164537+6715224&  GJ~617B  &  4.55&  1.0&  2.4    &  3.10& 41.9& 47.89  \\
 J16240913+4821112&   GJ~623  &  4.40&  1.3&  3.3    &  2.05&  9.3& 38.47  \\
 J16334161-0933116& HIP 81084 &  3.72&  1.4&  3.3    & 10.12& 56.2& 21.15  \\
 J16360563+0848491&  GJ~1204  &  5.32&  2.0&  3.0    &  4.08&173.6& 20.59  \\
 J16570570-0420559&           &  5.16&  1.6& 11.5    &  9.75&125.5&  7.57  \\
 J17093153+4340531&  GJ~3991  &  5.33&  2.1&  $<$ 2.0&  3.01& 11.0& 81.55  \\
 J17195422+2630030&  GJ~669A  &  4.96&  2.4&  3.2    &  6.92& 70.1& 22.48  \\
 J17362594+6820220&  GJ~687   &  4.61&  1.5 & $<$ 2.0&  2.44 &   -  & 48.5 \\
 J17375330+1835295&  GJ~686&   4.00&  0.9 & $<$ 2.0&  2.58 &   -  & 70.7 \\
 J17435595+4322441&  GJ~694&   4.55&  1.5 & $<$ 2.0&  3.39 &  6.8 & 70.7 \\
 J17574849+0441405&           &  4.99&  1.5&  3.1    &  2.46&238.8& 25.21  \\
 J17575096+4635182&  GJ~4040  &  4.74&  1.5&  2.0    &  3.16& 56.4& 36.94  \\
 J18021660+6415445&           &  5.72&  1.2& 13.2    &  6.01&    -&  7.66  \\
 J18061809+7249162&           &  3.58&  1.2&  $<$ 2.0&  3.24&    -& 71.02  \\
 J18151241-1924063&           &  4.34&  0.6&  2.1    &  3.62& 28.2& 38.04  \\
 J18172513+4822024&           &  4.42&  0.9&  3.1    &  7.15&    -& 21.09  \\
 J18415908+3149498& GJ~4070   &  4.55&  1.4&  $<$ 2.0&  2.68& 75.3& 40.72  \\
 J18424498+1354168&  GJ~4071  &  5.22&  2.5&  4.2    &  9.26&203.3& 14.49  \\
 J18424666+5937499&  GJ~725A  &  4.50&  1.3&  2.1    &  2.06&  9.2& 50.75  \\
 J18424688+5937374& HD 173740 &  4.78&  1.0&  2.4    &  2.45&  8.0& 31.99  \\
 J18441139+4814118&           &  4.73&  1.3&  2.7    & 69.57&    -& 21.23  \\
 J19071320+2052372&           &  4.25&  1.1&  2.1    &  1.71& 76.7& 39.67  \\
 J19082996+3216520&  GJ~4098  &  4.76&  1.5&  2.1    &  2.88&  9.1& 36.19  \\
 J20450949-3120266&   AU Mic  &  4.23&  2.1&  8.5    & 15.41& 71.9& 12.34  \\
 J20523304-1658289& HIP 103039&  5.32&  1.8&  $<$ 2.0&  3.15&  4.4& 35.12  \\
 J20564659-1026534&  GJ~811.1 &  4.65&  1.1&  2.0    &  2.99&101.6& 40.79  \\
 J21015865-0619070&   GJ~816  &  4.56&  1.4&  $<$ 2.0&  3.10& 49.7& 43.70  \\
 J21513828+5917383&           &  4.85&  1.74& $<$ 2.0&  3.48&  9.4& 67.22  \\
 J21514831+1336154&           &  5.48&  1.7&  2.1    &  3.66&    -& 28.02  \\
 J22022935-3704512&  GJ~4248  &  5.02&  1.5&  2.1    &  2.69& 16.2& 44.19  \\
 J22270871+7751579&           &  6.03&  2.3&  2.2    &  3.49&415.2& 28.56  \\
 J22554384-3022392& HIP 113221&  4.12&  1.3&  2.5    &  3.32& 79.5& 32.80  \\
 J20303207+6526586&  GJ~793   &  4.58&  1.7 & $<$ 2.0&  3.88 & 33.6 & 38.9 \\
 J23172807+1936469&           &  4.88&  0.7&  6.6    &  9.26&    -& 10.52  \\
 J23213752+1717284&    GJ~4333&  5.19&  2.1&  $<$ 2.0&  2.87&  7.9& 62.83  \\
 J23380819-1614100&   GJ~4352 &  4.29&  1.0&  2.1    &  1.90& 28.2& 43.34  \\
 J23415498+4410407&  GJ~905   &  6.40&  2.1&  $<$ 2.0&  3.24&104.1& 71.02  \\
 J23430628+3632132&   GJ~1289 &  5.45&  2.3&  2.7    &  5.40&128.0& 23.13  \\
 J23491255+0224037&  GJ~908   &  3.99&  1.1 & 2.2    &  2.33 &   -  & 35.7 \\
 J23583264+0739304&  GJ~4383  &  4.68&  1.4&  3.2    &  2.91& 12.4&  38.5  \\
\hline
\end{longtable}


\bsp	
\label{lastpage}
\end{document}